\documentclass[useAMS,usenatbib]{mn2e}
\usepackage{graphicx}
\usepackage{amsmath}
\usepackage{amssymb}
\usepackage{color}

\def \etal {et~al.~}

\newcommand{\be}{\begin{equation}}
\newcommand{\ee}{\end{equation}}

\newcommand{\ifm}[1]{\relax\ifmmode#1\else$\mathsurround=0pt #1$\fi}
\newcommand{\kms}{\ifmmode\,{\rm km}\,{\rm s}^{-1}\else km$\,$s$^{-1}$\fi}

\newcommand{\ltsima}{$\; \buildrel < \over \sim \;$}
\newcommand{\lsim}{\lower.5ex\hbox{\ltsima}}
\newcommand{\gtsima}{$\; \buildrel > \over \sim \;$}
\newcommand{\gsim}{\lower.5ex\hbox{\gtsima}}

\definecolor{green}{rgb}{0,0.5,0}
\definecolor{darkred}{rgb}{0.6,0,0}
\definecolor{grey}{rgb}{0.4,0.5,0.7}

\def\gtsima{$\; \buildrel > \over \sim \;$}
\def\ltsima{$\; \buildrel < \over \sim \;$}
\def\gta{\lower.7ex\hbox{\gtsima}}
\def\lta{\lower.7ex\hbox{\ltsima}}

\def\M11{M_{11}}
\def\V100{V_{100}}
\def\R1{R_{Mpc}}
\def\T6{T_6}

\begin{document}

\title{NIHAO XIX: How supernova feedback shapes the galaxy baryon cycle}
\pagerange{\pageref{firstpage}--\pageref{lastpage}} \pubyear{2018}
\date{Submitted for publication in MNRAS, 11 February 2018.}

\author[{\'E}. Tollet \etal]{{\'E}douard~Tollet$^{1,2}$\thanks{edouard.tollet@obspm.fr}, 
Andrea~Cattaneo$^{1}$,
Andrea~V.~Macci{\`o}$^{3,4}$,
\newauthor
Aaron~A.~Dutton$^3$ and 
Xi~Kang$^5$
\vspace*{0.2cm}
\\
$^1$ Observatoire de Paris, GEPI and LERMA, PSL University, 61 Avenue de l'observatoire, 75014 Paris, France\\
$^2$ Universit{\'e} Paris Diderot, Sorbonne Paris Cit{\'e}, Paris, France\\
$^3$ New York University of Abu Dhabi, P.O. Box 129188, Saadiyat Island, Abu Dhabi, United Arab Emirates\\
$^4$ Max Planck Institute f\"{u}r Astronomie, K\"{o}nigstuhl 17, D-69117 Heidelberg, Germany\\
$^5$ Purple Mountain Observatory, the Partner Group of MPI f{\"u}r Astronomie, 2 West Beijing Road, Nanjing 210008, China
}
\maketitle 

\label{firstpage}


\begin{abstract}

We have used the NIHAO simulations to explore how supernovae (SNe) affect star formation in galaxies.
We find that SN feedback operates on all scales from the interstellar medium (ISM) to several virial radii.
SNe regulate star formation by preventing condensation of {\sc Hi} into {\sc H}$_2$ and by moving cold neutral gas to the hot {\sc Hii} phase.
The first effect explains why the cold neutral gas in dwarf galaxies forms stars inefficiently. The second maintains the hot ISM of massive galaxies ({\sc Hii} vents out at lower masses).
At $v_{\rm vir}\gsim 67{\rm\,km\,s}^{-1}$, the outflow rate follows the relation: $\dot{M}_{\rm out}=23\,(v_{\rm vir}/67{\rm\,km\,s}^{-1})^{-4.6}\,{\rm SFR}$.
$20\%$ to $70\%$ of the gas expelled from galaxies escapes from the halo (ejective feedback) but outflows are dominated by cold swept-up gas, most of which
falls back onto the galaxy on a $\sim 1\,$Gyr timescale. This `fountain feedback' reduces the masses of galaxies by a factor of two to four, since gas spends
half to three quarter of its time in the fountain.
Less than $10\%$ of the ejected gas mixes with the hot circumgalactic medium and this gas is usually not reaccreted.
On scales as large as $6r_{\rm vir}$, galactic winds divert the incoming gas from cosmic filaments and prevent if from accreting onto galaxies (pre-emptive feedback). 
This process is the main reason for the low baryon content of ultradwarves.

\end{abstract} 



\section{Introduction}

Matching the abundances of galaxies and dark matter (DM) haloes \citep{leauthaud_etal12,papastergis_etal12,behroozi_etal13,moster_etal13,tollet_etal17},
the Tully-Fisher relation \citep{geha_etal06,pizagno_etal07,springob_etal07}
and the rotation curves of spiral galaxies \citep{persic_etal96}
show that a halo of mass $M_{\rm vir}\sim 10^{11}\,M_\odot$ contains a galaxy with a typical stellar mass of $M_\star\sim 3\times 10^8\,M_\odot$.
Many dwarf galaxies are rich in cold neutral gas due to inefficient star formation (e.g., \citealp{boselli_etal14};
also see the simulations by \citealp{munshi_etal13} and \citealp{hopkins_etal14}).
Its presence can push the baryonic galaxy mass up to $M_{\rm gal}\sim 10^9\,M_\odot$ \citep{papastergis_etal12}.
However, even with gas, the galaxy mass is still an order of magnitude lower than the one expected from the universal baryon fraction $f_{\rm b}=0.16$.

The most common explanation is that the rest of the baryons have been blown out by supernova (SN) feedback \citep{dekel_silk86}.
However, a purely ejective scenario requires mass-loading factors of $\eta\sim 50$. For $1\,M_\odot$ that forms stars, $50\,M_\odot$ of gas need to be expelled, and
the wind speed must be larger than the escape speed from the halo to prevent  gas from coming back.
These conditions are not easily met. 
Observed stellar outflows have typical mass-loading factors of order unity \citep{bouche_etal12,cicone_etal14,schroetter_etal15,schroetter_etal16} 
if not smaller \citep{kacprzak_etal14,pereira-santaella_etal16,falgarone_etal17}.

Hydrodynamic simulations, too, have for a long time struggled to produce systems with mass-loading factors $\eta\gg 1$.
Only in the present decade have they begun to form realistic disc galaxies and to reproduce the observed  $M_\star-M_{\rm vir}$ relation
\citep{guedes_etal11,brook_etal12,hopkins_etal14,shen_etal14,vogelsberger_etal14,schaye_etal15,crain_etal15,wang_etal15}.

Simulations by \citet{brook_etal14}, \citet{nelson_etal15} and \citet{mitchell_etal18} have shown
that the combined action of heating by a photoionising UV background and outflows can prevent gas from falling into 
the shallow gravitational potential wells of haloes with $M_{\rm vir}\ll 10^{11}\,M_\odot$.
However, galaxies with $M_{\rm vir}\gsim 10^{11}\,M_\odot$ do accrete a baryon mass of order $f_{\rm b}M_{\rm vir}$ and expel from the halo less than half that value.
Hence, massive spirals must expel a lot of gas that remains within the virial radius.
This gas is potentially available for reaccretion, and
\citet{brook_etal14} find that almost $40\%$ of the stellar mass at $z=0$ is composed of reaccreted baryons.

\citet{brook_etal14}'s $40\%$ figure is based on defining outflows as gas that flows out of a spherical surface of radius $r \sim 0.1r_{\rm vir}$.
\citet{nelson_etal15}, \citet{angles-alcazar_etal17} and \citet{mitchell_etal18} adopted similar criteria\footnote{ \citet{angles-alcazar_etal17} use two stellar half-mass radii,
which contain $80\%$ of the total stellar mass on average. For most galaxies, this criterion gives a $r<0.1r_{\rm vir}$. This may explain why
\citet{angles-alcazar_etal17} find higher recycled fractions ($60\%$ to $90\%$) than both \citet{brook_etal14} and \citet{christensen_etal16}.}.
However, gas may be heated by SNe, vent out of galactic discs, cool, and fall back onto galaxies without crossing this surface,
so that the real contribution of ejected and reaccreted baryons may be even larger.

\citet{christensen_etal16} added a density and temperature criteria to decide which gas particles belong to galaxies,
and they found that between $20\%$ and $70\%$ of the ejected gas was later reaccreted.
By tracking gas particles, they could also measure the average reaccretion timescale, which they found to be of the order of $1\,$Gyr.
However, they did not follow the evolution of the ejected gas in density and temperature.

Our work is based on a Numerical Investigation of a Hundred Astrophysical Objects (NIHAO, \citealp{wang_etal15}).
Besides the size our sample, its novelty is that we track the motions of gas particles in both physical and thermodynamic phase space (that is, on a density-temperature diagram).
Observed outflows have a multiphase structure with different speeds for gas that emits in CO, H$\alpha$ and X-rays 
(e.g., \citealp{mckeith_etal95,lehnert_etal99,strickland_heckman09,beirao_etal15,leroy_etal15} for the winds in M82).
Our approach allows us to separate the different thermodynamic phases of simulated outflows and to follow their individual reaccretion timescales.
We shall show that this separation is important for a meaningful comparison with outflow rates measured from observations.

The primary goal of our study is to disentangle the relative contributions of the different mechanisms and pathways through which stellar feedback reduces
the stellar masses of galaxies, since stellar feedback is not only a multiphase but also a multiscale phenomenon.
Our study has also the secondary goal to improve the description of feedback in semianalytic models of galaxy formation.

The article begins by presenting the NIHAO simulations (Section 2) and by describing the procedure used to assign particles to the different thermodynamic phases present within  haloes (Section 3).
Then, it presents our results for the galaxy baryon cycle, which we separate into
accretion (Section 4), star formation (Section 5), ejection (Sections~6 and 7) and recycling (Section 8).
Section~9 summarises the conclusions of the article.

\section{NIHAO simulations}

The simulations used for this work are from \citet{wang_etal15}'s Numerical Investigation of a Hundred Astrophysical Objects (NIHAO).
They were run with cosmological parameters from the \citet{Planck14}: 
Hubble constant $H_0=67.1\,\rm km\,s^{-1}\,\rm Mpc^{-1}$, 
matter density $\Omega_{\rm m}=0.3175$, 
dark-energy density $\Omega_{\rm \Lambda}=0.6824$, 
baryon density $\Omega_{\rm b}=0.049$, 
power-spectrum normalisation $\sigma_{\rm 8}=0.8344$, and power-spectrum slope $n=0.9624$. 

From three cubic N-body volumes of side lengths $15$, $20$ and $60\,h^{-1}$Mpc \citep{dutton_maccio14}, \citet{wang_etal15} selected 
$\sim$ 90 haloes with masses ranging from $10^{9.5}\,M_\odot$ to $10^{12.3}\,M_\odot$ discarding all haloes with a companion of comparable mass within $3r_{\rm vir}$ at any time.
The chosen haloes were resimulated at higher resolution with baryons.

At the highest level of refinement, the force softening length is $\simeq 50-80\,\rm pc$ for the gas particles and $\simeq 100-200\,\rm pc$ for the dark matter particles. The mass of the corresponding gas particles is of the order of $10^3\,\rm M_\odot$ (see Table 1 in \citealp{wang_etal15} for more details about the resolution of the simulations).

The initial conditions for the zoom simulations were computed using a modified version of the {\sc grafic2} code \citep{bertschinger_11,penzo_etal14}. 
The level of refinement was chosen to keep a roughly constant numerical resolution across the probed mass range. 
The hydrodynamic simulations were run with the {\sc gasoline2} smoothed-particle-hydrodynamics (SPH) code by \cite{wadsley_etal04,Wadsley2017}.

{\sc esf-gasoline2} contains a subgrid model for turbulent mixing of metals and energy \citep{wadsley_etal08}.
The calculation of heating and cooling includes
the effects of a photoionising UV background on metal-line cooling \citep{shen_etal10}.
\citet{ritchie_thomas01}'s implementation of SPH allows a more accurate description of the multiphase structure of the gas and of the destruction of cold clumps.
Limiting the timestep  when it is much longer than the one for neighbouring particles (as proposed by \citealp{saitoh_makino09}) improves accuracy in the description of strong shocks.
Finally, the number of neighbours used for SPH has been increased from thirty-two to fifty.

Gas is available for star formation when it has temperature $T<T_{\rm th}=15000\,\rm K$ and number density $n>n_{\rm th}=10.3{\rm\,cm}^{-3}$. 
Gas that satisfies these conditions is converted into stars at the rate per unit volume:
\begin{equation}
\label{eq_sf}
\dot{\rho}_\star = \epsilon_{\rm sf}{\rho\over t_{\rm dyn}}
\end{equation}
where  $\epsilon_{\rm sf}=0.1$ is the efficiency of star formation, $\rho$ is the density of the gas available for star formation,
and $t_{\rm dyn}$ is the dynamical time (\citealp{stinson_etal13} for technical details). Star formation is computed using timesteps of $0.8\,\rm Myr$.

The NIHAO simulations contains two types of stellar feedback:
pre-SN feedback (radiative feedback, winds from massive OB stars), also referred to as early stellar feedback (ESF), and SN blastwaves.

Assuming a \citet{chabrier03} initial mass function and the relation of \citet{torres_etal10} between the mass sand the luminosity of an individual star, 
a stellar population releases a luminous energy
of $2\times 10^{50}{\rm\,erg}\,M_\odot^{-1}$, most of which comes from short-lived massive stars.
Pre-SN feedback is modelled assuming that a fraction $\epsilon_{\rm ESF}$ of this energy is deposited as thermal energy in the interstellar medium (ISM),
while cooling continues to be allowed \citep{stinson_etal13}. In the NIHAO simulations, $\epsilon_{\rm ESF}=0.13$ \citep{wang_etal15}\footnote{For comparison with previous work, the MaGICC simulations \citep{stinson_etal13} used a similar model with
$T_{\rm th}=10^{4}\,\rm K$, $n_{\rm th}=9.3{\rm\,cm}^{-3}$,  $\epsilon_{\rm sf}=0.017$ and $\epsilon_{\rm ESF}=0.1$.}.


SN feedback is implemented using the blastwave formalism described in \citet{stinson_etal06} but without
 the superbubble feedback of \citet{keller_etal14}.
The SN rate associated with each stellar particle is computed from its mass and age assuming a SN energy of $10^{51}\,$erg for all 
stars between $8\,M_\odot$ and $40\,M_\odot$.
For a \citet{chabrier03} initial stellar mass function, this assumption gives one SN every $120\,M_\odot$ of stars.
This energy is distributed to the surrounding gas. To prevent its immediate radiation by the artificially dense surrounding gas
(the gas would not be so dense if the multiphase structure of the ISM were properly resolved), radiative cooling is switched off for $\sim 30\,\rm Myr$ for
some of the gas particles within the blastwave radius (\citealp{stinson_etal06} for details of how these particles are selected).


\section{Phases and components}

We classified gas particles into five categories: cold gas within the galaxy (the cold interstellar medium, ISM), cold gas outside the galaxy but within the halo
(the cold circumgalactic medium, CGM), hot gas within the galaxy (the hot ISM), hot gas outside the galaxy but within the halo (the hot CGM), and gas outside the halo
(the intergalactic medium, IGM).
This classification keeps into account both a particle's position in physical space (inside the galactic radius, between the galactic radius and the virial radius, outside the virial radius; Section~3.1)
and its location in the thermodynamic phase-space (that is, on a density-temperature diagram; Section~3.2).

\subsection{Virial radius and galactic radius}
\label{radius_computation}

The first step of our analysis is to identify the centre, the virial quantities and the radius for each simulated galaxy.

We use the software {\sc Pynbody} \citep{pontzen_etal13}
to identify the main halo in each NIHAO simulation. Its centre is determined by the AHF halo finder \citep{knollmann_knebe09}.
The virial radius $r_{\rm vir}$ is the radius within which the mean density equals $\Delta_{\rm c}\rho_{\rm c}$, where $\rho_{\rm c}$ is the critical density of the Universe and
$\Delta_{\rm c}$ is the critical overdensity contrast computed with the fitting formulae by \citet{bryan_norman98}, which give $\Delta_{\rm c} \simeq 109$ at $z=0$. 

To separate the cold ISM from the cold CGM, we define the galactic radius $r_{\rm g}$ as
the radius $r_{\rm HI}$ within which the surface density $\Sigma_{\rm gas}$ of the cold gas ($T<15,000$)
is larger than the threshold for star formation in
\citet{kennicutt98}'s law, $\Sigma_{\rm th} = 9\,M_\odot{\rm pc}^{-2}$.

The galactic radius $r_{\rm g}$ differs from $r_{\rm HI}$ only when $r_{\rm HI}$ decreases
($r_{\rm g}$ is not allowed to decrease) 
or when $r_{\rm HI}$ is smaller than the half-mass radius of the
stars, $r_{\rm half-mass}$, in which case we set $r_{\rm g}=r_{\rm half-mass}$, but that happens rarely and only in extremely gas poor objects. 
We also prevent the growth of $r_{\rm g}$ when $r_{\rm HI}$ increases by
more than a factor of three from one output timestep to the next as this is invariably a sign of an ongoing merger.
We impose these conditions to prevent the possibility that gas particles in the disc
may be counted as expelled and reaccreted 
simply because of stochastic fluctuations in the value of $r_{\rm HI}$. 

\begin{figure}
\includegraphics[width=1.\hsize,angle=0]{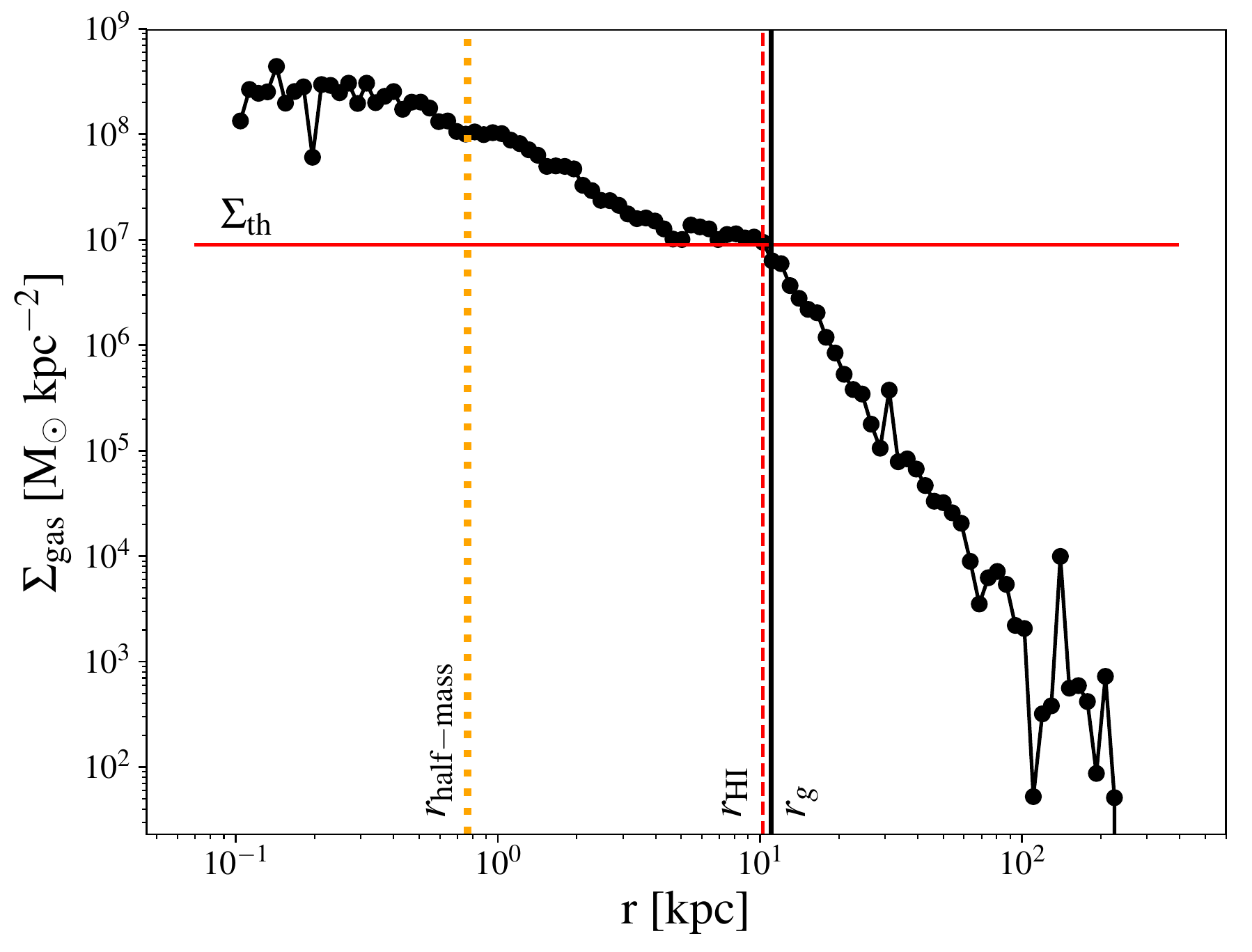} 
\caption{Gas surface density profile $\Sigma_{\rm gas}$ (black circles) and
  galactic radius $r_{\rm g}$ (vertical black solid line) at $z=0$ for the NIHAO galaxy g1.12e12. 
$\Sigma_{\rm gas}$ is the mean surface density of gas with $T<15,000\,$K. It is computed in circular annuli
looking at the galaxy face on (i.e., from the direction of the angular momentum of the stars within $r_{\rm vir}$).
$\Sigma_{\rm th}=9\,M_\odot{\rm pc}^{-2}$ (horizontal red solid line) is the minimum surface
density for star formation \citep{kennicutt98} and $r_{\rm HI}$
(vertical red dashed line) is the outermost radius for which $\Sigma_{\rm
  gas}>\Sigma_{\rm th}$ (the calculation is performed using a hundred radii equally spaced in log$\,r$ between $0.1\,$kpc and $r_{\rm vir}$)
The vertical orange dotted line shows the half-mass radius of the stars, $r_{\rm half-mass}$.
We require that $r_{\rm g}\ge{\rm max}(r_{\rm HI},r_{\rm half-mass})$ and that the value of $r_{\rm g}$
should not decrease.
}
\label{fig_gas_profile}
\end{figure}

\begin{figure}
\includegraphics[width=1.\hsize,angle=0]{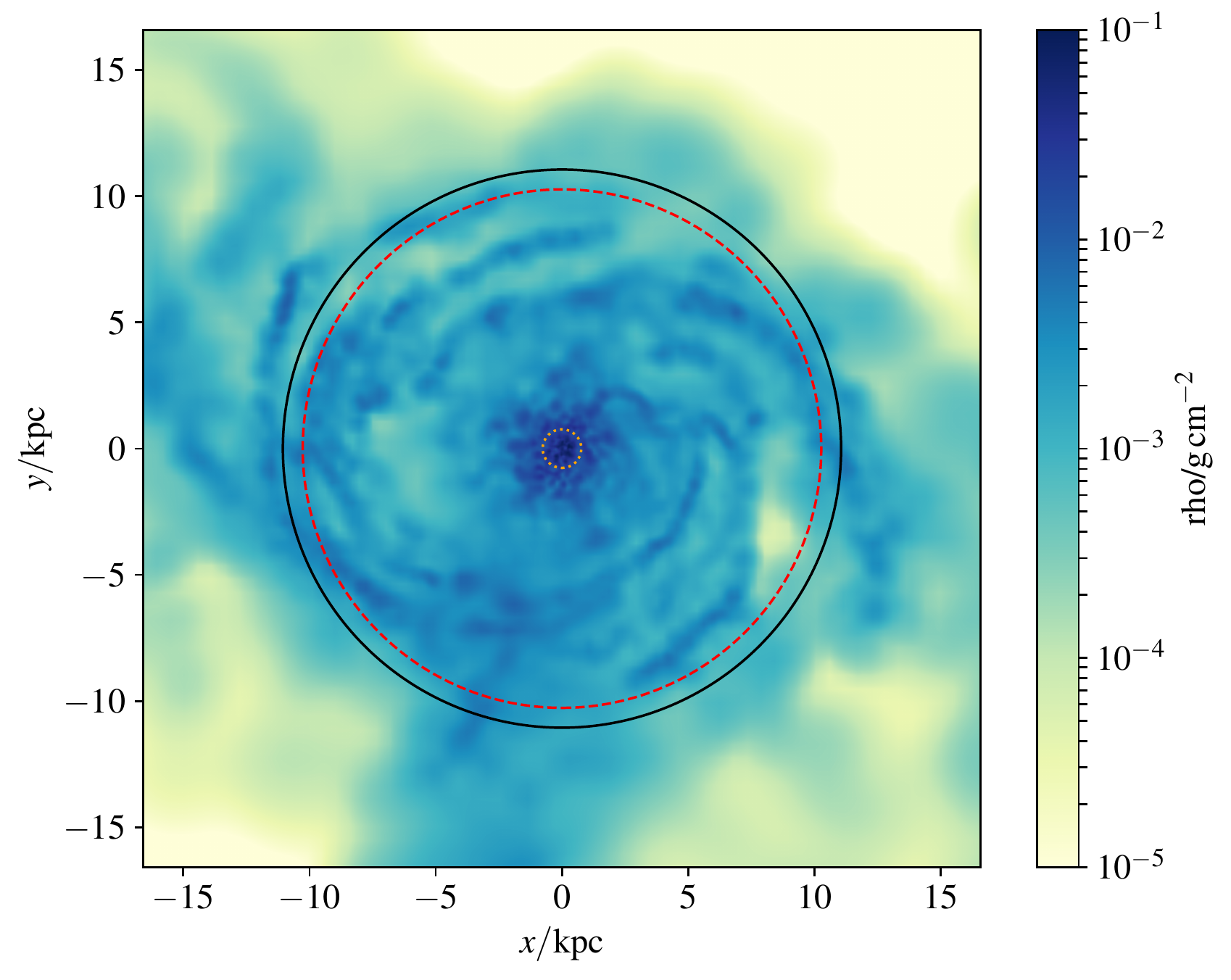} 
\caption{Cold-gas ($T<15,000\,$K) distribution in a face-on image of the NIHAO galaxy g1.12e12.
The black solid circle, the red dashed circle and the orange dotted circle correspond to $r_{\rm g}$, $r_{\rm HI}$ and $r_{\rm half-mass}$, respectively.
The halo hosting this galaxy has $R_{\rm vir} = 274\,\rm kpc$.
}
\label{fig_galaxy_image}
\end{figure}

\begin{figure}
\includegraphics[width=1.\hsize,angle=0]{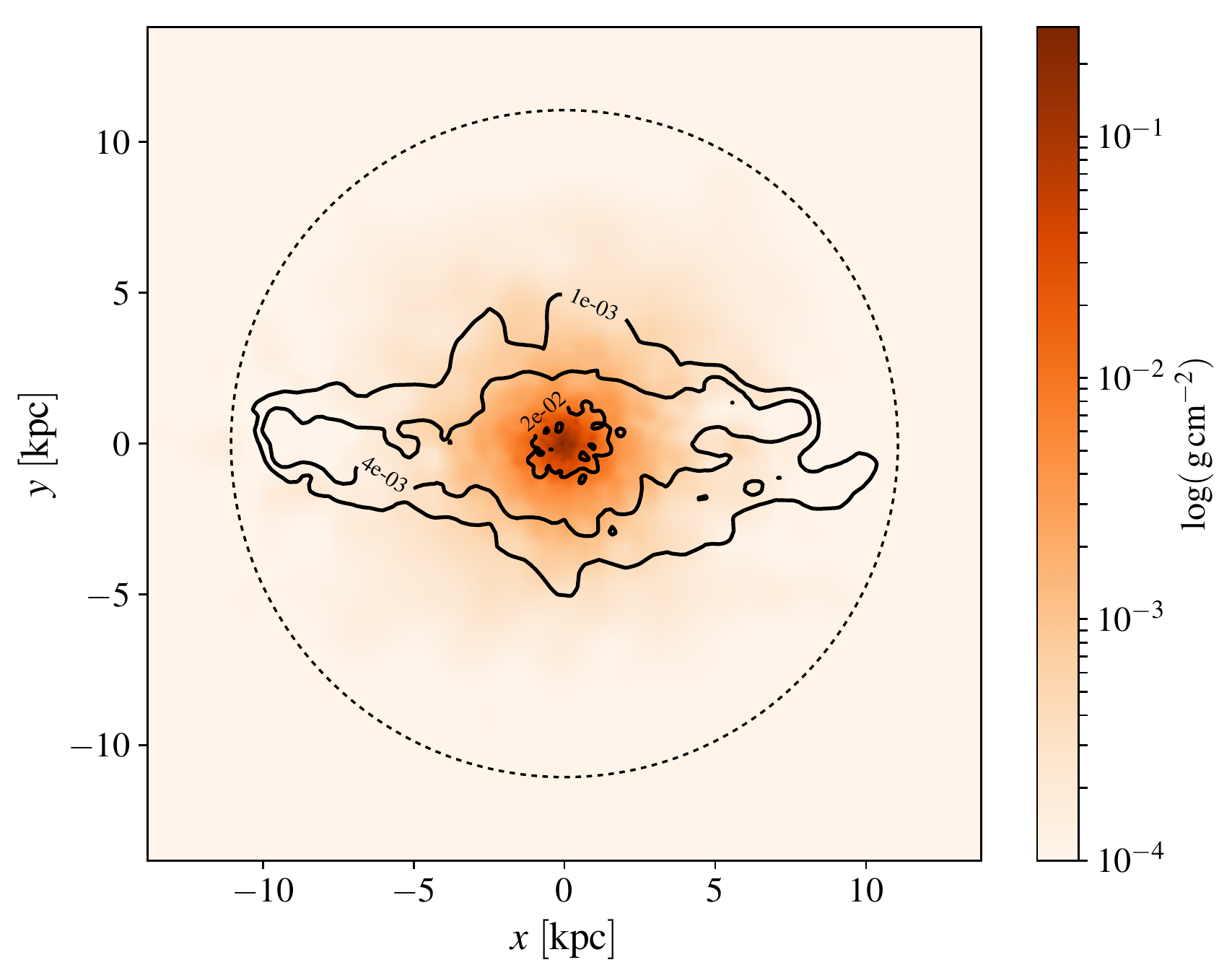} 
\caption{Spatial distribution of the cold and the hot ISM in the NIHAO galaxy g1.12e12.
The galaxy is viewed edge-on. The dashed circle has radius $r_{\rm g}$ and is centred on the galactic centre.
The colour scale shows the projected density distribution of the hot ISM (gas with $T>15,000\,$K and $p>p_{\rm ISM}$).
The hot ISM is concentrated in the central part of the galaxy but the image also shows evidence for outflow.
The contours show the projected density distribution of the cold ISM in units of ${\rm g\,cm}^{-2}$.
They follow the plane of the disc but they also show evidence that some cold gas has been lifted up by the hot ISM that vents out of the galaxy
(see how the contours follow the hot ISM in the central region).}
\label{fig_galaxy_wind_edgeon}
\end{figure}

Figs.~\ref{fig_gas_profile} to \ref{fig_galaxy_wind_edgeon} are all for the same NIHAO galaxy g1.12e12, which exemplifies a typical massive spiral (it is the second most massive galaxy in the NIHAO series).
We have given its identifier not only to be able to refer to it later on but also to enable comparison with other NIHAO publications.
Fig.~\ref{fig_gas_profile} illustrates the calculation of $r_{\rm g}$ graphically 
Fig.~\ref{fig_galaxy_image} shows the distribution of cold gas in g1.12e12 when the galaxy is viewed face-on;
the red dashed circle corresponds to $r_{\rm HI}$ and shows that the outer edge of the {\sc Hi} disc has been correctly identified.
Fig.~\ref{fig_galaxy_wind_edgeon} shows the distribution of cold gas (contours) and hot gas (colour scale) edge-on; one can see that the hot gas is strongly concentrated within the galaxy's inner region,
and that the cold gas is in a disc that thickens at the centre due to entrainment of cold gas by the outflowing hot component.



\begin{figure*}
\begin{center}$
\begin{array}{c}
\includegraphics[width=0.9\hsize]{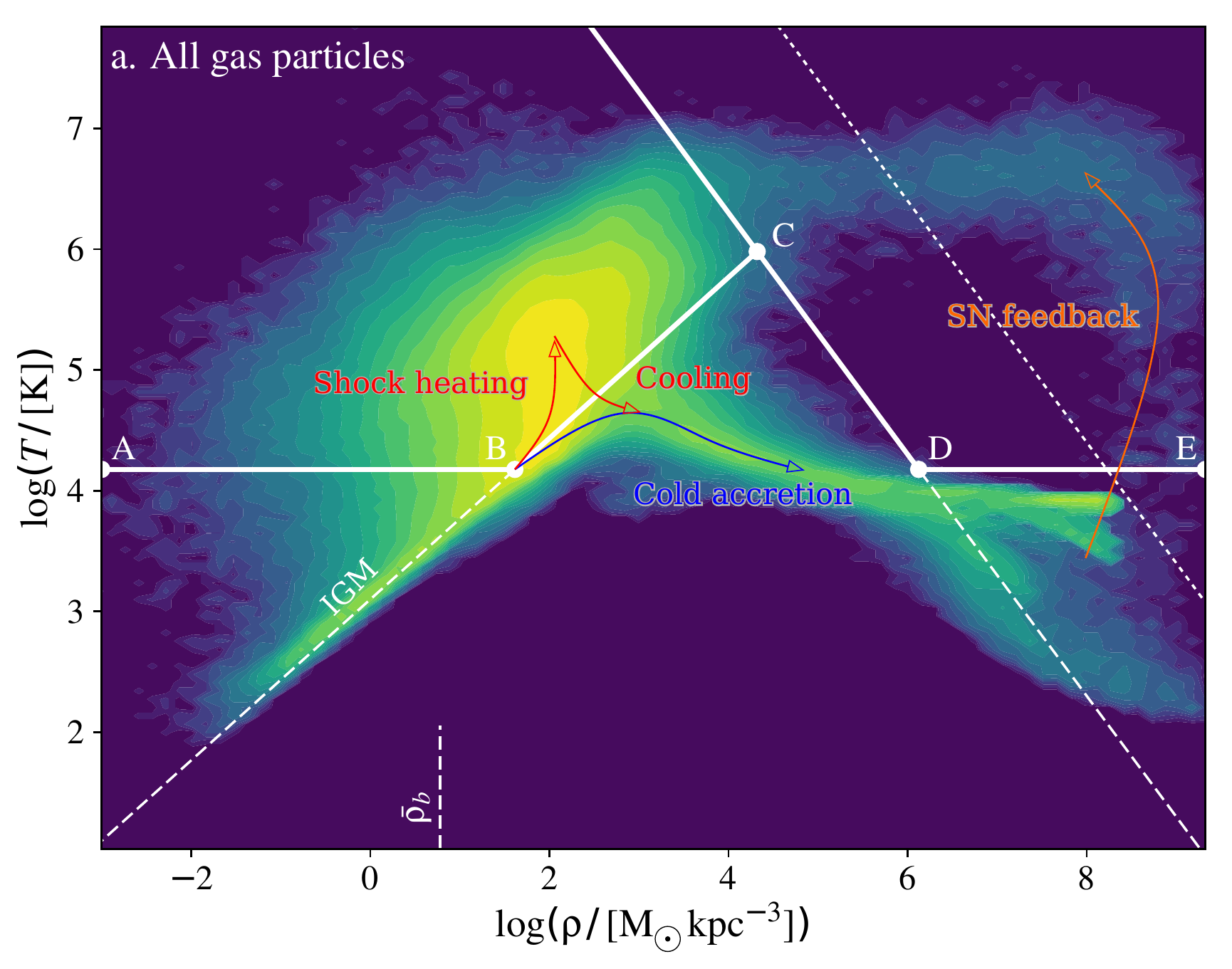} \\
\includegraphics[width=0.5\hsize]{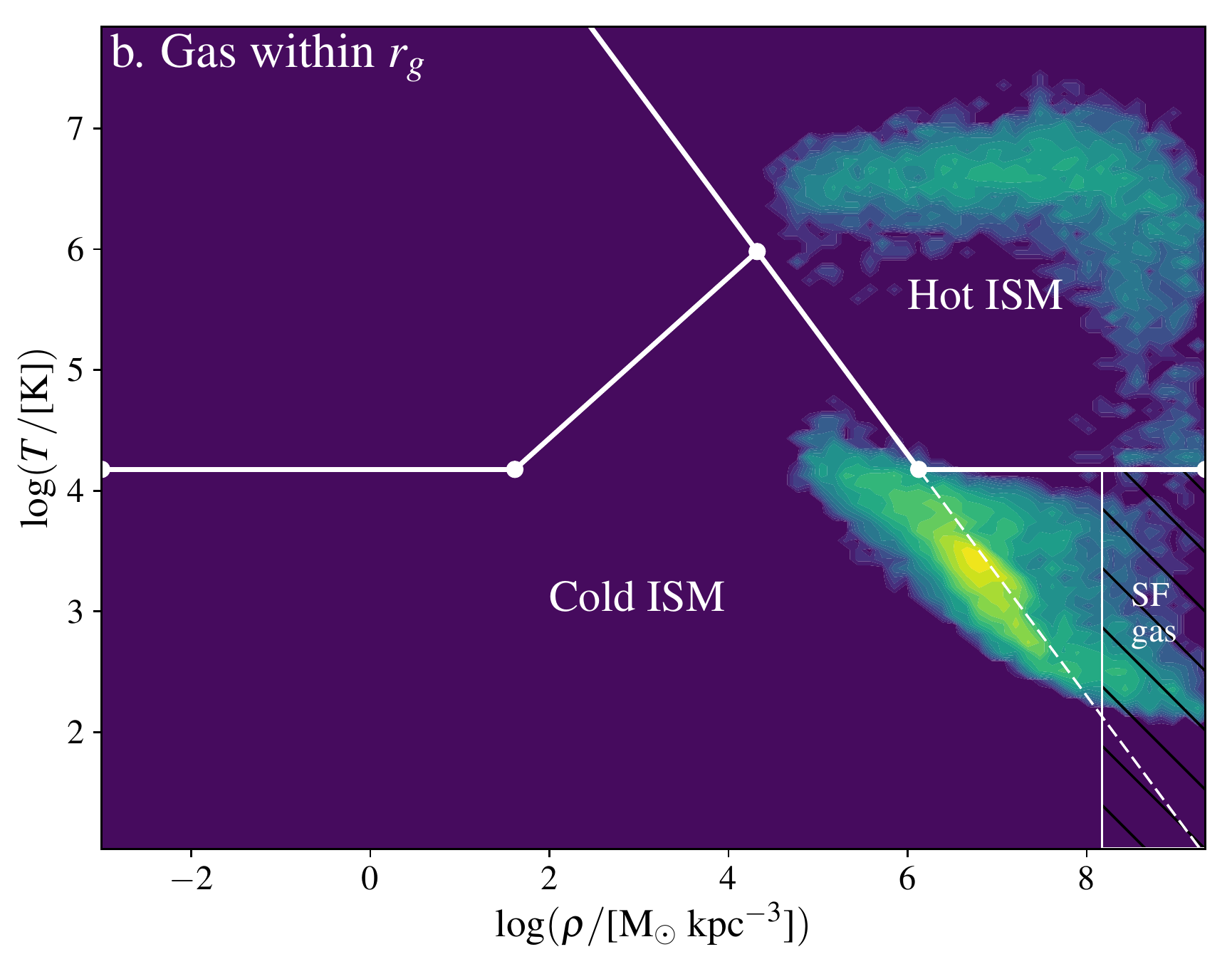}
\includegraphics[width=0.5\hsize]{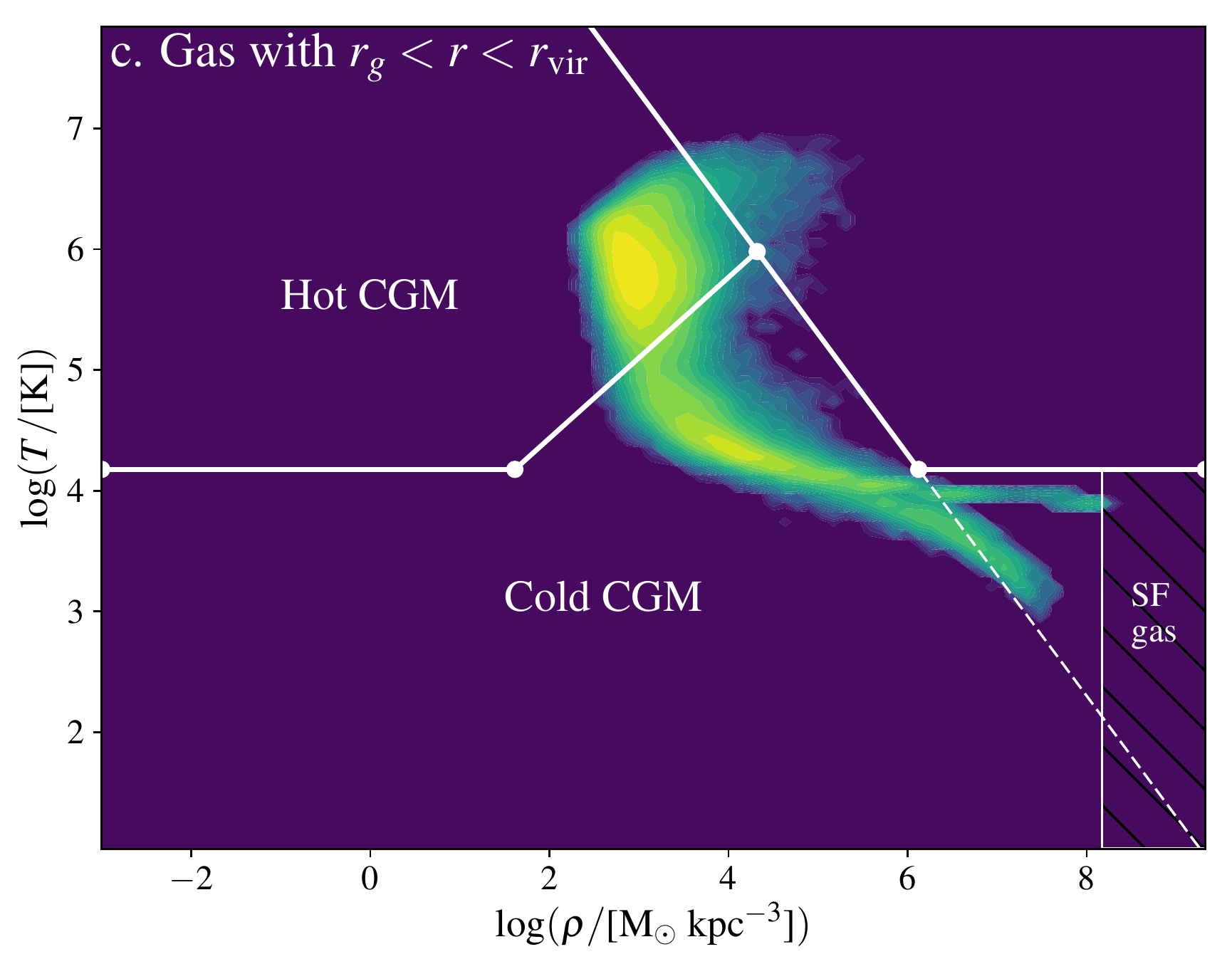}
\end{array}$
\end{center}
\caption{Density-temperature diagram for g1.12e12 ($M_{\rm vir}=1.12\times 10^{12}\,M_\odot$). The three panels refer to: 
a) all the gas in the simulated region,
b) gas within the galaxy ($r\le r_{\rm g}$), and
c) gas within the halo but outside the galaxy ($r_{\rm g}<r<r_{\rm vir}$).
The white solid lines separate the different thermodynamic
phases and are the same in all three panels.
AB and DE correspond to the isotherm $T=15,000\,$K.
BC is an adiabat with entropy equal to the one of the IGM (Eq.~\ref{adiabat_BC}).
The isobar CD and the white dotted-dashed line in panel a correspond to
$p=p_{\rm ISM}\simeq 2\times 10^{-13}{\rm\,dyn\,cm}^{-2}$ and $p\simeq 2\times 10^{-11}{\rm\,dyn\,cm}^{-2}$, respectively.
The hatched rectangles in the bottom right corners of panels b and c 
highlight the cold dense star-forming gas.
They show that this gas is almost entirely within the galaxy.
The vertical white dashed line in panel a shows the mean baryonic density of the Universe, $\bar{\rho}_{\rm b}$.
}
\label{fig_rhoT} 
\end{figure*}

\subsection{Gas phases}

Fig.~\ref{fig_rhoT}a shows the density-temperature distribution for all gas particles in g1.12e12.
Figs.~\ref{fig_rhoT}b and~\ref{fig_rhoT}c are identical to Fig.~\ref{fig_rhoT}a  
except that they show only particles with $r<r_{\rm g}$ and $r_{\rm g}<r<r_{\rm vir}$, respectively.

The gas along the white dashed line $T\propto\rho^{2/3}$ in the bottom left part of
Fig.~\ref{fig_rhoT}a belongs to the IGM because it is  still outside $r_{\rm vir}$
(hence its absence in Figs.~\ref{fig_rhoT}b,c). 
This gas has such low density and temperature that it cannot radiate
efficiently. Hence, its equation of state is virtually adiabatic.

When gas particles from the IGM reach the virial radius, they can take two paths \citep[e.g.,][]{keres_etal05}:
they can streams onto the central galaxy in cold flows (cold accretion) or
they can be shock-heated, form a hot low-density atmosphere (the hot CGM) and later cool onto the galaxy (Fig.~\ref{fig_rhoT}a).

Galaxy g1.12e12 displays an atmosphere of shock-heated gas because it is a massive spiral with $M_{\rm vir}=1.29\times 10^{12}\,M_\odot$.
We chose a massive galaxy as an example precisely so that we would be able to see all thermodynamic phases on a density-temperature diagram.
The hot CGM is virtually absent in haloes with $M_{\rm vir}<10^{11}\,M_\odot$, where all the baryons are accreted cold.
However, the rest of our considerations from Fig.~\ref{fig_rhoT} do extend to lower masses.


Comparing Figs.~\ref{fig_rhoT}b and~\ref{fig_rhoT}c shows that the hot high-density gas in the top right part of the density-temperature diagram
and the hot low-density gas in top left part are different thermodynamic phases with different spatial distributions. 
The hot low-density gas is mainly at $r>r_{\rm g}$. The hot dense gas is confined to $r<r_{\rm g}$
and most of it is within the galaxy itself, although some of it is in a wind (Fig.~\ref{fig_galaxy_wind_edgeon}).
Hence, we refer to the hot low-density gas and the hot dense gas as the hot CGM and the hot ISM, respectively.\footnote{The hot ISM is identified on the basis of a purely thermodynamic criterion. Hence it includes both the genuine hot ISM and a hot dense wind component (the gas in the top right corner of Fig.~\ref{fig_rhoT}c).}
We emphasize that the latter is artificial and exists only because {\sc Gasoline} suppresses radiative cooling of gas particles heated by SN for $\sim 30\,$Myr \citep{stinson_etal06}. 
Hence, its only use is to identify gas that has been recently heated  by  SNe.
Separating cold gas accreting onto a galaxy from cold gas in the galaxy is less straightforward but Figs.~\ref{fig_rhoT}b,c show that the star-forming
gas is almost entirely within the galaxy.

Having identified the five phases that compose the gas in the NIHAO
simulations
(the IGM, the cold CGM, the cold ISM, the hot ISM and the hot CGM), we are now ready to set
out precise criteria to assign a phase to each particle.

An obvious choice is to use a temperature criterion (the maximum
temperature for star formation) to separate cold gas from hot
gas.
All the hot CGM (Fig.~\ref{fig_rhoT}c) and the hot ISM (Fig.~\ref{fig_rhoT}b) are
above the lines AB and DE, corresponding to $T=15,000\,$K.
The problem of using a temperature criterion alone is that this choice would attribute part
of the cold gas to the hot CGM, since cold streams within haloes are in temperature range
 $T\sim 10,000-20,000\,$K.

To separate the cold filaments from the hot
atmosphere, we observe that the adiabat BC corresponds to
the entropy of IGM. Cooling pushes
gas to the right of this adiabat. Shock-heating pushes it to the left.
Hence, it is natural to use this adiabat at $T>15,000\,$K
to separate the cold CGM from the hot CGM.

An adiabat is specified by its temperature at a given density. Let this density be
the mean baryonic density of the Universe $\bar{\rho}_{\rm b} = \Omega_{\rm b}\rho_{\rm c}\simeq 6\,M_\odot{\rm\,kpc}^{-3}$.
The adiabat that fits best the temperature-density relation around $\bar{\rho}_{\rm b}$ is:
\begin{equation}
T=4200{\,\rm K}\times 10^{-0.25a(z)-0.86a^2(z)}\left(\rho\over\rho_{\rm b}\right)^{2\over 3},
\label{adiabat_BC}
\end{equation}
where $z$ is redshift and $a(z)=z/(1+z)$.

\begin{figure*}
\begin{center}
\includegraphics[width=0.9\hsize]{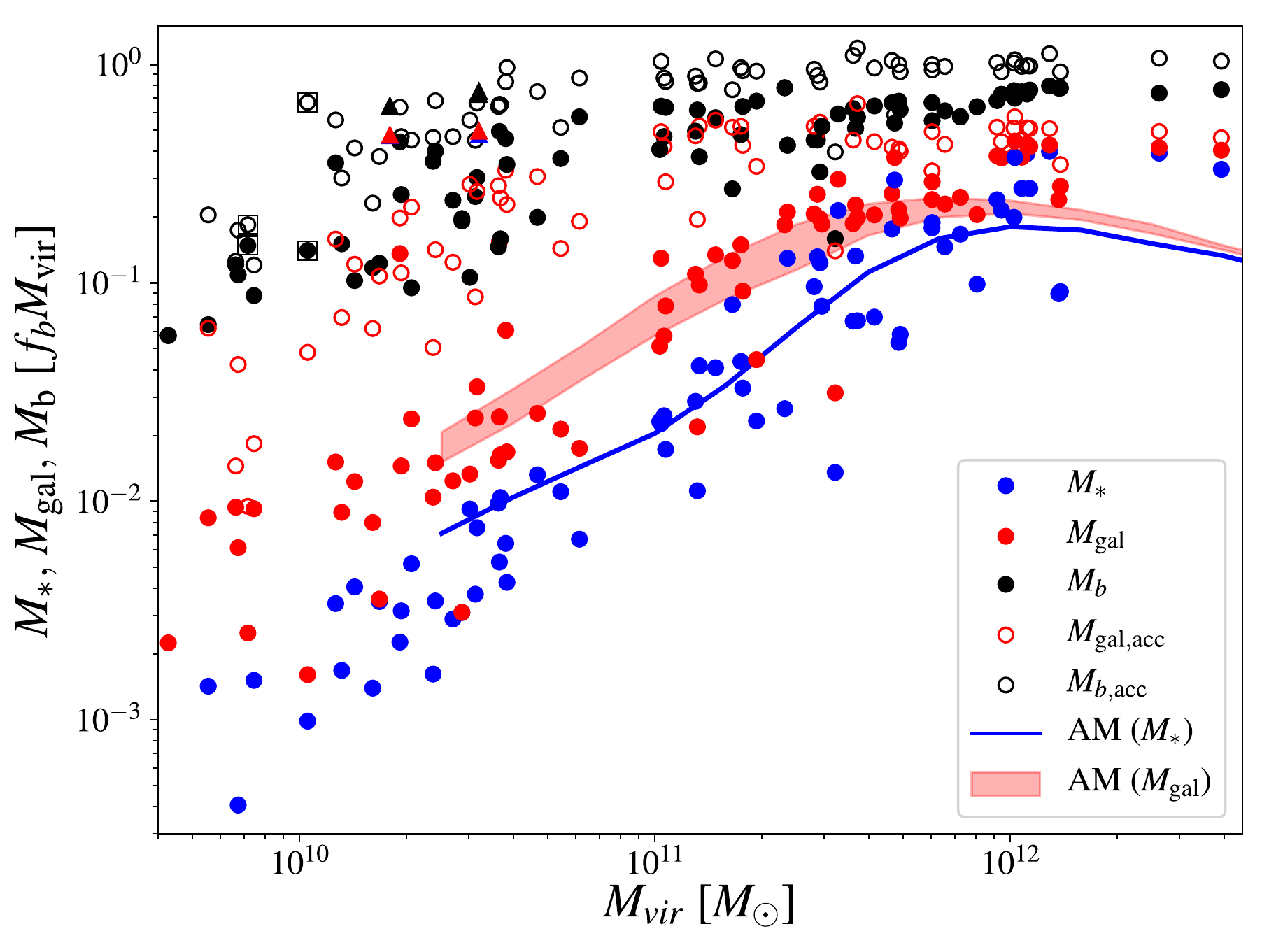}
\end{center}
\caption{
Stellar masses (filled blue circles), galaxy masses (stars plus cold galactic
gas; filled red circles) and total baryonic masses (filled black circles) within $r_{\rm vir}$ at $z=0$
as a function of $M_{\rm vir}$. All masses are expressed in units of $f_{\rm b}M_{\rm vir}$.
$M_{\rm gal,acc}$ ($M_{\rm b,acc}$) is the total mass of the baryons particles identified as belonging to the galaxy (the halo) at any output timesteps
(open red and black circles, respectively).
The filled triangles are the same as the filled circles for a couple of galaxies that have been resimulated without stellar feedback.
The circles enclosed in squares are a couple of objects that we have highlighted to compare their behaviours.
The blue curves show the stellar mass-halo mass relation from the AM studies of Papastergis et al. (2012; solid curve).
The red shaded area shows the galaxy mass-halo mass relation from the AM study of \citet{papastergis_etal12}, who computed the baryonic mass function of galaxies
using both optical and {\sc Hi} data.}
\label{fig_baryon_frac}
\end{figure*}

To separate cold gas from gas that is sufficiently hot to vent out of the galaxy, we use a pressure criterion.
The line CD is an isobar with pressure equal to the
pressure $p_{\rm ISM}\simeq 2\times 10^{-13}{\rm\,dyn\,cm}^{-2}$ of the cold dense star-forming ISM (i.e., the molecular gas).
Gas with higher temperature and thus higher pressure will leave this phase but will remain within the galaxy as cold neutral gas if $p_{\rm ISM}$ is not exceeded by a large factor.
Hence, gas with $T\le 15,000\,$K is still considered part of the cold ISM even if it has pressure $p>p_{\rm ISM}$.
The dotted-dashed line in Fig.~\ref{fig_rhoT}a shows that this choice is
equivalent to separating the cold ISM from the hot ISM at $p=100p_{\rm ISM}$.

We note that the ISM in the NIHAO simulations is very different from the one of \citet{mckee_ostriker77}, 
based on the assumptions of different phases in pressure equilibrium. 
The hot ISM is not in pressure equilibrium with the cold ISM. It is continually venting out of galaxies (and cooling back onto them, as we shall see in Section~9).

The distinction between cold gas in the galaxy and cold gas in the
halo is the only one that cannot be performed based on the
thermodynamic phase-space diagram alone and relies on our measurement of
$r_{\rm g}$.

Particles at the borders between phases may complicate our analysis because small fluctuations in density and temperature
may be misinterpreted as ejection and reaccretion.
For this reason, a particle is moved from one phase to another only when the threshold value has been exceed by $10\%$.
A particle initially outside the galaxy moves inside the galaxy when it moves to $r<0.9r_{\rm g}$.
A particle within the galaxy is counted as ejected only when it moves to $r>1.1r_{\rm g}$.
A particle crosses the lines AB or DE and moves to the hot CGM or ISM only when its temperature raises above $16,500\,K$.
A particle crosses the lines AB or DE and moves to the cold phase only when its temperature drops below $13,500\,$K.
The same $10\%$ elastic border is applied to the pressure $p_{\rm ISM}$ that separates the cold gas from the hot ISM (line CD).
The only exception is the adiabat BC, to which we apply a tolerance of $5\%$ only,
since many particles in the hot CGM have entropies just above the critical entropy that separates the hot CGM from the cold CGM.

\section{Baryon content of galaxies and haloes}

\begin{figure}
\includegraphics[width=\hsize]{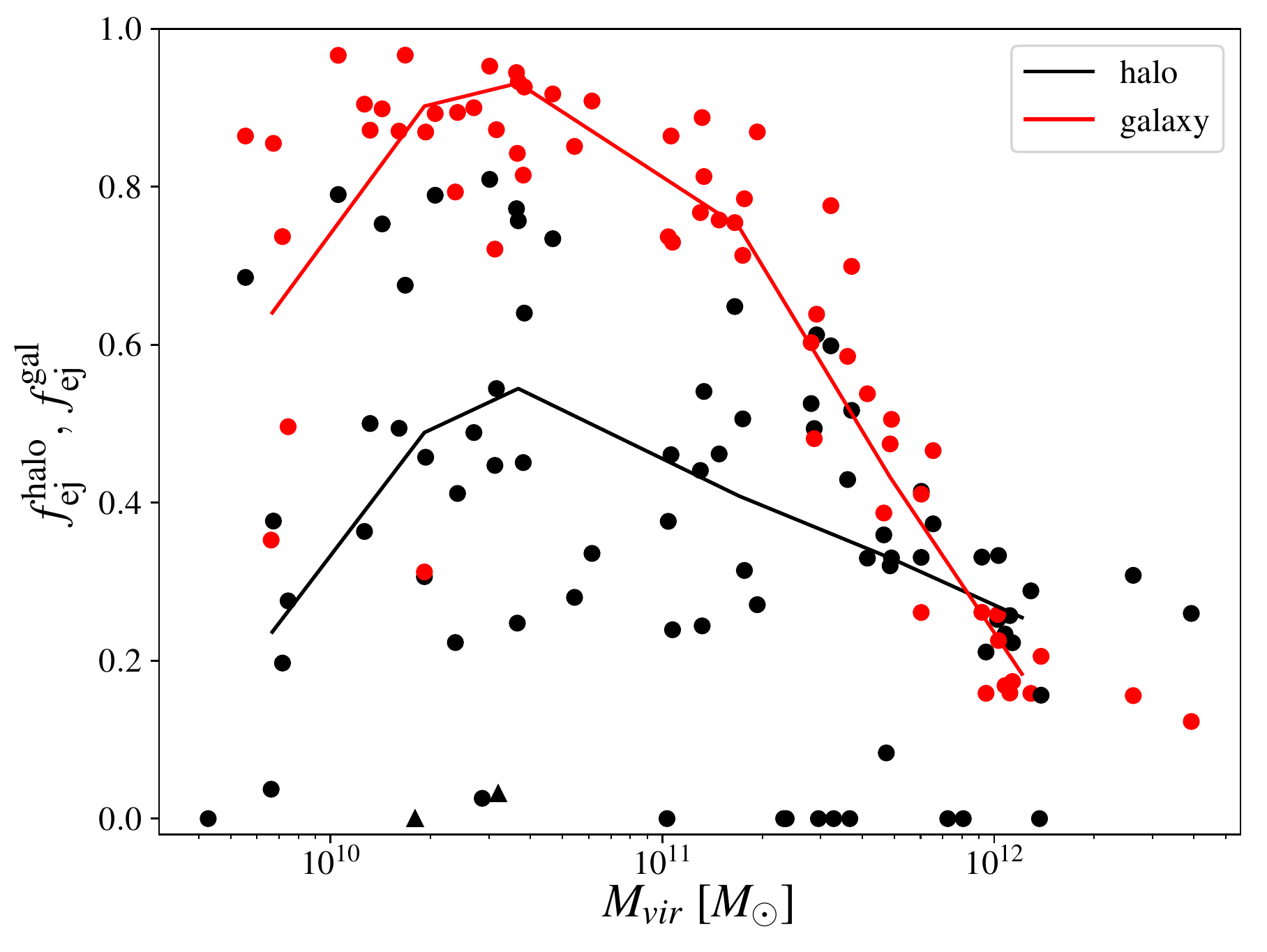}
\caption{Ejected mass fraction from the galaxy (red) and the halo (black).
The symbols correspond to individual galaxies. The curves are medians in bis of halo mass.
Ejected fractions are relative to the total baryonic masses accreted by galaxies or haloes over the lifetime of the Universe.
They have been computed counting all baryons that have been inside the galaxy or the halo at some point but are no longer there at $z=0$.
The triangles are two objects that have been resimulated without SN feedback.
Both are consistent with $f_{\rm ej}^{\rm halo}\simeq 0$.}
\label{fig_ej_fraction}
\end{figure}

Having introduced the NIHAO simulations (Section~2) and our method to analyse them (Section~3), we are now ready to present our results.
We start from the baryon fraction within galaxies and haloes (current section). Later we shall investigate the astrophysical processes that are responsible
for the baryon fractions we find.

Fig.~\ref{fig_baryon_frac} shows the stellar mass $M_\star$ (filled blue circles), the
baryonic mass $M_{\rm gal}$ within the galaxy (stars and cold gas, filled red circles), and the total
baryonic mass $M_{\rm b}$ within $r_{\rm vir}$ (filled black circles) for the NIHAO galaxies at $z=0$. All masses are shown as a function of $M_{\rm vir}$
and in units of $f_{\rm b}M_{\rm vir}$, where $f_{\rm b}=0.16$ is the cosmic baryon fraction.

The filled blue circles (our $M_\star - M_{\rm vir}$ relation) are in reasonable agreement with the results from abundance matching (AM)
by Papastergis et al. (2012; blue solid curve).
%
%
The only small tension with the observations is that our ultradwarves tend to be of lower mass than in AM studies 
(but notice that our procedure to estimate $r_{\rm g}$ lowers $M_\star$ in gas-poor galaxies, and that passive ultradwarves fall into this category).
Throughout this article, dwarves are galaxies with $3\times 10^7\,M_\odot\lsim M_\star\lsim 3\times 10^9\,M_\odot$ 
(corresponding to $3\times 10^{10}\,M_\odot\lsim M_{\rm vir}\lsim 3\times 10^{11}\,M_\odot$);
ultradwarves and normal spirals are galaxies below and above this range of masses, respectively.


\begin{figure}
\includegraphics[width=1.\hsize,angle=0]{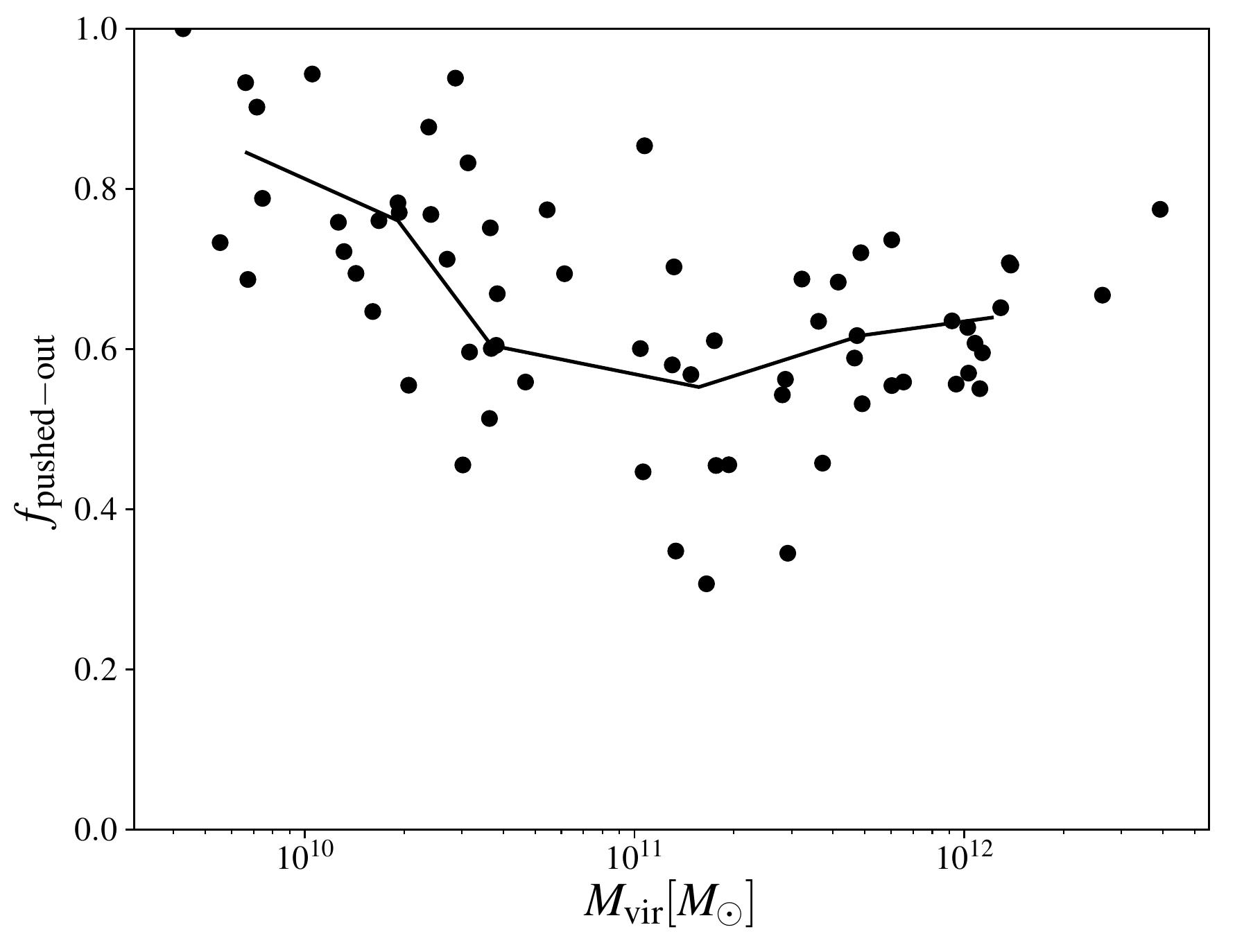}
\caption{Baryon fraction that has never been inside the galaxy relative to the total mass of the gas particles expelled from the halo
(symbols: individual galaxies; curve: median in bins of halo mass).
For $M_{\rm vir}\lsim 3\times 10^{10}\,M_\odot$, $f_{\rm pushed-out}$ is close to unity (median value $>0.8$).
Hence, most of the gas ejected from the halo comes from the CGM and not from the galaxy.
The larger $M_{\rm vir}$,  the more difficult for winds to entrain halo gas ($f_{\rm pushed-out}$ decreases).
The trend is reversed at high masses because dilute shock-heated gas is more easily pushed out than dense cold gas.}
\label{fig_attila}
\end{figure}

\begin{figure*}
\begin{center}$
\begin{array}{cc}
\includegraphics[width=0.541\hsize]{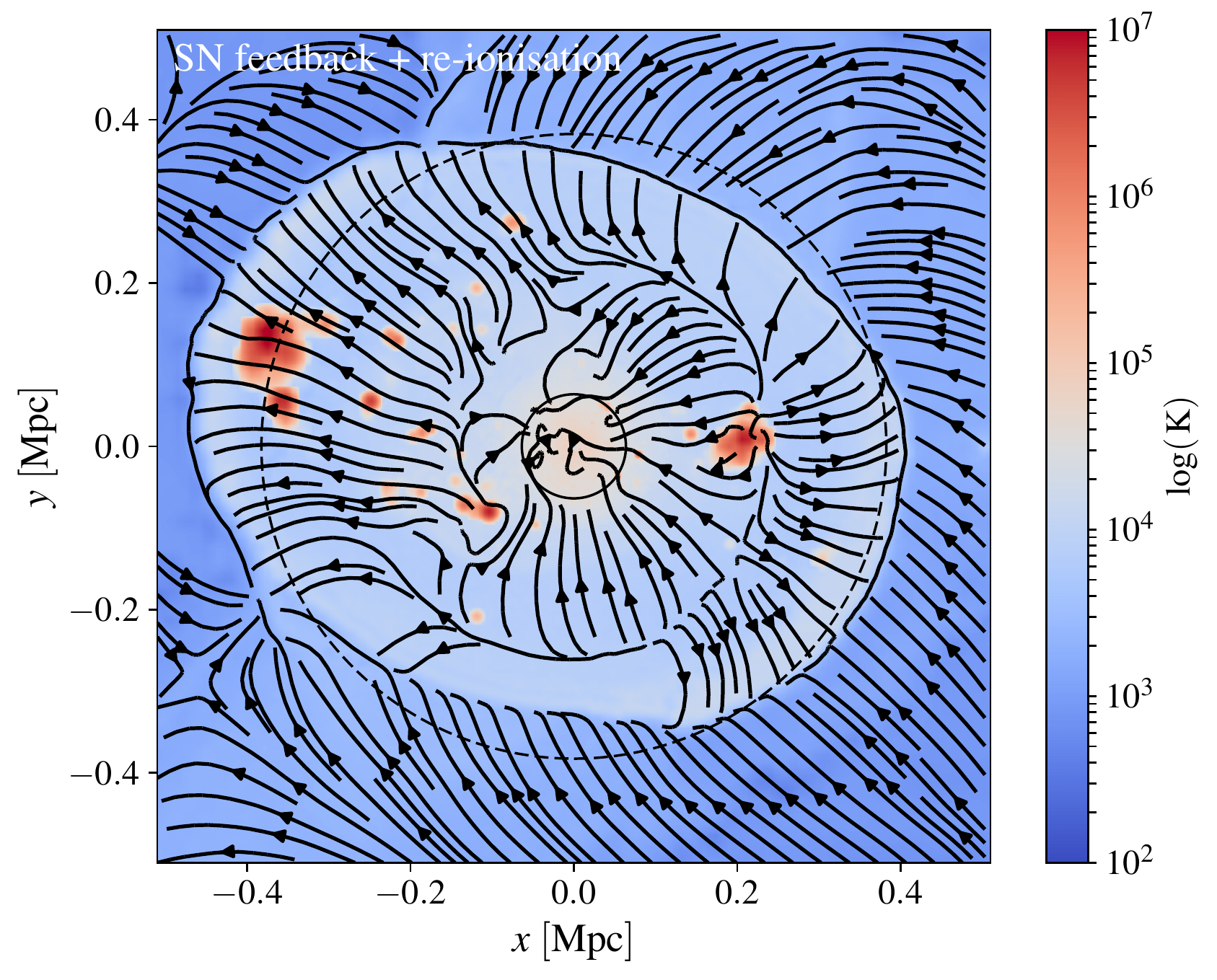}
\includegraphics[width=0.45\hsize]{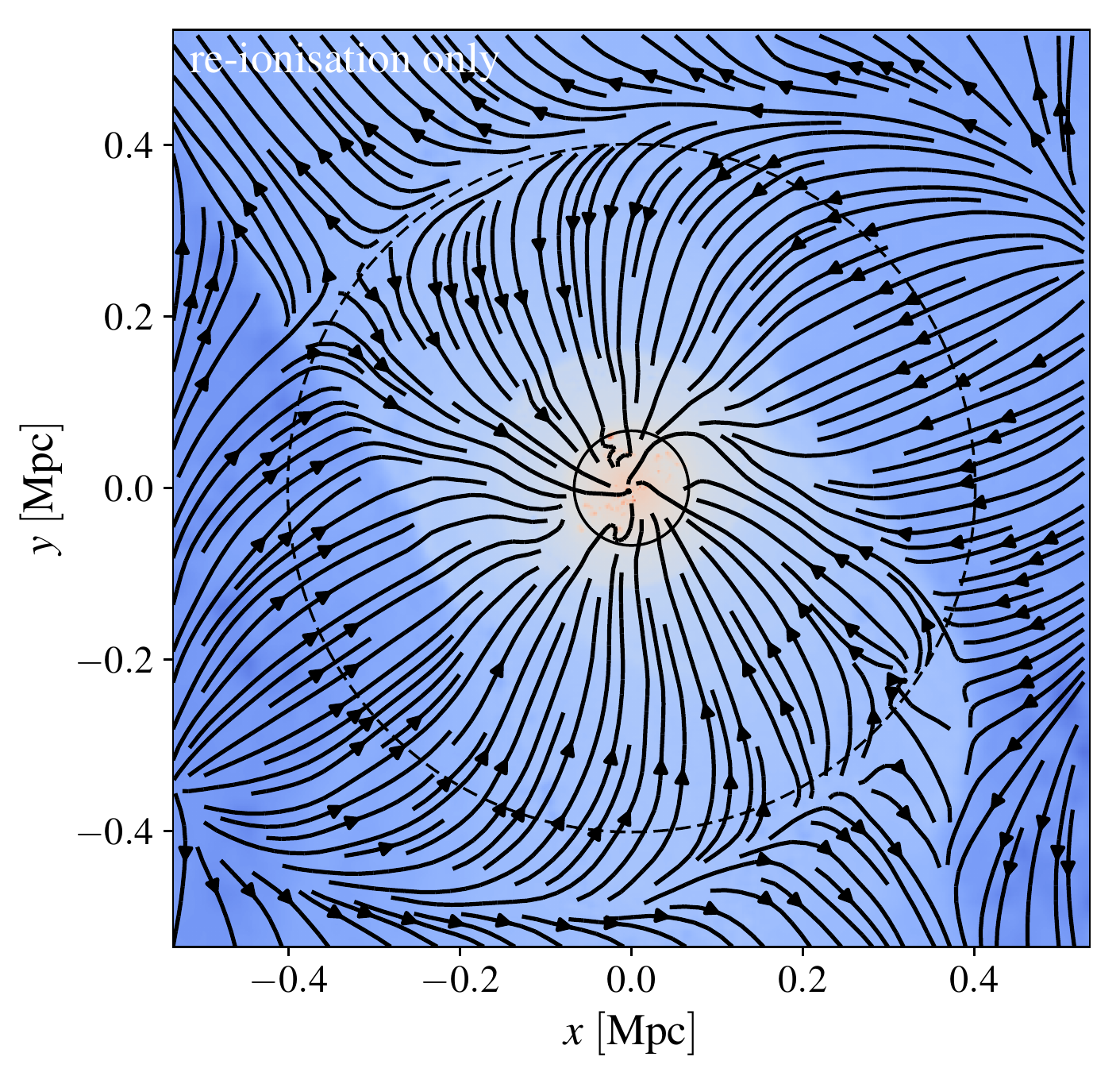}
\end{array}$
\end{center}
\caption{Temperature map and flow lines for the gas in the NIHAO simulation g1.5e10 ($M_{\rm vir}=1.6\times 10^{10}\,M_\odot$ at $z=0$), which
corresponds to the lower-mass triangle in Fig.~\ref{fig_baryon_frac}.
The left panel shows the standard NIHAO simulation inclusive of both stellar feedback and feedback from a photoionising UV background.
The right panel shows the same system resimulated with feedback from a photoionising UV background only.
The small circle at the centre shows the virial radius. The large dashed circle corresponds to $r=6r_{\rm vir}$.
Gas particles shown in these figures are part of the zoom-in simulated region, which extends up to $\sim 1\,\rm Mpc$ for this galaxy.
}
\label{fig_velocity}
\end{figure*}

\begin{figure}
\includegraphics[width=\hsize]{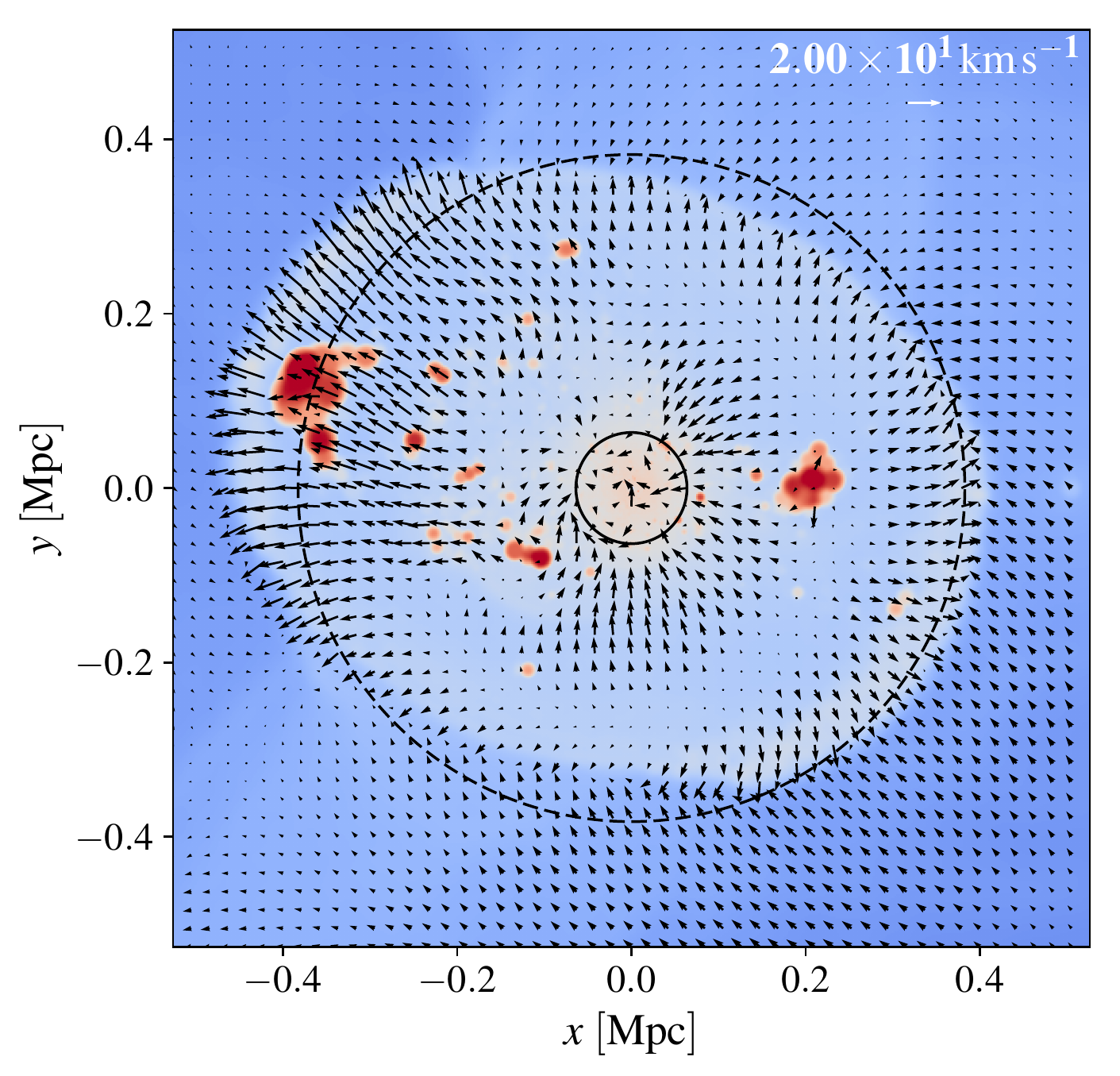}
\caption{Temperature map and velocity for the gas in the NIHAO simulation g1.5e10 (with stellar feedback and UV background), Fig.~\ref{fig_velocity} shows the corresponding flow lines.
The small circle at the centre shows the virial radius. The dashed circle corresponds to $r=6r_{\rm vir}$. Colour coding shows temperature and is the same as in Fig.~\ref{fig_velocity}.
The outflow speed is much higher at 10 o'clock ($20{\rm\,km\,s}^{-1}$ at $6r_{\rm vir}$)
than in the other directions.
This is also the direction along which the hot bubbles at $T\sim 10^7\,$K are coming out.}
\label{fig_velocity2}
\end{figure}

The filled red circles show the same relation for the baryonic galaxy mass
$M_{\rm gal}$ (the mass of stars and the cold ISM). Comparing them to the AM relation by Papastergis et al. (2012; red shaded area), who measured 
$M_{\rm gal}=M_\star+1.4M_{\rm HI}$ from a sample with {\sc Hi} data for each galaxy (the factor $1.4$ accounts for helium)
shows that the agreement with the total galaxy mass is also satisfactory. 
The simulations' ability to reproduce the stellar and baryonic  masses of galaxies simultaneously for a same data set 
proves that gas fractions are reproduced reasonably well, at least for $M_{\rm vir}\gsim 3\times 10^{10}\,M_\odot$.

The filled black circles show that the halo baryon content is  $60-80\%$ of the cosmic baryon fraction for normal spirals with $M_{\rm vir}>3\times 10^{11}\,M_\odot$
but drops to $10-20\%$ of the cosmic baryon fraction for ultradwarves with $M_{\rm vir}<2\times 10^{10}\,M_\odot$. 
As already reported by \citet{wang_etal17},
these findings are consistent with those by \citet{brook_etal14}, \cite{hopkins_etal14} and \cite{mitchell_etal18}, even though our baryon fractions tend to be slightly lower.

The open red circles replace $M_{\rm gal}$ with $M_{\rm gal,acc}$, the sum of the masses of all particles ever identified as belonging to the galaxy.
$M_{\rm gal,acc}$ includes  $M_{\rm gal}$, but it also includes the masses of all the particles that have been expelled from the galaxy.
When $M_{\rm gal,acc}/M_{\rm gal}\sim 10$, that means that $90\%$ of the baryons accreted by the galaxy have been ejected from the galaxy.

The impact of the expulsion of matter on the baryon content of galaxies and haloes can be quantified more precisely by introducing the ejected fractions:
\begin{equation}
f_{\rm ej}^{\rm gal}=1-{M_{\rm gal}\over M_{\rm gal,acc}},
\end{equation}
\begin{equation}
f_{\rm ej}^{\rm halo}=1-{M_{\rm b}\over M_{\rm b,acc}}.
\end{equation}
Fig.~\ref{fig_ej_fraction} (red curve) shows that $f_{\rm ej}^{\rm gal}$ decreases from about $95\%$ at $M_{\rm vir}\sim 4\times 10^{10}\,M_\odot$ to  about $20\%$ at $M_{\rm vir}\sim 10^{12}\,M_\odot$.
Low $f_{\rm ej}^{\rm gal}$ are also measured in ultradwarves
where there has not been enough star formation to produce massive outflows.

In dwarves, the median ejected fraction from the halo is about half the ejected fraction from the galaxy (one third in the case of ultradwarves).
The difference between $f_{\rm ej}^{\rm halo}$ and $f_{\rm ej}^{\rm gal}$ becomes smaller at higher masses.
At  $M_{\rm vir}> 8\times 10^{11}\,M_\odot$, nearly all galaxies have $f_{\rm ej}^{\rm halo}>f_{\rm ej}^{\rm gal}$ (Fig.~\ref{fig_ej_fraction}, black and red circles). This finding 
is only apparently paradoxical because {\it most of the baryons ejected from haloes do not come from galaxies}.
They come from halo gas that has been pushed out by galactic winds or unbound by the thermal energy that SNe have deposited into it (Fig.~\ref{fig_attila}).
The NIHAO sample selects galaxies that are fairly isolated (Section~2) but some of the gas that leaves the halo without having ever been part of the galaxy may also come from satellites that have come in and out of the virial sphere.

One may wonder to what extend the baryons that are ejected from haloes but do not come from galaxies may be in fact winds coming from small satellites. We have verified that such winds contribute to the outflow rate from the halo but that our qualitative results do not change if we remove DM substructures and the associate baryons from our analysis.

The existence of haloes with $M_{\rm b}/(f_{\rm b}M_{\rm vir})\sim 0.06$ (Fig.~\ref{fig_baryon_frac}) while $f_{\rm ej}^{\rm halo}\lsim 0.8$ for all objects (Fig.~\ref{fig_ej_fraction})
proves that ejective feedback is not the only mechanism that is responsible for the low baryon fractions of many NIHAO haloes.

To demonstrate that two different processes are at work, we have identified two haloes with $M_{\rm b}/(f_{\rm b}M_{\rm vir})\sim 0.1$.
They correspond to the  circles enclosed in squares in Fig.~\ref{fig_baryon_frac}.
The halo with $M_{\rm vir}\sim 1.5\times 10^{10}\,M_\odot$ has accreted $60-70\%$ of the universal baryon fraction but has lost $\sim 70\%$ of its baryons.
The other halo (the one with $M_{\rm vir}\sim 8\times 10^{9}\,M_\odot$) has accreted less than $20\%$ of the universal baryon fraction but has retained most of the accreted baryons.
Ejective feedback has played a major role in the first halo but not in the second one, where baryons never made it into the virial radius in the first place.
The presence of many haloes with $M_{\rm b,acc}\ll f_{\rm b}M_{\rm vir}$ (Fig.~\ref{fig_baryon_frac}, open circles)
demonstrates that this is not an isolated case and that
inability to accrete is a significant limiting factor for the growth of galaxies in haloes with
$M_{\rm vir}<3\times 10^{10}\,M_\odot$ \citep{nelson_etal15,wang_etal17,mitchell_etal18}.

To understand whether this inability is due to stellar feedback or heating by a photoionising UV background, we have resimulated two galaxies without stellar feedback (triangles in Fig.~\ref{fig_baryon_frac}).
Fig.~\ref{fig_velocity} shows the flow lines and the temperature distribution of gas in the less massive of the two galaxies 
($M_{\rm vir}=1.6\times 10^{10}\,M_\odot$)
with and without stellar feedback (left and right panels, respectively).
The right panel shows that reionisation feedback alone (the only feedback that is present when SN feedback and early stellar feedback are turned off)
has limited effect at $M_{\rm vir}\gsim 2\times 10^{10}\,M_\odot$. Hence, gas streams into the virial radius (dashed circle), accretes onto the galaxy (Fig.~\ref{fig_velocity}, right) and forms stars so efficiently that,
in Fig.~\ref{fig_baryon_frac}, the blue triangles ($M_\star$) hide the red triangles ($M_{\rm gal}$) almost entirely.
The left panel of Fig.~\ref{fig_velocity} shows that including stellar feedback changes the dynamics of the gas completely.
Stellar-driven outflows terminate with a shock front at  $r=6r_{\rm vir}$ (dashed circle) and prevent the infalling gas from coming in.
The only gas that accretes onto the galaxy is warm ($10^4{\rm\,K}<T<10^5{\rm\,K}$) gas that cools within a sphere of radius $4r_{\rm vir}$.
Fig.~\ref{fig_velocity2} shows that the hot outflowing gas drives the expansion of a shock front that propagates in the wind's direction.

In conclusion, without stellar feedback, the NIHAO galaxies have $M_\star>0.4f_{\rm b}M_{\rm vir}$ down to $M_{\rm vir}\sim 2\times 10^{10}\,M_\odot$ (Fig.~\ref{fig_baryon_frac}, blue triangles).
SNe reduces $M_\star$ by two orders of magnitudes but not through ejective feedback alone.
{\it One order of magnitude is gained at a much lower energy cost by preventing gas from coming while it is still far away.}
This pre-emptive feedback is dominant mechanism in ultradwarves with $M_{\rm vir}\lsim  10^{10}\,M_\odot$.

\section{Star formation}

The efficiency of star formation $\epsilon_{\rm sf}=0.1$  adopted in the NIHAO simulations (Section~2) is calibrated on the local \citet{kennicutt98} law and
we have already seen that this assumption leads to integrated SFRs in reasonable agreement with the observed stellar masses of galaxies (Fig.~\ref{fig_baryon_frac}).
Here, however, we want to check the NIHAO galaxies' star formation efficiencies directly by comparing the global star formation timescale 
\begin{equation}
t_{\rm sf}={M_{\rm gas}\over {\rm SFR}}
\label{tsf}
\end{equation}
with observations from the {\it Herschel} Reference Survey (HRS, \citealp{boselli_etal14}).

\begin{figure}
\includegraphics[width=1.\hsize,angle=0]{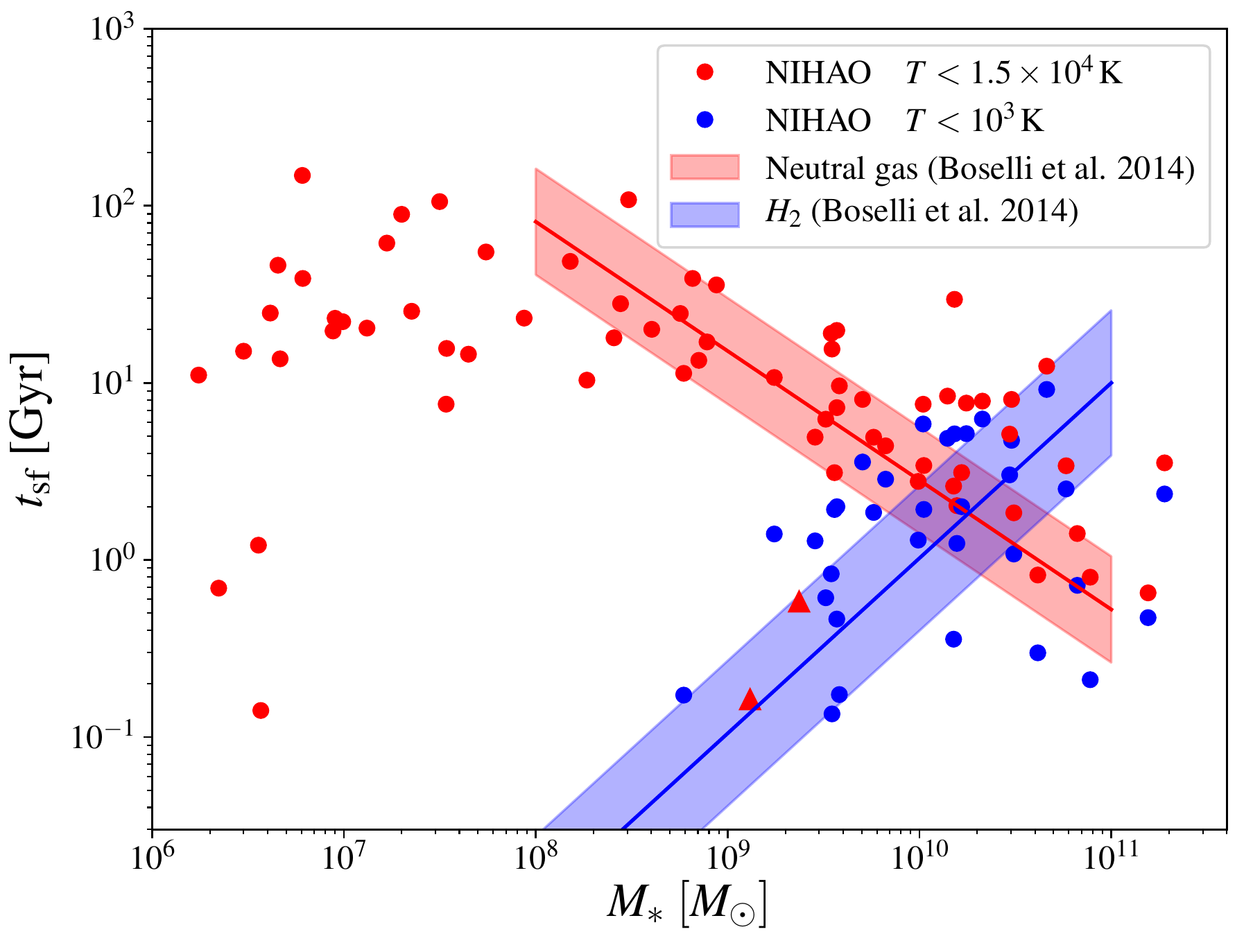}
\caption{The relation between star-formation timescale $t_{\rm sf} = M_{\rm gas}/{\rm SFR}$ and stellar mass $M_\star$ in the NIHAO simulations (symbols) and the HRS
(\citealp{boselli_etal14}; shaded areas). Red and blue refer to the total mass of cold neutral gas within a galaxy and to the mass of dense cold molecular mass, respectively.
In the NIHAO simulations, this gas is identified by using the condition $T<1,000\,$K.
In \citet{boselli_etal14}, the H$_2$ mass is measured from CO data assuming a CO/H$_2$ ratio that depends on luminosity \citep{boselli_etal02}.
The red triangles show the results for the simulations without SN feedback. In those objects nearly all the gas has $T<1,000\,$K.}
\label{fig_tsf}
\end{figure} 

In Fig.~\ref{fig_tsf}, the red circles show the $t_{\rm sf} - M_\star$ relation in the NIHAO simulations when $M_{\rm gas}$ 
(Eq.~\ref{tsf}) is the total mass of the cold ISM.
The red shaded area is a linear fit to the data for {\sc Hi}-and-CO-detected late-type gas-rich galaxies \citep{boselli_etal14}. In these observations,
$M_{\rm gas}=1.3(M_{\rm HI}+M_{\rm H_2})$, where the factor $1.3$ accounts for the presence of helium. 
The NIHAO simulations are in reasonable agreement with the data and recover the observational scaling relation $t_{\rm sf}\propto M_\star^{-0.7}$.

For $M_\star<10^9\,M_\odot$, $t_{\rm sf}$ becomes longer than the age of the Universe. Hence, inefficient accretion and
the expulsion of gas from gas galaxies and haloes are not the only mechanisms that slow down the conversion of baryons into stars.
Inefficient star formation plays a role, too.

The fraction of the available gas that can be converted into stars in a time $t$ is $f=1-{\rm exp}(-t/t_{\rm sf})$. When $t_{\rm sf}$ is much shorter than the age of the Universe, $f$ is of order unity,
but that is not true in galaxies with $M_\star\sim 10^8\,M_\odot$, where $t_{\rm sf}$ can take all values between $10$ and $200\,$Gyr (Fig.~\ref{fig_tsf}, red circles).
Taking a median value $t_{\rm sf}\sim 40\,$Gyr and assuming $t\sim10\,$Gyr 
limits the mass that can form stars to $\sim 20\,\%$ of the baryons available in the cold ISM.

{\it The inefficiency of star formation at low $M_\star$ derives from the cold ISM's inability to condense into molecular clouds}:
we have inspected the density-temperature diagram for low-mass galaxies and we have verified that dwarves with $t_{\rm sf}\gsim 10\,$Gyr contain very little gas with $T<1,000\,$K.
We refer to this phenomenon with the name of anti-condensation feedback.

The same conclusion can also be reached from Fig.~\ref{fig_tsf}. The star-formation timescale for gas with $T<1,000$ 
is always short and decreases at low $M_\star$ (in agreement with \citealp{boselli_etal14}).
Hence, the molecular gas always forms stars effectively. 

The proof that anti-condensation feedback is a genuine feedback effect is that nearly all the gas is converted into stars when stellar feedback is switched off
(in Fig.~\ref{fig_baryon_frac}, the blue triangles overlap with the red ones).

The data of \citet{boselli_etal14} do not extend below $M_\star=10^8\,M_\odot$. In fact, they contain very few galaxies with $M_\star < 10^9\,M_\odot$.
However, the NIHAO simulations predict that {\it the star-formation timescale for the {\sc Hi} gas cannot keep rising like} $t_{\rm sf}\propto M_\star^{-0.7}$ at low masses:
dwarves galaxies with  $M_\star<10^8\,M_\odot$ form so few stars that there are not enough SNe for effective feedback.
Ultradwarves with $M_\star<10^7\,M_\odot$ are predicted to exhibit highly stochastic SFRs with intense starbursts followed by long periods of quiescence.
Fig.~\ref{fig_tsf} shows that in these systems $t_{\rm sf}$ can take all values between $0.3$ and $200\,$Gyr.

\section{Mass loading}

\begin{figure}
\includegraphics[width=1.\hsize,angle=0]{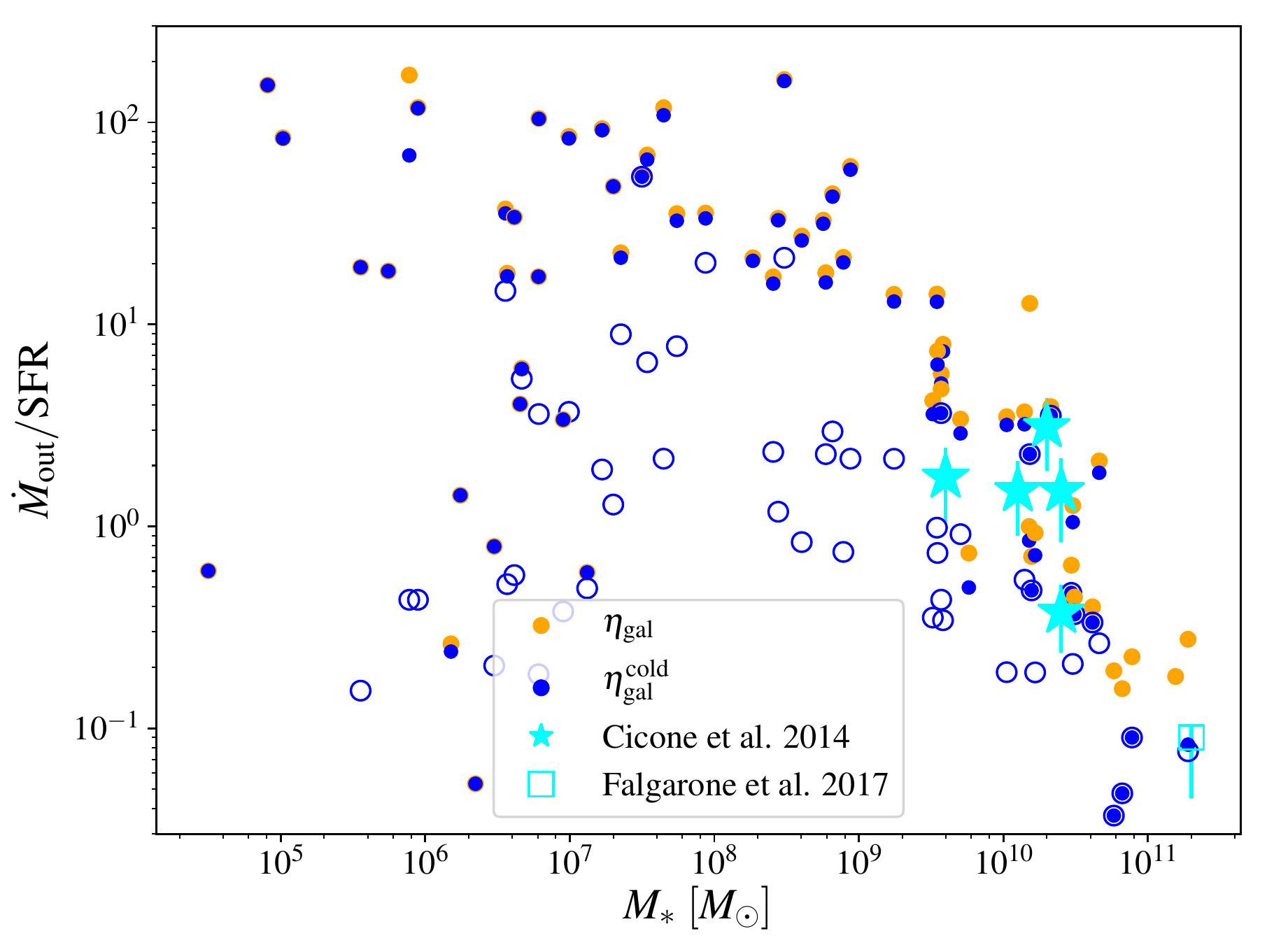}
\caption{Mass-loading factors for the total outflow rate ($\eta_{\rm gal}$, yellow circles) and the cold component only ($\eta_{\rm ga}^{\rm cold}$, blue filled circles) versus stellar mass $M_\star$ in the NIHAO simulations.
The blue open circles show $\eta_{\rm gal}^{\rm cold}$ when we consider only gas with $T<10^3\,$K.
The stars with error bars are the data for five nearby galaxies (M 82, NCG 3256, NGC 3628, NGC 253, NGC 2146).
They are based on CO observations of molecular outflows \citep{cicone_etal14}.
The open cyan square is obtained by stacking six submillimetre galaxies at $z\sim 2.3$ \citep{falgarone_etal17}.
The NIHAO simulations predict mass-loading factors that decrease with $M_\star$ at $M_\star>10^8\,M_\odot$ and stochastic mass-loading factors at lower masses.
}
\label{fig_eta_obs}
\end{figure}

\begin{figure*}
\begin{center}$
\begin{array}{cc}
\includegraphics[width=0.5\hsize]{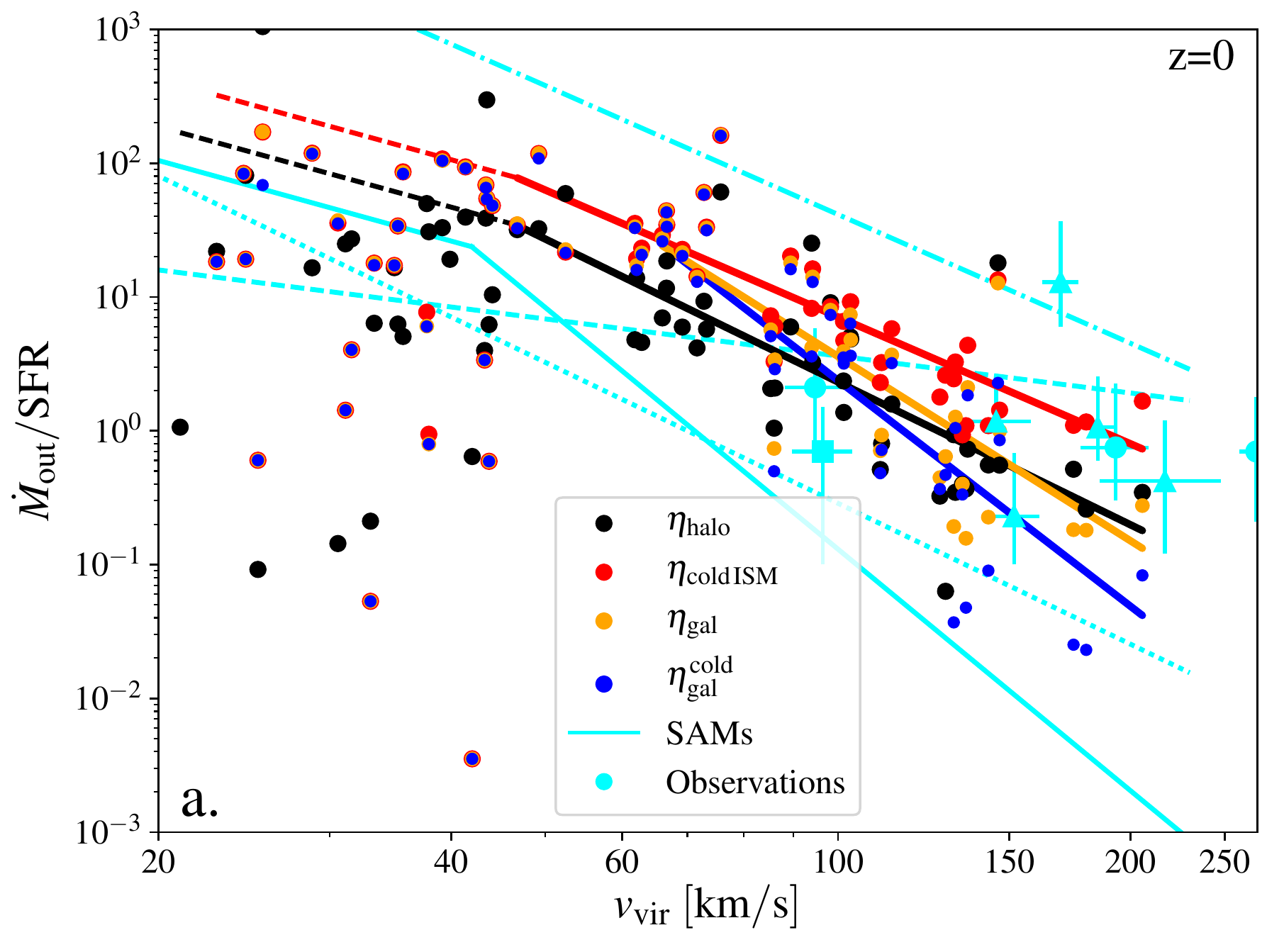} &
\includegraphics[width=0.5\hsize]{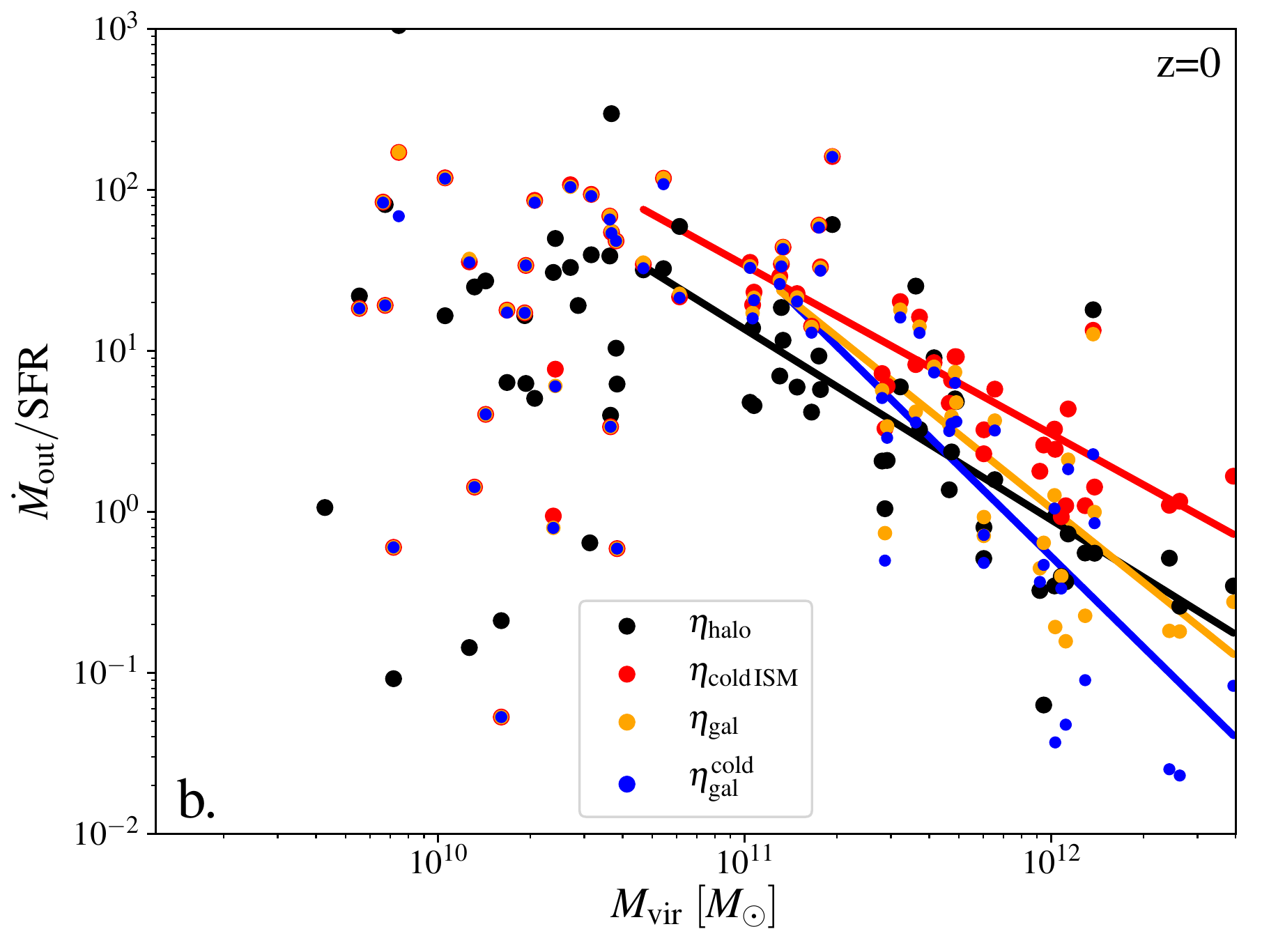} \\
\includegraphics[width=0.5\hsize]{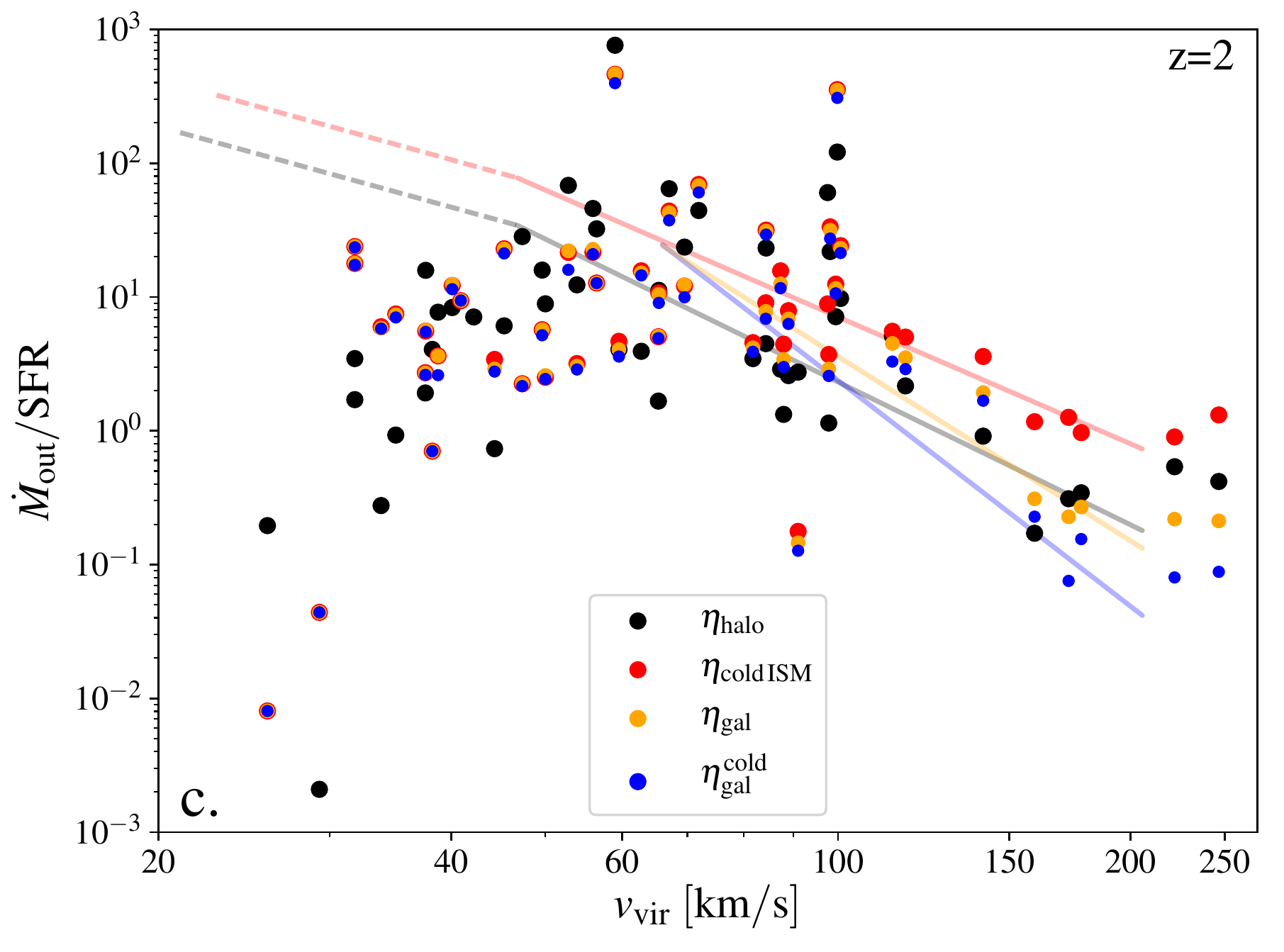} &
\includegraphics[width=0.5\hsize]{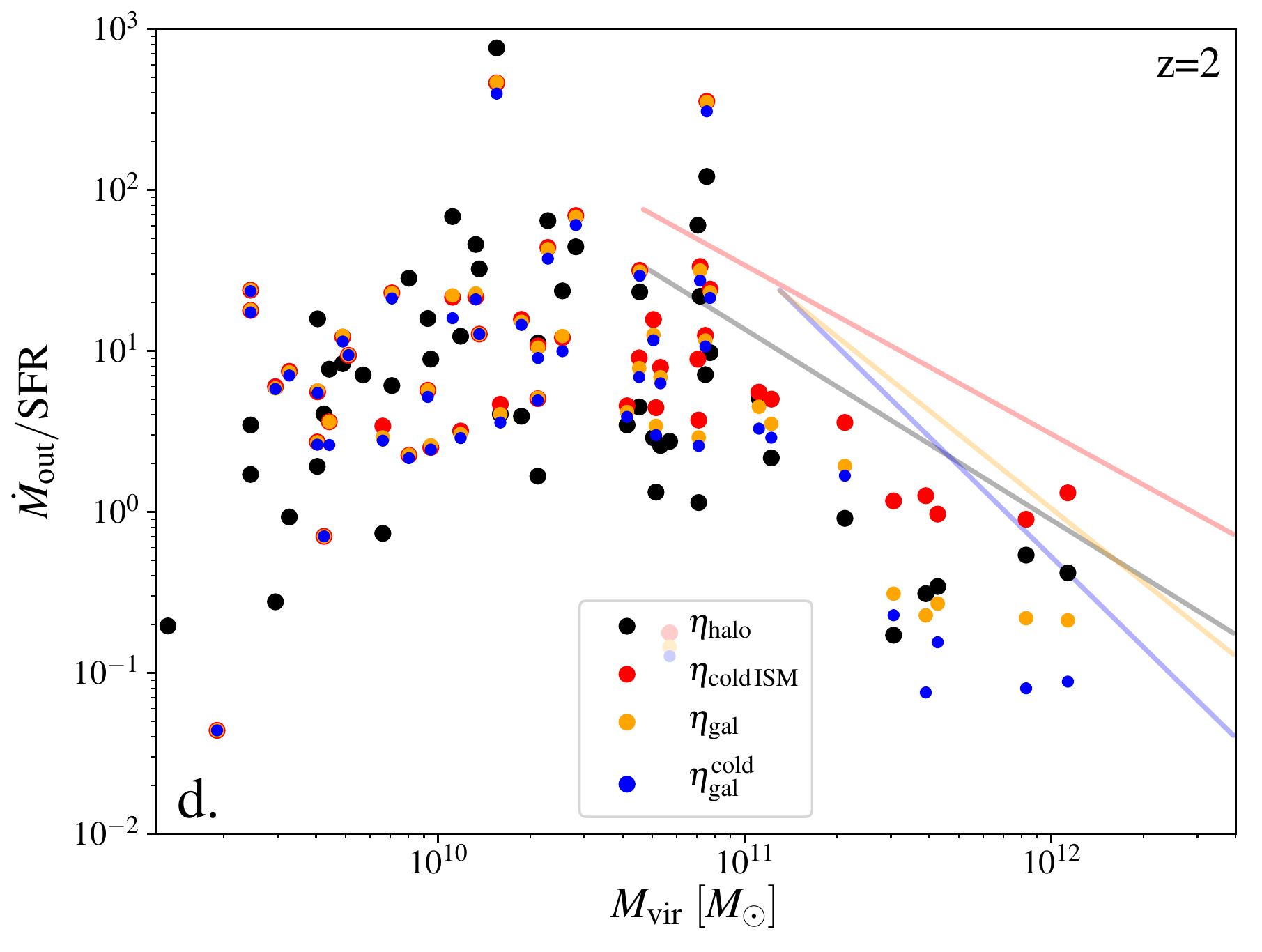}
\end{array}$
\end{center}
\caption{The mass-loading factors $\eta_{\rm cold\, ISM}$ (red symbols), $\eta_{\rm gal}$ (yellow symbols), $\eta_{\rm gal}^{\rm cold}$ (blue symbols) and  $\eta_{\rm halo}$ (black symbols).
Mass-loading factors are shown as a function of both virial velocity $v_{\rm vir}$ (left) and virial mass $M_{\rm vir}$ (right).
The upper and lower panels show the results of the NIHAO simulations at $z=0$ and $z=2$, respectively.
The thick red, blue, yellow and black solid lines are log-log linear least-square fits to the red, blue, yellow and black symbols at $z=0$, respectively.
The red dashed line and the black dashed line correspond to relations of the form $\eta\propto v_{\rm vir}^{-2}$ (as expected if feedback had a constant efficiency, e.g., $100\%$).
They have been repeated with a smaller linewidth on the panels at $z=2$ to show the evolution of $\eta(M_{\rm vir})$ and $\eta(v_{\rm vir})$ with redshift.
The cyan lines in the top left panel show the $\eta_{\rm gal}$ - $v_{\rm vir}$ relations in four semianalytic models (SAMs): 
\citet{guo_etal11}'s version of the Munich model (dotted line), \citet{henriques_etal13}'s version of the Munich model (dashed line), the Durham model (\citealp{lacey_etal16}; dotted-dashed line)
and GalICS~2.0 (\citealp{cattaneo_etal17}; solid line).
The cyan points with error bars in Fig.~\ref{fig_eta}a are the observations from Bouch{\'e} et al. (2012, triangles), Kacprzak et al. (2014, square) and Schroetter et al. (2015, circles). \citet{schroetter_etal15} observe galaxies at $z\sim 0.8$, \citet{kacprzak_etal14} at $z\sim 0.2$ and \citet{bouche_etal12} at $z\sim 0.1$.
}
\label{fig_eta}
\end{figure*}

In Section~4, we have seen that $90\%$ of the baryons that accrete onto a dwarf galaxy ($M_\star\sim 10^8\,M_\odot$) are expelled from the galaxy and that this fraction goes down to $\sim 20\%$
for massive spirals such as Andromeda and the Milky Way.
We have also seen that this is not the only reason why dwarf galaxies have very low $M_\star/M_{\rm vir}$ mass ratios. Inability to accrete (Section~4) and inefficient star formation (Section~5) play a role, too.
We now want to take our analysis one step forward and establish a direct link between star formation and outflows.
Our goal is to extract predictions that may be compared to observations and also to derive physically motivated prescriptions that may improve the description of feedback in semianalytic models of galaxy formation.

Let us start by introducing the mass-loading factor:
\begin{equation}
\eta={1\over{\rm SFR}}{\Delta M_{\rm out}\over\Delta t}, 
\label{eta_def}
\end{equation}
where $\Delta t$ is the time between the consecutive output times $t$ and $t+\Delta t$,
and where $\Delta M_{\rm out}$ is the mass of the gas particles ejected between $t$ and $t+\Delta t$.

In this article, we consider four different definitions of ejected particles:
\begin{enumerate}
\item particles that are in the cold ISM at $t$ but no longer belong to it at $t+\Delta t$;
\item particles that are in the cold ISM at $t$ but have moved to $r>r_{\rm g}$ at $t+\Delta t$;
\item particles that are in the cold ISM at $t$ but have moved or are in the process of moving to the cold CGM at $t+\Delta t$;
\item particles that are within the virial radius at $t$ but have moved to $r>r_{\rm vir}$ at $t+\Delta t$.
\end{enumerate}
Each definition serves a different purpose. The first gives the rate at which gas ceases to be available for star formation.
The second gives the rate at which gas flows out of the galaxy. The third is useful to compare with observations that trace the cold molecular component\footnote{Definition (iii)
includes hot ISM's particles in transit from the cold ISM to the cold CGM but few particles take this route.
When they do, the time spent in the intermediate hot-ISM phase is very short.
Therefore, particles identified with definition (iii) are effectively cold at all times.}.
The fourth gives the rate at which gas leaves the halo.
We use $\eta_{\rm cold\,ISM}$, $\eta_{\rm gal}$, $\eta_{\rm gal}^{\rm cold}$ and $\eta_{\rm halo}$, respectively, to denote the mass-loading factors computed with these four definitions.
Our definitions are such that, for each individual galaxy, $\eta_{\rm gal}^{\rm cold}\le\eta_{\rm gal}\le\eta_{\rm cold\,ISM}$ by construction.

\begin{table}
\begin{center}
\caption{Best-fit parameters for Eq.~(\ref{eta_fit}) at $z=0$.}
\begin{tabular}{ l c c}
\hline
\hline 
$\eta$   & $\eta_0$ & $\alpha$ \\
\hline
$\eta_{\rm cold\,ISM}$     & 7 &  -3.2$\pm$ 0.3\\
$\eta_{\rm gal}$          & 4 &  -4.6$\pm$ 0.5\\
$\eta_{\rm gal}^{\rm cold}$& 2  & -5.6$\pm$ 0.5\\
$\eta_{\rm halo}$         & 2 & -3.5$\pm$ 0.5\\
\hline
\hline
\label{table_eta}
\end{tabular}
\end{center}
\end{table}

Fig.~\ref{fig_eta_obs} compares the mass-loading factors from the NIHAO simulations with those estimated by \citet{cicone_etal14} for M82, NGC~253, NGC~2146, NGC~3256 and NGC~3628
(we took the sample by Cicone et al. and removed all objects in which they found evidence for an active galactic nucleus).
The greatest difficulty of this comparison was to find reliable masses for such nearby objects. Their apparent sizes are so large that the SDSS and 2MASS miss a lot of the light from the outer regions.
We therefore reverted to extinction-corrected B-, V- and R-band photometry from the  Third Reference Catalogue of Bright Galaxies
\citep{devaucouleurs_etal91} and converted luminosities into masses using stellar mass-to-light ratios from {\sc pegase} \citep{fioc_rocca97}, which kept colours into account.
The good agreement of $\eta_{\rm gal}$ (yellow circles) with the observations becomes excellent when the comparison is based on the cold component $\eta_{\rm gal}^{\rm cold}$ (filled blue circles) only,
as it should be, since the points by \citet{cicone_etal14} are based on  {\sc CO} data.
We stress that no free parameter was tuned to obtain this result.

The point from \citet{falgarone_etal17} refers to a different redshift range ($2<z<2.6$ instead of the local Universe) and carries considerable uncertainty\footnote{\citet{falgarone_etal17}
used ALMA data to measure mass-loading factors for six submillimetre galaxies with SFR$= 300 - 13000\,M_\odot{\rm\,yr}^{-1}$ at $2<z<2.6$ and found values of $\sim 0.1$ for all of them (within the large uncertainties of their
measurements). We have no information about the masses of these galaxies but we have estimated their order of magnitude from the SFR - $M_\star$ 
at $z\sim 2$ \citep{daddi_etal07,tasca_etal15}. $M_\star\sim 2\times 10^{11}\,M_\odot$ is the average stellar mass of a galaxy with  SFR$= 300\,M_\odot{\rm\,yr}^{-1}$ at $z=2$.
This calculation is more likely to overestimate $M_\star$ than underestimate it.}.
We have added it nevertheless because our  $\eta_{\rm gal}^{\rm cold}$ - $M_\star$ relation at $z=2.3$ is not significantly different from the one at $z=0$
and this point is our only constraint for $\eta_{\rm gal}^{\rm cold}$ at high masses. \citet{falgarone_etal17}'s data point
shows that the steep decrease of $\eta_{\rm gal}^{\rm cold}$ with $M_\star$ is real and that the NIHAO simulations reproduce the correct relation over two orders of magnitude in stellar mass.

The mass-loading factor $\eta_{\rm gal}^{\rm cold}$ depends on our definition of cold gas. The filled blue circles in Fig.~\ref{fig_eta_obs} consider all the gas identified as belonging to the cold ISM in Fig.~\ref{fig_rhoT}b. As the mass-loading factor from Cicone et al. (2014) and Falgarone et al. (2017) are inferred from Co and CH+ tracers, respectively one may wish to apply a more restrictive criterion and consider cold only the molecular gas ($T < 1000\,\rm K$).
When we do it the blue filled circles become the blue open circles, which are in good agreement with the observations at high masses but seems to underestimate mass-loading factor at lower masses.
One should notice, however, that only massive NIHAO galaxies exhibit a dense molecular phase distinct from the atomic gas. In lower mass galaxies, the cold ISM does not contain enough particles for this phase to be properly resolved.

One should also be aware that Cicone et al. measured outflow rates by assuming that the outflowing clouds populate the spherical (or multi-conical) region affected by outflows uniformly. This assumption is reasonable if outflows are continually refilled with clouds ejected from the disc but it overestimates $\eta_{\rm gal}^{\rm cold}$ for stochastic outflows (which we find to be prevalent in dwarf galaxies but not in massive objects, see Section 7).

The comparison with observations has made us confident that the NIHAO simulations approximate the Universe reasonably well.
With this confidence, we are now ready to study the dependence of $\eta$ on  the virial velocity $v_{\rm vir}$ and the virial mass $M_{\rm vir}$ for each of our four definitions of the mass-loading factor. 
The red, yellow, blue and black symbols in Fig.~\ref{fig_eta}a correspond to $\eta_{\rm cold\,ISM}$, $\eta_{\rm gal}$, $\eta_{\rm gal}^{\rm cold}$ and $\eta_{\rm halo}$, respectively,
while the red, yellow, blue and black solid lines are log-log linear least-square fits of the form:
\begin{equation}
\eta = \eta_0\left({v_{\rm vir}\over 100{\rm\,km\,s}^{-1}}\right)^\alpha ,
\label{eta_fit}
\end{equation}
to the symbols of matching colours. 

Table~\ref{table_eta} gives the best-fit parameters $\eta_0$ and $\alpha$ for each definition of $\eta$  when we stop the fits at 
$v_{\rm vir}=45{\rm\,km\,s}^{-1}$.
Below this virial velocity, the scatter in $\eta$ becomes huge and the fits would be meaningless.
This scatter is due to the stochastic SFR histories of dwarves, characterised by intermittent bursts of star formation
 interspersed between periods of quiescence
(there can be no outflows without type II SNe and there can be no type II SNe without star formation).

The yellow line and the blue line in Fig.~\ref{fig_eta}a have been computed fitting all the points at  $v_{\rm vir}>45{\rm\,km\,s}^{-1}$ 
but we have stopped them at $v_{\rm vir}=67{\rm\,km\,s}^{-1}$ because
it would be meaningless to extrapolate the fits where $\eta_{\rm gal}^{\rm cold}>\eta_{\rm gal}$ (the cold component cannot larger than the total outflow rate)
or where $\eta_{\rm cold\,ISM}<\eta_{\rm gal}$
(including gas that is heated but remains within the galaxy cannot give lower outflow rates than excluding it).
The steeper decline of the blue line compared to the yellow line and of the yellow line compared to the red line shows that at low masses most of the mass is the cold component,
while in massive galaxies plenty of gas is heated by SNe but does not escape.

The cyan points with error bars in Fig.~\ref{fig_eta}a compare the results of the NIHAO simulations to measurements from {\sc MgII} absorption lines by Bouch{\'e} et al. (2012, triangles), Kacprzak et al. (2014, square) and Schroetter et al. (2015, circles), which trace warm ionised gas with outflow speeds of a few hundred kilometres per seconds which are intermediate between those of cold molecular gas and hot X-ray gas.
We also remark that the red line in Fig.~\ref{fig_eta}b is consistent with the $\eta_{\rm c}$ - $M_{\rm vir}$ relation at $z=0$ in \citet{angles-alcazar_etal17} and that the black line is consistent with the $\eta_{\rm loss}$ - $M_{\rm vir}$ relation
at $z=0$ in the same article. \citet{angles-alcazar_etal17} define $\eta_{\rm c}$ as the ratio of the total mass ejected in winds from the ISM (regardless of its fate after ejection) to the total stellar mass formed in situ
from early times down to $z=0$, and $\eta_{\rm loss}$ as the same ratio counting only the particles that have been ejected and never reaccreted.

Besides the sporadic star formation of dwarves discussed above (also see Section~5), there is another physical reason why $\eta$ cannot not keep growing as $v_{\rm vir}^\alpha$ with $\alpha\ll -2$ at low $v_{\rm vir}$ \citep{cattaneo_etal17}. 
Outflows require speeds greater than the escape speed $v_{\rm esc}$ and this sets an upper limit to outflow rate:
\begin{equation}
\dot{M}_{\rm out,\, max}= {2\epsilon_{\rm SN}E_{\rm SN}\Psi_{\rm SN}\over v_{\rm esc}^2}{\rm SFR},
\label{eta_max}
\end{equation}
\citep{silk03}, where $\Psi_{\rm SN}$ is the number of SNe per unit stellar, $E_{\rm SN}$ is the energy released by one SN, and $\epsilon_{\rm SN}$ is the efficiency with which the energy from SNe is converted into kinetic energy.
As $v_{\rm esc}\propto v_{\rm vir}$, Eq.(\ref{eta_max}) implies $\eta_{\rm max}=\dot{M}_{\rm out,\,max}/{\rm SFR}\propto v_{\rm vir}^{-2}$ if we assume that $\epsilon_{\rm SN}$ is constant.

In reality, there is no reason why $\epsilon_{\rm SN}$ should be constant. On the contrary, there are physical reasons why $\epsilon_{\rm SN}$ should decrease with $v_{\rm vir}$.
One of them is that massive haloes contain massive atmospheres of shock-heated gas \citep{dekel_birnboim06}.
The efficiency of SNe is limited by the radiative efficiency of the heated gas, which is higher if the gas  is denser.
SNe deposit their energy in the ISM, where even the dilute hot component is relatively dense.
Only by expanding faster than it can cool can gas retain enough energy to produce an outflow.
The hot CGM of massive galaxies confine the outflowing gas and slow down its expansion. Hence, the outflowing gas radiates a much larger fraction of its internal energy.

In Section~7, we shall present another physical reason why $\epsilon_{\rm SN}$ should decrease with $v_{\rm vir}$. 
However, the NIHAO simulations themselves show that this must be the case, since all fits in Table~\ref{table_eta} decrease much faster than $v_{\rm vir}^{-2}$.
So do semianalytic models of galaxy formation, since nearly all of them require $\alpha\ll -2$ to be consistent with the slope of the stellar mass function of galaxies at low masses
(see, e.g., \citealp{guo_etal11}, \citealp{lacey_etal16}, and \citealp{cattaneo_etal17}, but also \citealp{henriques_etal13}; cyan lines in  Fig.~\ref{fig_eta}a\footnote{
Fig.~\ref{fig_eta}a shows that the normalisation of $\eta$ varies wildly from one semianalytic model to another.
{\sc GalICS~2.0} \citep{cattaneo_etal17} reproduces the stellar mass function of galaxies with a lower normalisation than other models because in {\sc GalICS~2.0} the ejected gas cannot be reaccreted.}).

Still, $\epsilon_{\rm SN}$ cannot keep growing indefinitely at low $v_{\rm vir}$ because it cannot become larger than unity.
Hence, there must be a virial velocity below which $\eta\le\eta_{\rm max}\propto v_{\rm vir}^{-2}$ for all galaxies.
\citet{cattaneo_etal17} predicted this behaviour based on this consideration in the context of the {\sc GalICS~2.0} semianalytic model and argued that it could explain the upturn
of the stellar mass function of galaxies at low masses, which is observed both in the local Universe (e.g., \citealp{yang_etal09,baldry_etal12,moustakas_etal13}) 
and at higher redshifts (e.g., \citealp{tomczak_etal14}).

The NIHAO simulations confirm this prediction.
All galaxies with  $v_{\rm vir}<45{\rm\,km\,s}^{-1}$ lie below the red dashed line $\eta_{\rm gal}\simeq 90(v_{\rm vir}/45{\rm\,km\,s}^{-1})^{-2}$ in Fig.~\ref{fig_eta}a.
This line corresponds to $\epsilon_{\rm SN}=1$ for $E_{\rm SN}\Psi_{\rm SN}=8\times 10^{48}{\rm\,erg\,M}_\odot^{-1}$ (Section~2) and $v_{\rm esc}(r_{\rm g})\simeq 2.1v_{\rm vir}$, which is a reasonable estimate for the escape
speed from galaxies. Adding early stellar feedback increases the energy injected into the ISM by a factor of four.
However, the fact that we find $\epsilon_{\rm SN}=1$ for a reasonable escape speed when we consider the contribution of SN feedback alone
confirms the supposition that energy from early stellar feedback is dissipated much more efficiently and that its main role is to dilute the ISM in which SNe are about to explode.

The escape speed from the haloes is higher than the one from the galaxy. Hence, the outflow rate of the halo should be lower than the outflow rate from the galaxy by a factor $v_{\rm esc}^2(r_{\rm g})/v_{\rm esc}^2(r_{\rm vir})$
Assuming an NFW profile \citep{navarro_etal97}, the escape speed to $r>r_{\rm vir}$ from the centre of the halo is:
\begin{equation}
v_{\rm esc}(r_{\rm vir}) = \sqrt{2[c-{\rm ln}(1+c)]\over{\rm ln}(1+c)-c/(1+c)}v_{\rm vir},
\label{v_esc}
\end{equation}
where $c$ is the concentration of the NFW profile. The fitting formulae by \citet{dutton_maccio14} give $c\simeq 14$ for a halo with $v_{\rm vir}=45{\rm\,km\,s}^{-1}$ in a cosmology with $\Delta_{\rm c}=109$ at $z=0$ (Section~3.1).
For this concentration, Eq.~(\ref{v_esc}) gives $v_{\rm esc}(r_{\rm vir})\simeq 3.6v_{\rm vir}$. Hence, we expect $\eta_{\rm halo}/\eta_{\rm gal}\sim (2.1/3.6)^2\simeq 0.3$ at low virial velocities.
Fig.~\ref{fig_eta}a shows that this very crude estimate is not far off from reproducing correctly the ratio of the black dashed line to the red dashed line.

\begin{figure*}
\begin{center}$
\begin{array}{cc}
\includegraphics[width=0.36\hsize]{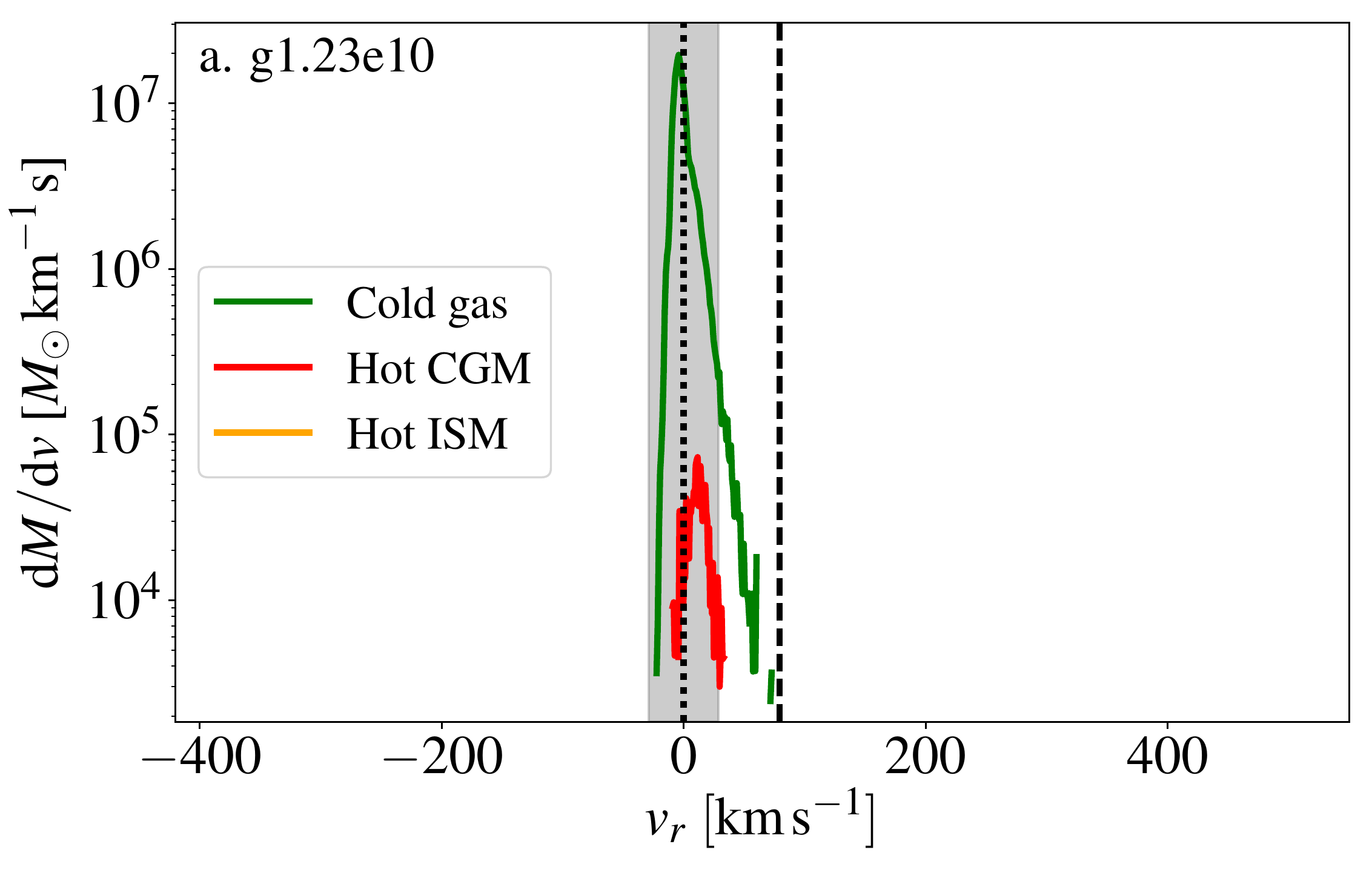} &
\includegraphics[width=0.36\hsize]{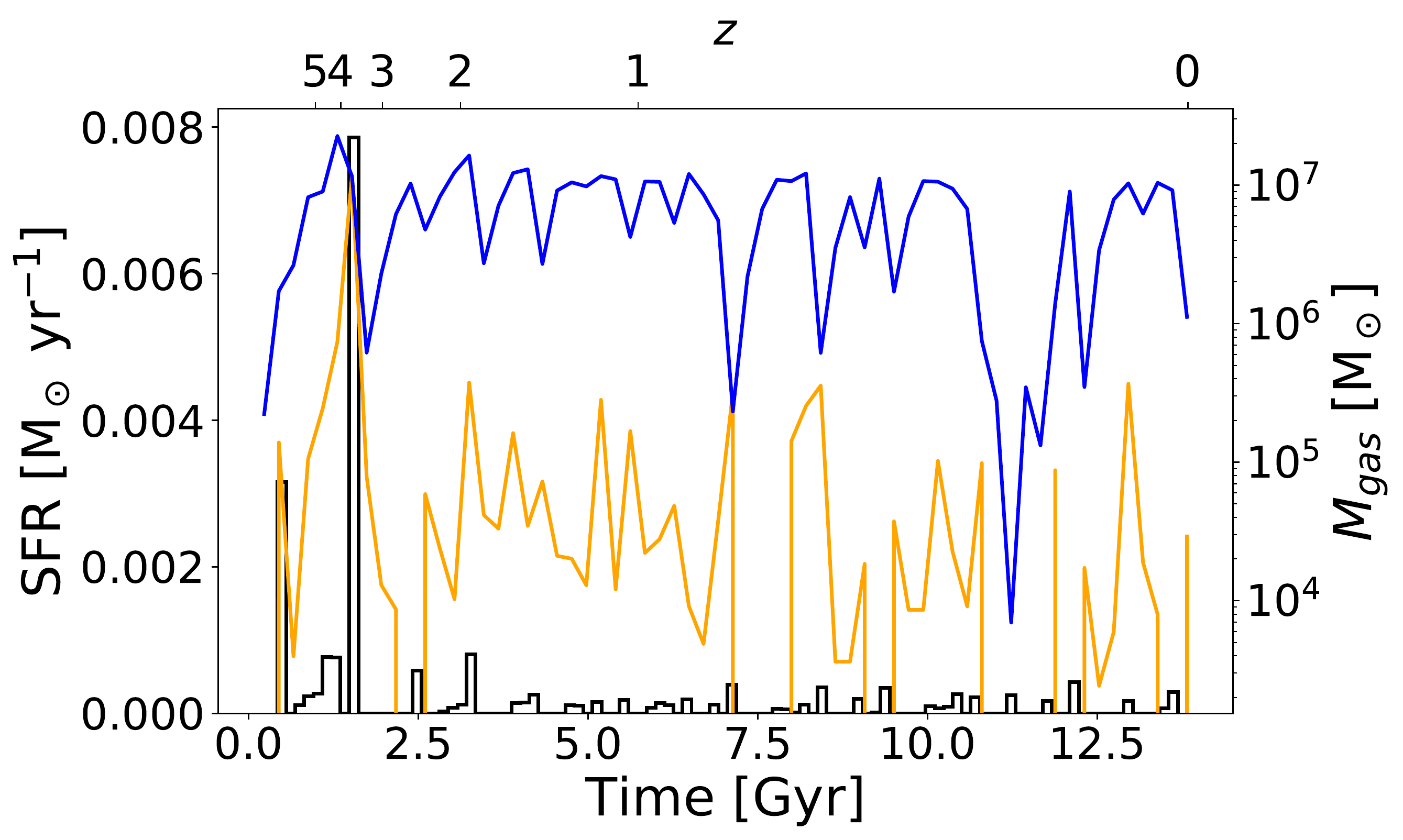} \\
\includegraphics[width=0.36\hsize]{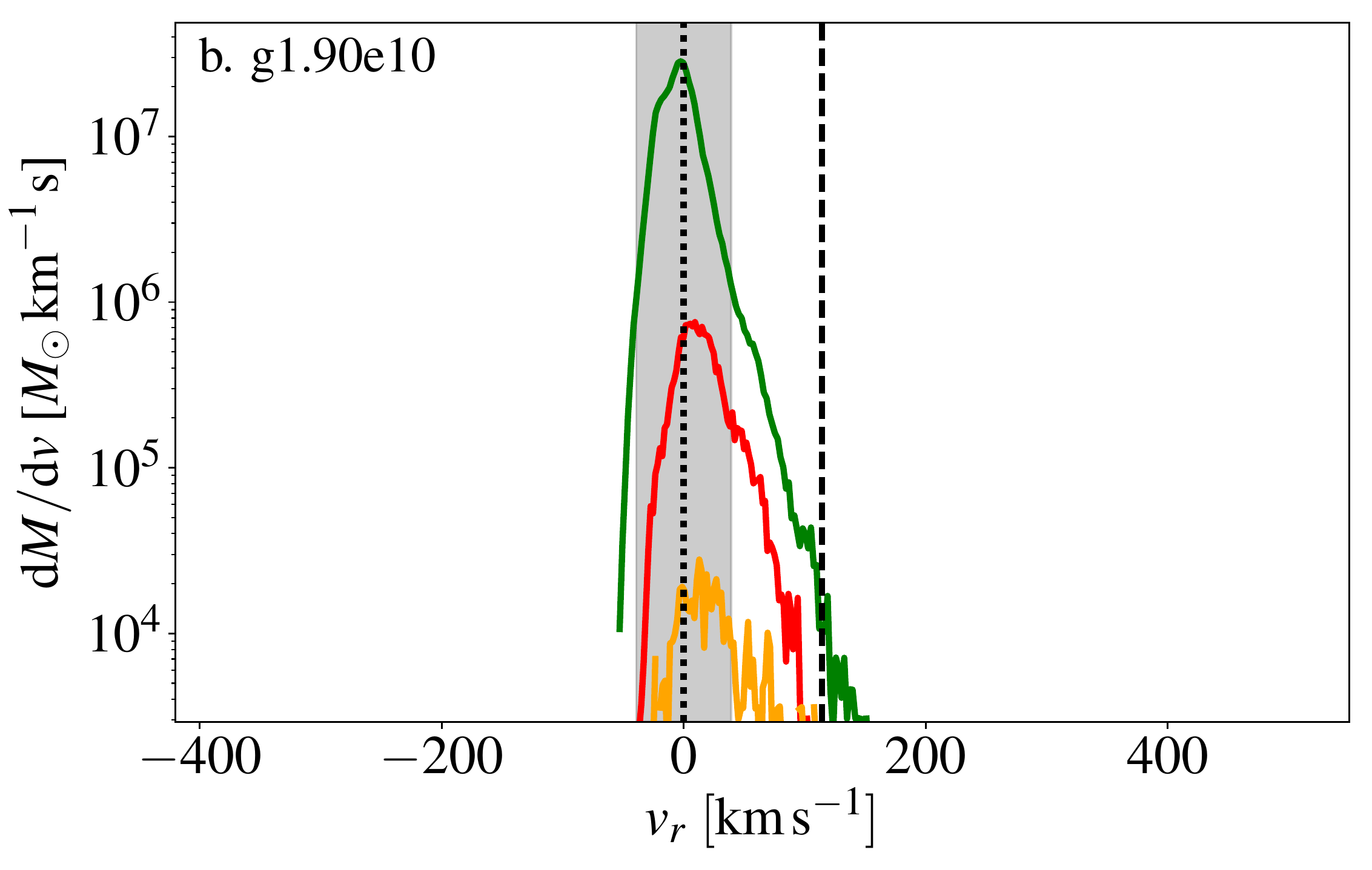} &
\includegraphics[width=0.36\hsize]{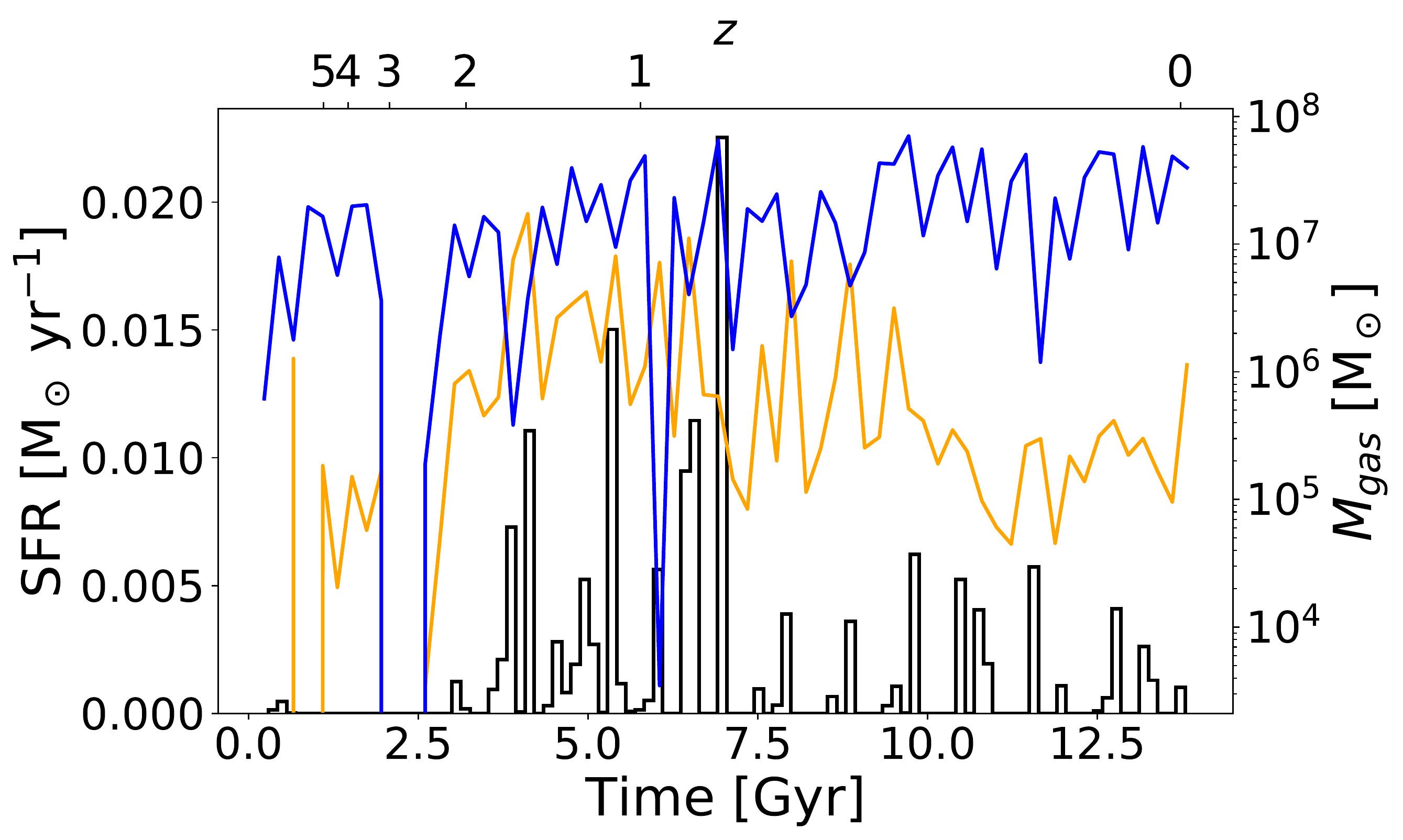} \\ 
\includegraphics[width=0.36\hsize]{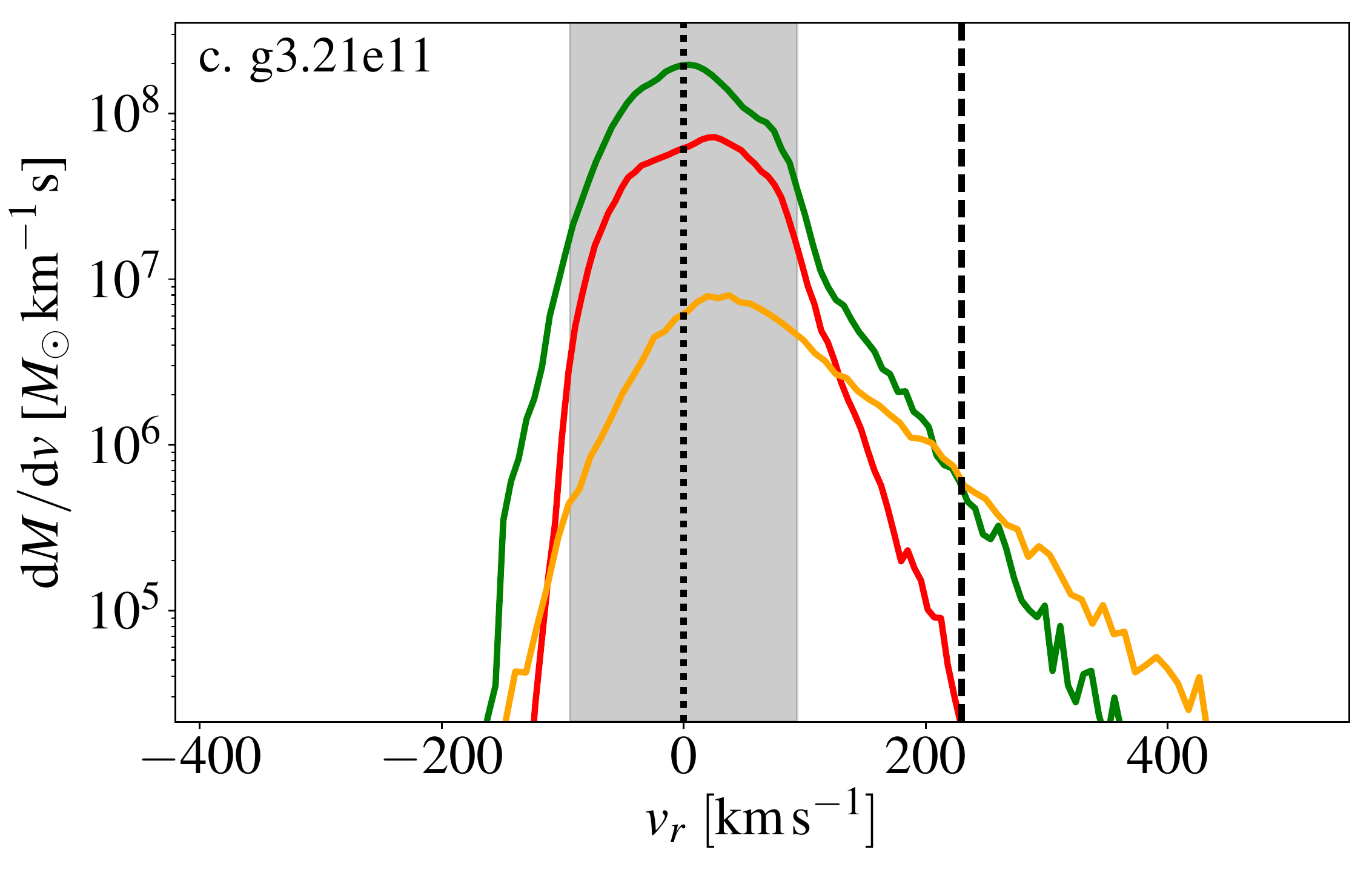} &
\includegraphics[width=0.36\hsize]{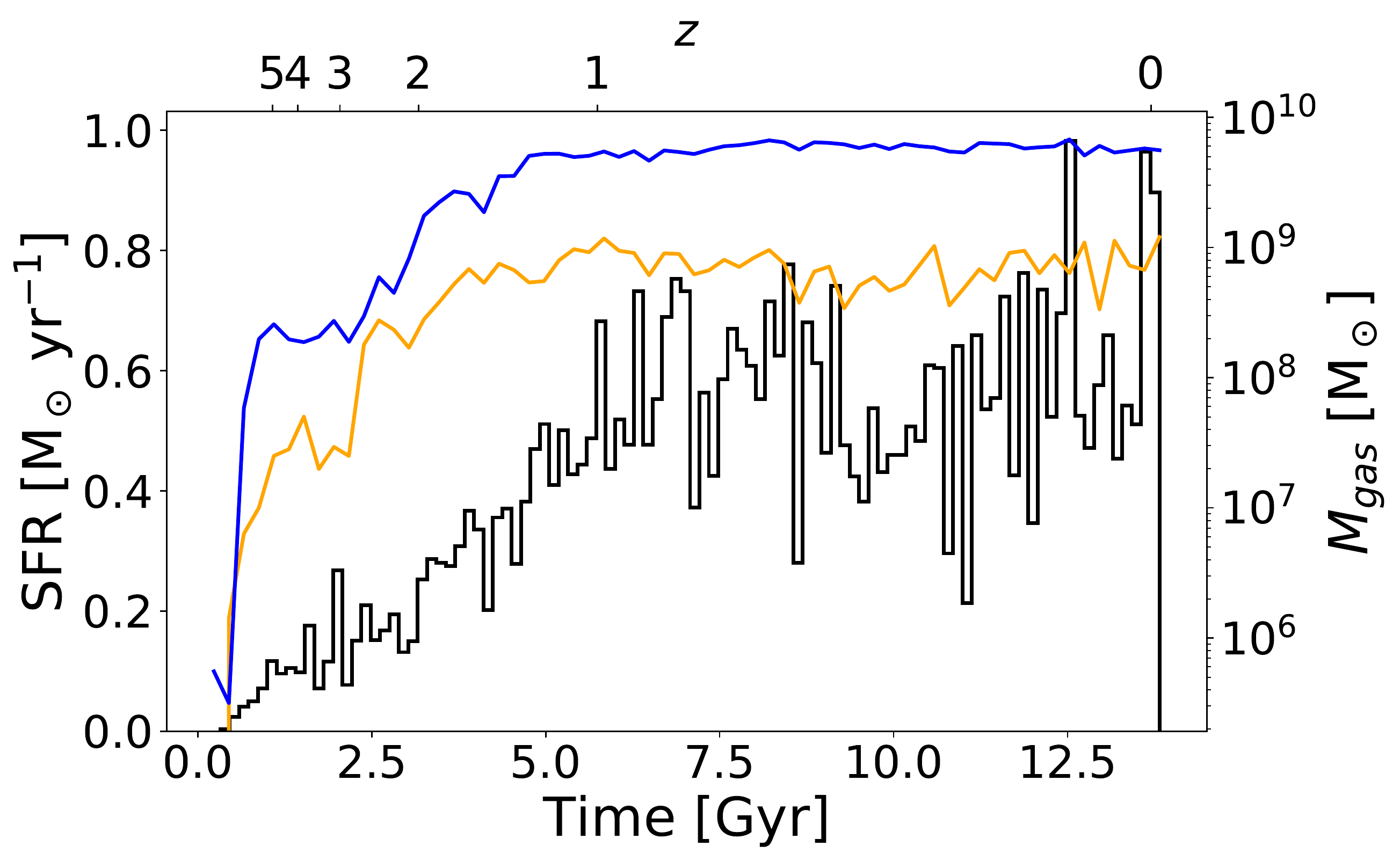} \\
\includegraphics[width=0.36\hsize]{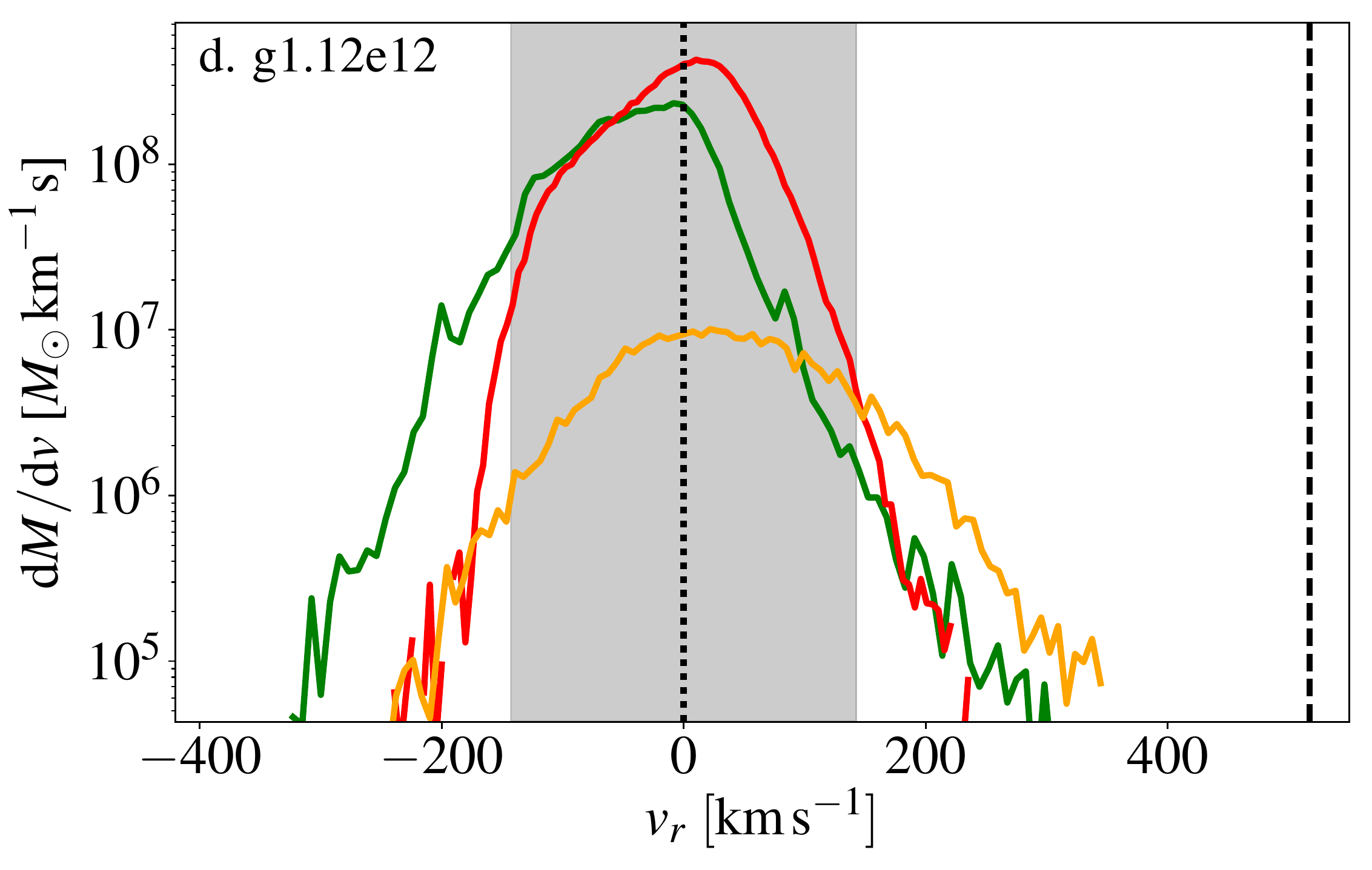} &
\includegraphics[width=0.36\hsize]{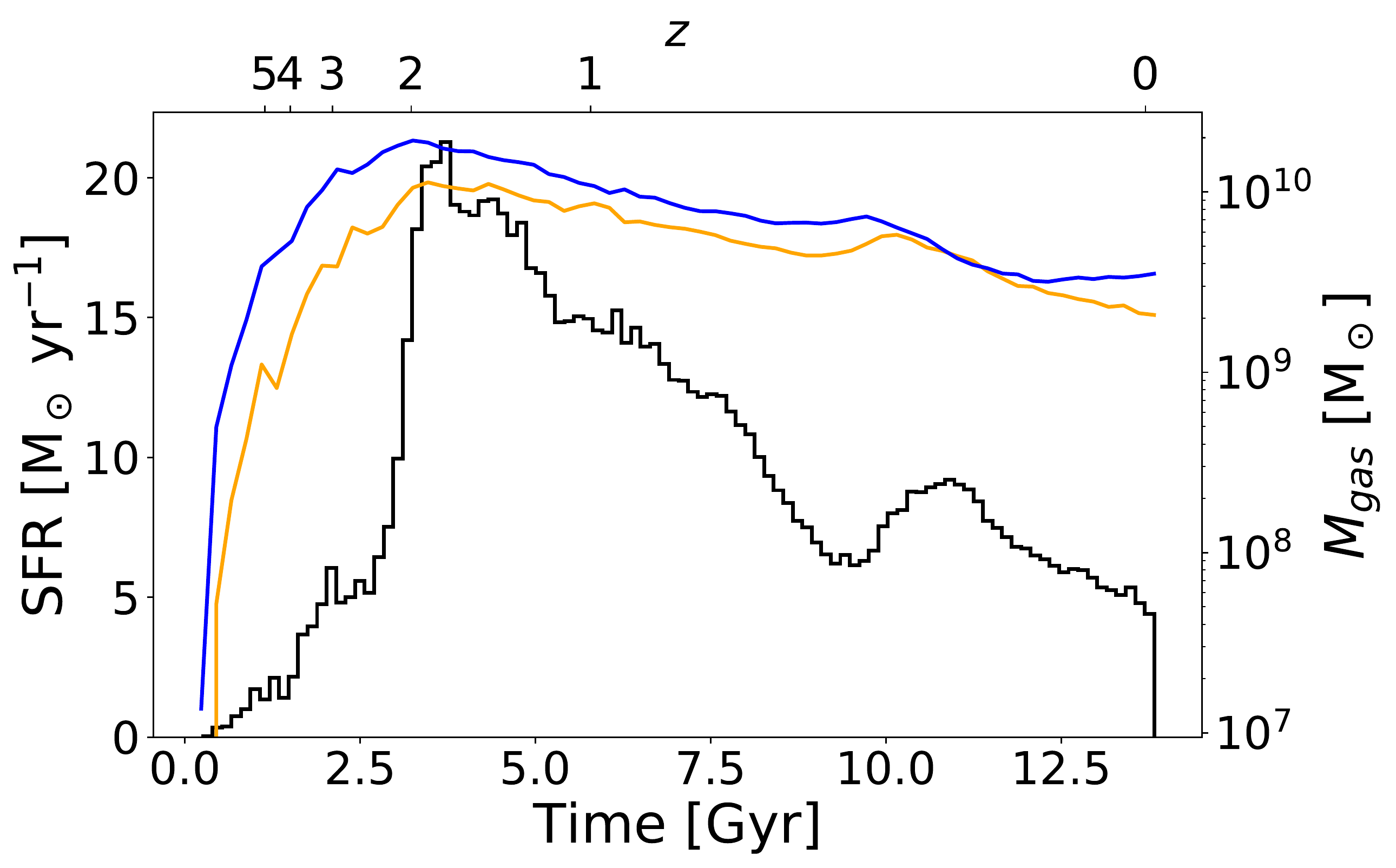}
\end{array}$
\end{center}
\caption{Left: radial-velocity distribution of the cold gas (ISM and CGM, green curves), the hot ISM (yellow curves) and the hot CGM (red curves) for the four objects in Table~\ref{table_vr}
arranged by growing $M_{\rm vir}$.
The normalisations of the curves reflect the masses in the three phases (the ordinate range is not the same in all panels. The vertical dashed lines show the escape speeds
$v_{\rm esc}$ from the centres of the haloes.
The shaded areas corresponds $-v_{\rm vir} < v_r < v_{\rm vir}$.
Right: corresponding SFR histories (black histograms) and evolution of the mass in the hot ISM (yellow curves) and in the cold CGM (blue curves). If the SFR rises, so does the mass in the hot ISM.
Notice that we use a linear scale SFRs and a logarithmic one for $M_{\rm hot\,ISM}$.}
\label{fig_vr_distribution}
\end{figure*}

\begin{table*}
\begin{center}
\caption{Virial masses, stellar masses, escape speeds, and mean radial velocities of the cold gas, the hot CGM and the hot ISM for the four NIHAO galaxies in Fig.~\ref{fig_vr_distribution}.
All speeds are in $\rm km\,s^{-1}$.}
\begin{tabular}{ l l l l l l l}
\hline
\hline 
Name & $M_{\rm vir}$ [$M_\odot$] & $M_{\rm *}$ [$M_\odot$] & $v_{\rm esc}$ & $\bar{v}_r^{\rm cold}$ & $\bar{v}_r^{\rm hot\, ISM}$ & $\bar{v}_r^{\rm hot\,CGM}$\\
\hline
g1.23e10 & $1.05\times 10^{10}$ & $1.60\times 10^{6}$ & $79.4$ & $-0.85$ &  not detected & $12.0$\\
g1.90e10 & $2.70\times 10^{10}$ & $1.20\times 10^{7}$ & $114$ & $-2.61$ & $23.1$         & $15.2$ \\
g3.21e11 & $3.61\times 10^{11}$ & $3.73\times 10^{9}$ & $230$ & $6.67$ & $57.0$          & $13.6$ \\
g1.12e12 & $1.29\times 10^{12}$ & $7.95\times 10^{10}$ & $517$ & $-41.0$ & $32.4$        & $-3.66$ \\
\hline
\hline
\label{table_vr}
\end{tabular}
\end{center}
\end{table*}


Fig.~\ref{fig_eta}a shows the relation between $\eta$ and $v_{\rm vir}$. As $M_{\rm vir}\propto v_{\rm vir}^3$,
it is not surprising that we find a similar relation between $\eta_{\rm galaxy}$ and $M_{\rm vir}$ (Fig.~\ref{fig_eta}b).
However, the proportionality constant between $M_{\rm vir}$ and $v_{\rm vir}^3$ depends on redshift.
Hence, the dependences of $\eta$ on $v_{\rm vir}$ and $M_{\rm vir}$ cannot both remain the same at $z=0$ and $z=2$. At least one has to evolve.
The one that evolves less is the more fundamental one.
Figs.~\ref{fig_eta}c,d show that the $\eta$ - $v_{\rm vir}$ relation is more or less the same at at $z=2$ and $z=0$, whereas, for a same $M_{\rm vir}$,
$\eta$ is lower at $z=2$, since the escape speed for a same virial mass is higher at high redshift.
This finding confirms the theoretical expectation that the gas's ability to escape is linked to the depth of the gravitational potential well more than it is to halo mass.

We remark that the picture presented in this section is in stark contrast with the one by Muratov et al. (2015, FIRE), who claimed that the logarithmic slope of $\eta(v_{\rm vir})$
is steeper for ultradwarves ($\alpha=-3.2$), shallower for dwarves ($\alpha=-1$), and then very steep again for massive spirals. \citet{muratov_etal15} attribute the change of exponent
from $\alpha=-3.2$ to $\alpha=-1$ around $v_{\rm vir}=60{\rm\,km\,s}^{-1}$ to a transition from energy-driven to moment-driven winds.
Had the NIHAO simulations included momentum feedback (e.g., radiation pressure) and not only energy feedback (thermal feedback from SNe and earlier phases of stellar evolution),
we might have discovered an intermediate range of $v_{\rm vir}$ where $\eta(v_{\rm vir})$ is shallower than what we find, although it is equally possible that scatter might have blurred any such effect
(we have already remarked that  Fig.~\ref{fig_eta}b is consistent with a later FIRE publication by \citealp{angles-alcazar_etal17}). 
We consider, however, that the exponent $\alpha=-3.2$ at  $v_{\rm vir}<60{\rm\,km\,s}^{-1}$ is unphysical,
and that it derives from defining as outflowing any particle with positive radial velocity, even if its speed is lower than the escape speed from the galaxy,
whereas the behaviour that we find is much more consistent with the episodic nature of star formation and outflows in ultradwarves.

We conclude this section by comparing the lessons from mass-loading factors and ejected fractions.
In Section~4, we saw that $f_{\rm ej}^{\rm gal}$ decreases with $M_{\rm vir}$ faster than $f_{\rm ej}^{\rm halo}$ (Fig.~\ref{fig_ej_fraction}).
Hence, we may have expected that $\eta_{\rm cold\,ISM}$ would decrease with $v_{\rm vir}$ faster than $\eta_{\rm halo}$
($\eta_{\rm cold\,ISM}$ is the relevant quantity for this comparison because the calculation of  $f_{\rm ej}^{\rm gal}$ is based on $M_{\rm gal}$,
the total mass of stars and cold neutral gas within a galaxy).
On the contrary, the $\alpha$ parameters for $\eta_{\rm cold\,ISM}$ and $\eta_{\rm halo}$ are very similar  (Table~\ref{table_eta})
and $\eta_{\rm halo}\sim \eta_{\rm galaxy}/3$ on average. If anything, it is $\eta_{\rm halo}$ that decreases slightly faster (compare the red line and the black line in Fig.~\ref{fig_eta}a).
This apparent paradox arises because  $f_{\rm ej}^{\rm gal}$ is linked to the mass $M_{\rm gal,acc}-M_{\rm gal}$ of the baryons that are no longer in the galaxy at $z=0$ but have been inside it at some earlier epoch,
while $\eta_{\rm cold\,ISM}$ is linked to the instantaneous rate $\dot{M}_{\rm out,\,cold\,ISM}$ at which gas leaves the cold ISM. If gas that leaves the cold ISM never came back, then:
\begin{equation}
M_{\rm gal,acc}-M_{\rm gal} = \int \dot{M}_{\rm out,\,cold\,ISM}{\rm\,d}t,
\end{equation}
but we shall see that this is not the case (Section~8). Hence, the different behaviour of $f_{\rm ej}^{\rm gal}$ and $\eta_{\rm cold\,ISM}$ must be due to reaccretion:
$f_{\rm ej}^{\rm gal}$ decreases more rapidly than $\eta_{\rm cold\,ISM}$ because, in massive galaxies, a lot of gas is ejected and reaccreted.
This gas contributes to $\eta_{\rm cold\,ISM}$ but not to $f_{\rm ej}^{\rm gal}$.


\section{Outflow speed and multiphase structure}

In Section~6, we investigated the rate at which gas is expelled from galaxies and we saw that separating the cold component from the total outflow rate is important for an accurate comparison with the observations (Fig.~\ref{fig_eta_obs}).
Here, we analyse the outflow speed of each component by studying its distribution of radial velocity $v_r$,
so that we can separate inflows ($v_r < -v_{\rm vir}$), outflows ($v_r > v_{\rm vir}$), and hot gas in hydrostatic equilibrium or cold galactic gas ($|v_r|\ll v_{\rm vir}$).

Fig.~\ref{fig_vr_distribution} shows the distribution of radial velocity $v_r$ for the cold gas (cold ISM plus cold IGM), the hot ISM, and the hot CGM of the four galaxies listed in Table~\ref{table_vr}.
Galaxies g1.23e10 and g1.90e10 typify the bimodality of star formation in ultradwarves, which can be very inefficient (as in g1.23e10) or very intense (as in g1.90e10).
Galaxies g3.21e11 and g1.12e12 are representative of massive dwarves/low-mass spirals and massive spirals, respectively, which constitute a more homogeneous population than ultradwarves,
even though significant differences from one galaxy to another are observed.

The first result that emerges from Fig.~\ref{fig_vr_distribution} is that {\it the presence of a hot ISM is linked to recent star formation}.
The passive ultradwarf g1.23e10 displays neither outflows nor a hot ISM.
In contrast, a hot ISM is present in the other three galaxies, all of which display star formation and outflows.

Secondly, when dwarf galaxies do exhibit outflows (as in g1.90e10), they are normally dominated by the cold component.
At higher masses, the hot dense component carries most of the gas with $v_r>200{\rm\,km\,s}^{-1}$.
This is the same result that we had already found under a different guise in Figs.~\ref{fig_eta_obs} and~\ref{fig_eta}.
Notice, however, that the bulk of the hot dense gas is not outflowing (all galaxies in Table~\ref{table_vr} have $\bar{v}_r^{\rm hot\,ISM}<0.2 v_{\rm esc}$).
This is why we call this phase hot ISM and not wind.

Finally, the distributions of outflow speeds in g3.21e11 and g1.12e12 are not very different, but the escape speeds are. The vertical dashed lines in Fig.~\ref{fig_vr_distribution} show
the escape speeds from $r_{\rm vir}$ computed with Eq.~\ref{v_esc} but the argument that follows would still apply if we considered escape speeds from $r_{\rm g}$ instead.
In g3.21e11, the gas fraction with $v_r>v_{\rm esc}$ is small (relative to the total mass of gas)\footnote{The observation that the mass in the wind is always small compared to the mass of the galaxy
does not preclude the possibility that most of the baryons may be ejected. There are many more healthy people than sick ones. Yet, most of us will fall ill and die at some point.} 
but still significant.
In g1.12e12, no outflow is detected at $z=0$. 
{\it In massive galaxies, SNe waste a lot of energy accelerating particles that will not escape}.
Since the efficiency of SNe that is relevant for the calculation in Eq.~(\ref{eta_max}) refers to the kinetic energy of particles with $v_r>v_{\rm esc}$,
this is another reason why $\epsilon_{\rm SN}$ decreases with $v_{\rm vir}$
in addition to the appearance of confining atmospheres at high masses discussed in Section~6
(comparing Fig.~\ref{fig_vr_distribution}c and Fig.~\ref{fig_vr_distribution}d shows that the hot CGM is more important in g1.12e12 than it is in g3.21e11).

\section{The baryon cycle}

Having discussed accretion (Section~4), star formation (Section~5) and outflows (Sections~6 and~7),
we are now ready to address the fate of the ejected gas and close the loop of the galaxy baryon cycle.

\begin{figure}
\includegraphics[width=1.\hsize,angle=0]{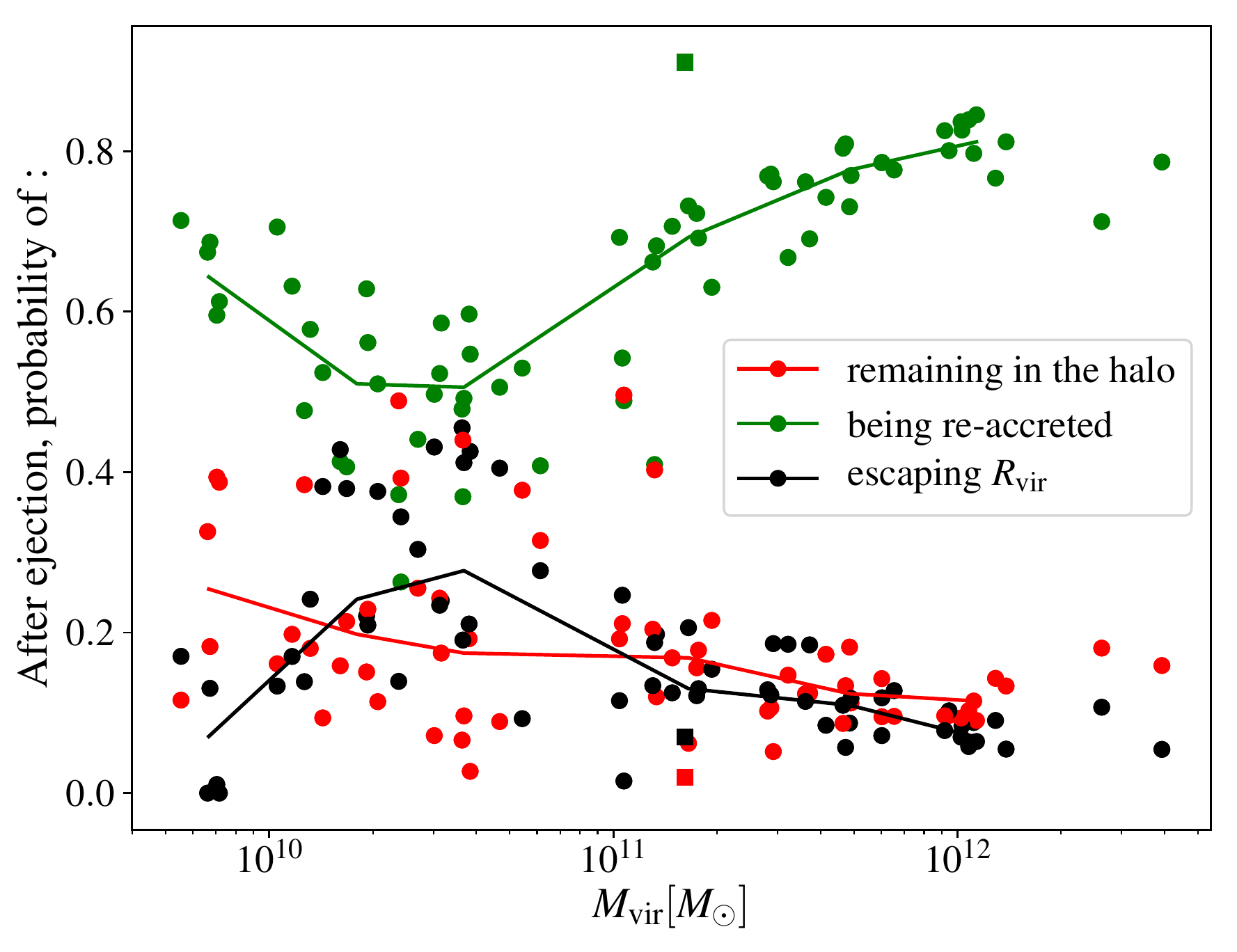}
\caption{Fraction of the ejected mass from the galaxy that escapes from the halo and never comes back (black symbols), remains in the halo (red symbols) and accretes back onto the galaxy (green symbols).
Symbols refer to individual NIHAO galaxies. Their sum is unity. The curves are median values in bins of $M_{\rm vir}$.
Each curve corresponds to the symbols of the same colour.
}
\label{fig_reaccreted_fraction}
\end{figure}

Fig.~\ref{fig_reaccreted_fraction} anticipates our most basic result.
In dwarves with $M_{\rm vir}\sim 2 - 4\times 10^{11}\,M_\odot$, out of $100\,M_\odot$ expelled from the galaxy, $\sim 30\%$ escapes from the halo and never comes back,
$\sim 20\%$ remains inside the halo until $z=0$, and $\sim 50\%$ accretes back onto the galaxy, although some of them will be blown out again.
In massive galaxies with $M_{\rm vir} > 3\times 10^{11}\, M_\odot$, the reaccreted fraction rises to $\sim 80\% $ and less than $10\% $ of the gas is expelled from the halo.
Thus, reaccretion is statistically the single most likely outcome at all halo masses but especially at the highest.
Our recycled fractions (green curve in Fig.~\ref{fig_reaccreted_fraction}) are consistent with the recycled fractions at $z=0$ from Fig.~9 of \citet{angles-alcazar_etal17}).

We have stressed the possibility that some of the reaccreted gas may be ejected again because all figures in this paragraph are for a single ejection event.
If a particle is ejected several times, the probability that it is expelled from the halo once will be larger than the $10\%$ to $30\%$ figure above.

\begin{figure*}

\begin{center}$
\begin{array}{cc}
\includegraphics[width=0.5\hsize]{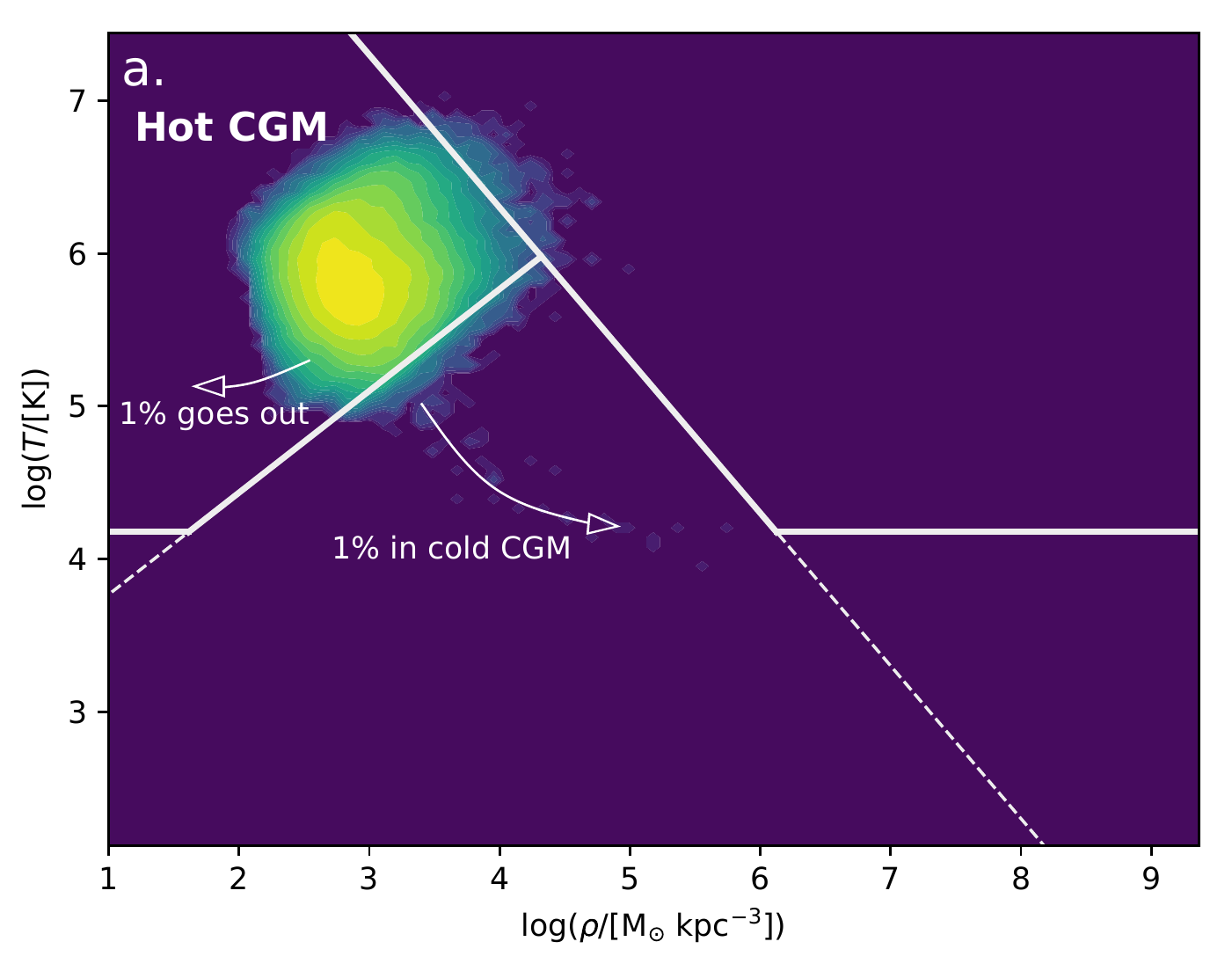} &
\includegraphics[width=0.5\hsize]{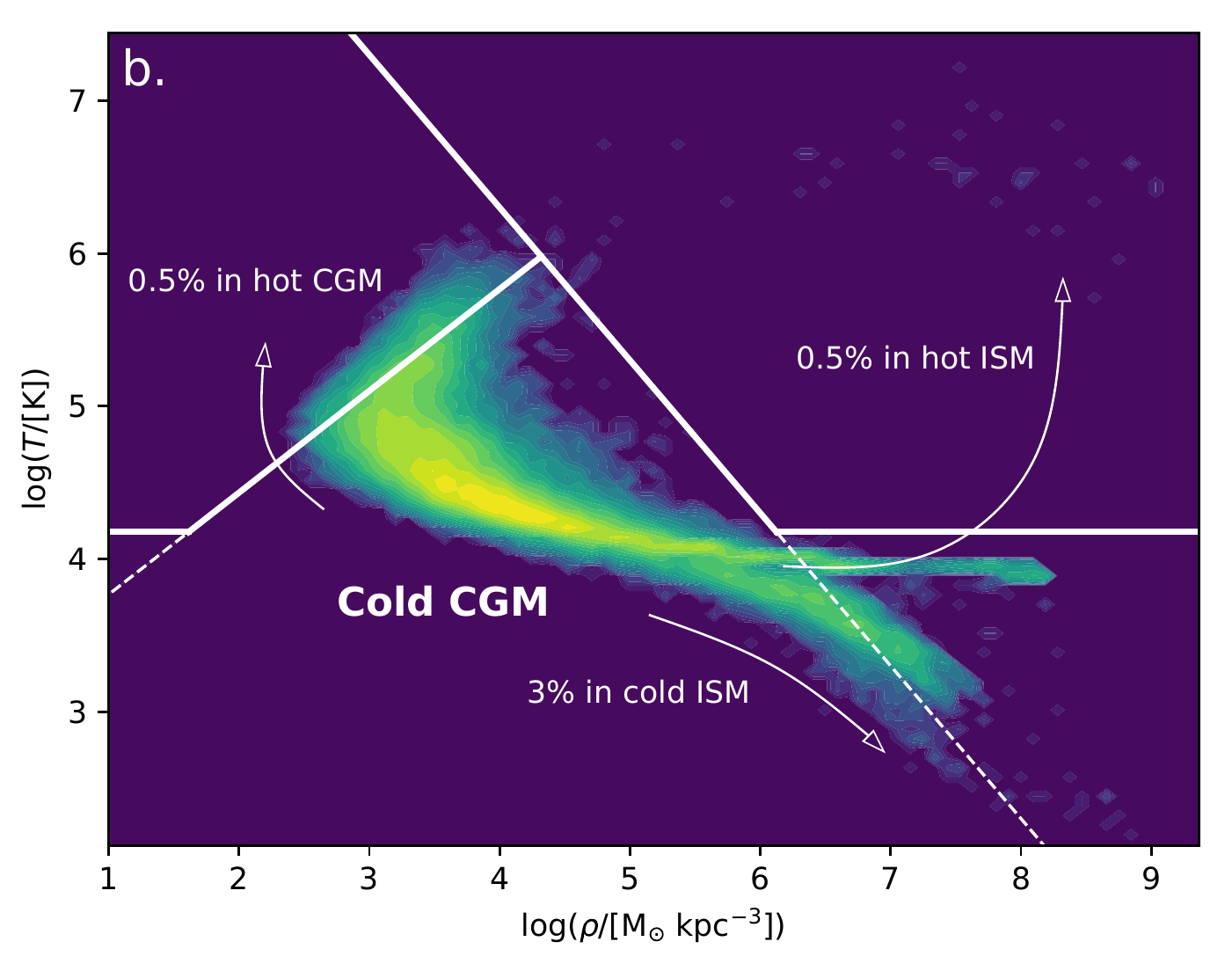} \\
\includegraphics[width=0.5\hsize]{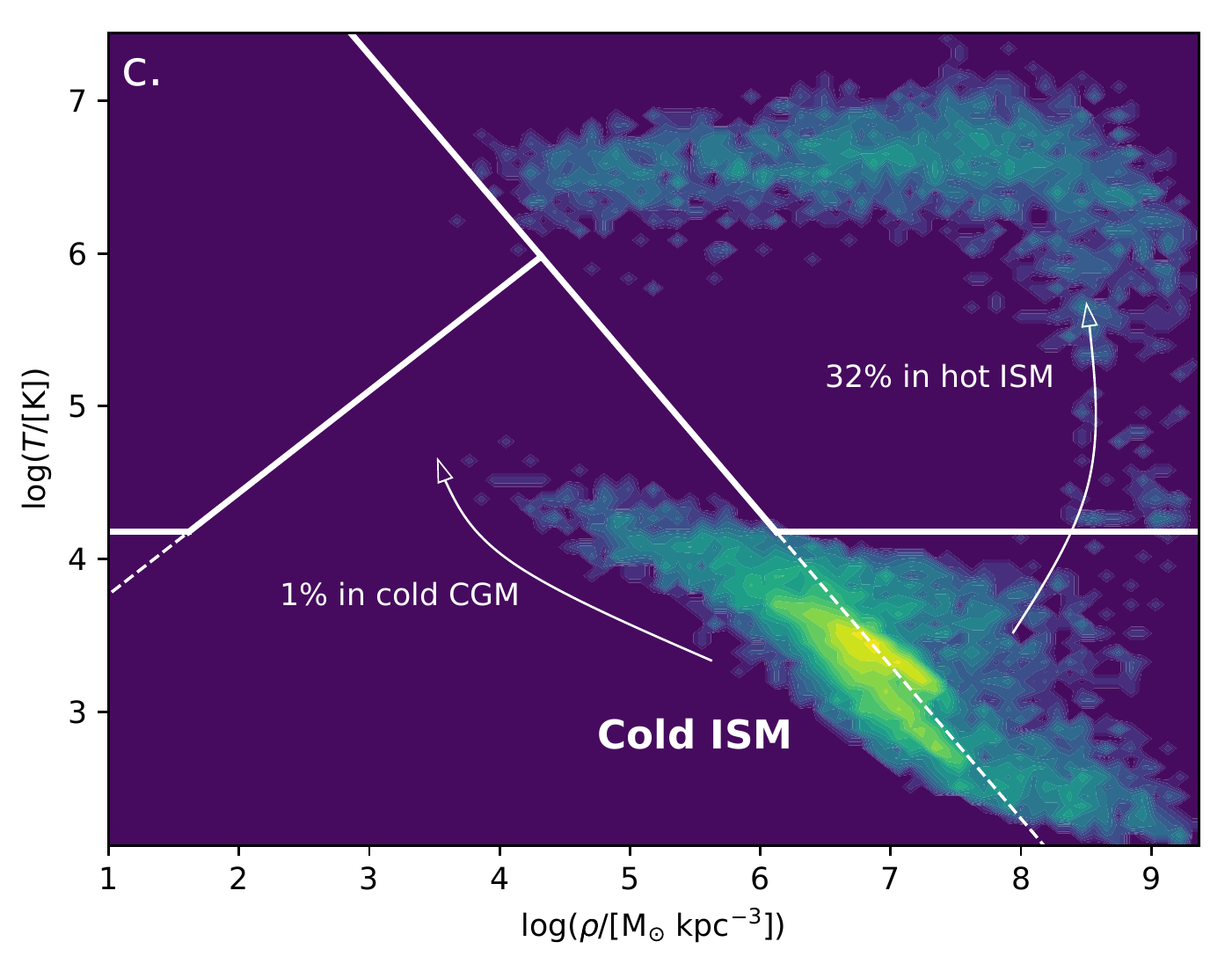} &
\includegraphics[width=0.5\hsize]{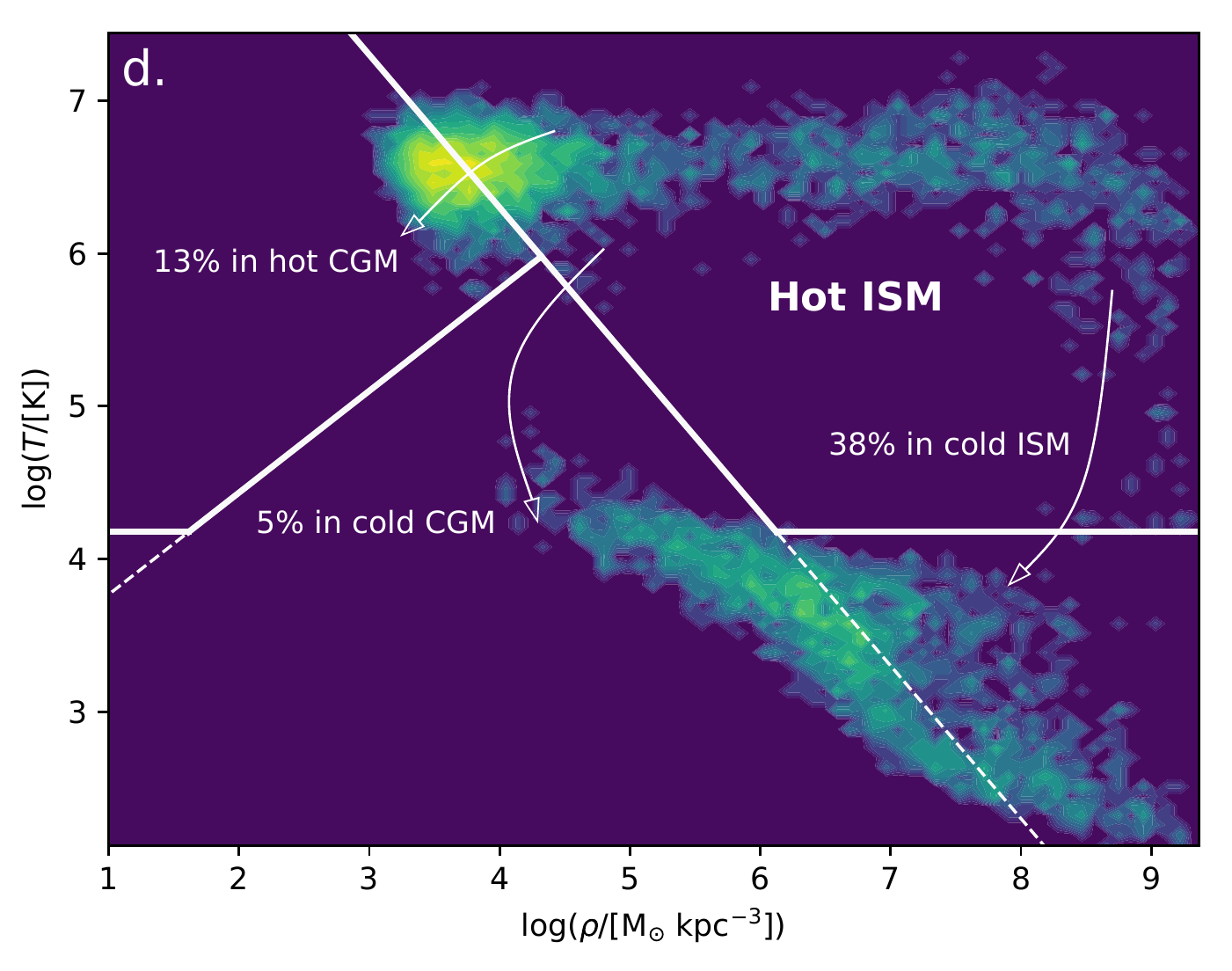} 
\end{array}$
\end{center}

\caption{Panels a, b, c and d show the position on a temperature-density diagram at $z=0$ of particles that were in the hot CGM, the cold CGM, the cold ISM and the hot ISM at $z=0.05$, respectively.
This figure is for the NIHAO simulation g1.12e12.
Percentages are relative to the mass of the phase from which a transition starts and to the time interval between $z=0.05$ and $z=0$ ($0.216\,$Gyr).
}
\label{fig_rhoT_motions}
\end{figure*}

To go beyond this final outcome, and study the pathways through which gas is expelled and reaccreted, we begin with our favourite galaxy g1.12e12 and consider all particles within the virial radius
at the penultimate output time $z=0.05$ (Fig.~\ref{fig_rhoT_motions}), starting from those in the hot CGM.

The hot CGM that cools in the time interval $\Delta t=0.216\,$Gyr between $z=0.05$ and $z=0$ is about $1\%$ of the mass in the hot CGM (Fig.~\ref{fig_rhoT_motions}a).
From this, one infers a cooling time of more than $20\,$Gyr, much longer than the one expected from the density, temperature and metallicity of the hot CGM
(the cooling function by \citealp{sutherland_dopita93} gives $\sim 0.7{\rm\,Gyr}$).
The difference arises because the cooling time that we infer from  Fig.~\ref{fig_rhoT_motions}a is not $(3/2)nkT/{\cal C}$, where ${\cal C}$ is the power radiated per unit volume, but 
rather   $(3/2)nkT/{\cal C-H}$, where ${\cal H}$ is the thermal power that SNe transfer to hot CGM per unit volume.

To test this explanation, we have computed the total power $L_{\rm X}$ radiated by CGM and we have compared it to the power $P_{\rm SN}=E_{\rm SN}\Psi_{\rm SN}{\rm SFR}$ from SNe, where
$E_{\rm SN}=10^{51}\,$erg, and $\Psi_{\rm SN}=1/(120\,M_\odot)$
(we consider the power from SNe only and neglect the contribution from pre-SN feedback because we assume that the latter is dissipated already at the level of the ISM).
For galaxy g1.12e12 at $z=0$, we find $L_{\rm X}=1.2\times 10^{42}{\rm\,erg\,s}^{-1}$ and $P_{\rm SN}=1.6\times 10^{42}{\rm\,erg\,s}^{-1}$.
Both quantities are likely to be overestimated:
the NIHAO simulations use the cooling function by \citet{shen_etal10}, which considers the presence of an ultraviolet background and gives lower cooling rates than the one by \citet{sutherland_dopita93};
the thermal power that SNe inject into the CGM will be lower than $P_{\rm SN}$ because some of the energy from SNe, too, may be dissipated at the level of the ISM and not reach the CGM.
Furthermore, $P_{\rm SN}/L_{\rm X}$ varies stochastically from one snapshot to the next.
However, within these uncertainties, our calculations support the conclusion that in massive spirals, such as g1.12e12, the cooling rate and the heating rate comparable, and thus balance each other off on average.
This explains the very low cooling rate that we infer from Fig.~\ref{fig_rhoT_motions}a.
We remark that this maintenance feedback is equivalent to the one by radio galaxies in clusters (e.g., \citealp{birzan_etal04}, \citealp{rafferty_etal06}, \citealp{cattaneo_etal09}, and references therein), only
spiral galaxies have more star formation and lower-mass black holes than elliptical galaxies, hence it is SNe that heat the gas.

Physically, SNe heat the hot CGM via two mechanisms: the hot ISM vents out of the galaxy, mixes with the hot CGM, and transfers its thermal energy to it; SN-driven outflows from the galaxy shock-heat the hot CGM. The cold CGM is located mainly in the plane of the disc and galactic wind mixes with the cold CGM only after radiating their thermal energy (in the descending part of the fountain). Hence, SN heating couples much more effectively to the hot CGM phase than to the cold CGM phase.

The prevalent motion in the cold CGM is accretion onto the galaxy  (Fig.~\ref{fig_rhoT_motions}b).
Particles that leave the cold ISM are, in their vast majority, particles that have been heated by SNe and transferred to the hot ISM (Fig.~\ref{fig_rhoT_motions}c).
The hot ISM is a highly unstable phase that needs constant replenishment.
By $z=0$, more than half of the particles in the hot ISM at $z=0.05$ have either cooled ($43\%$) or they have moved to the hot CGM ($13\%$; Fig.~\ref{fig_rhoT_motions}d). 
Hence, cooling is thrice more probable than becoming part of the hot CGM
(twice more probable if we consider only cooling in situ, that is, within the ISM).

\begin{figure}
\includegraphics[width=1.\hsize,angle=0]{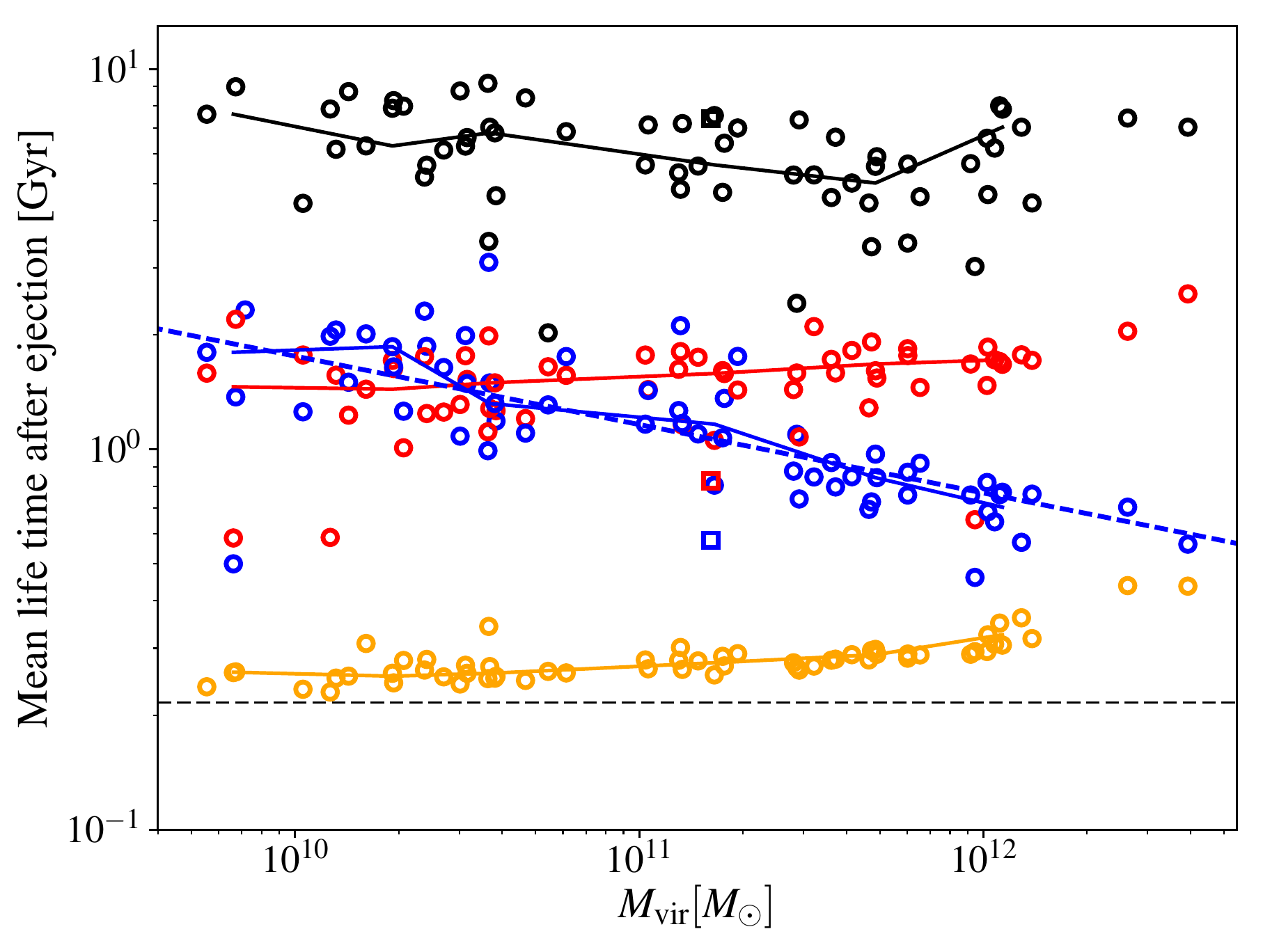}
\caption{Mean mass-weighted lifetime in the IGM (gas outside the virial radius, black symbols), the hot CGM (red symbols), the cold CGM (blue symbols) and the hot ISM (yellow symbols)
for particles that enter these phases after leaving the cold ISM (particles are not required to reaccrete onto the galaxy to be counted for this figure).
The solid lines show the median behaviour of the symbols of matching colour.
The dotted black line shows the time resolution of our analysis ($0.216\,$Gyr at all $z$). 
The open squares show the same results for a  simulation that was analysed with a time resolution of $0.013\,$Gyr. 
In this simulation, the lifetime in the hot ISM is $37\,$Myr.
The blue dotted line is a fit of the blue circles.
}
\label{fig_mean_time}
\end{figure}

These results are based on instantaneous mass exchange rates for one galaxy in our sample, which is not a random object but the second most massive galaxy of the NIHAO series.
To extend and generalise these findings to the entire NIHAO series,
we have computed the mean lifetimes in the hot ISM, the cold CGM, the hot CGM and the IGM ($r>r_{\rm vir}$) 
for particles that entered these phases after been ejected from the cold ISM (Fig.~\ref{fig_mean_time}).
By construction, these lifetimes cannot be longer than the age of the Universe and they cannot be shorter than the time interval $\Delta t=0.216\,$Gyr between two consecutive outputs (dashed horizontal line).

Fig.~\ref{fig_mean_time} shows that the hot ISM is so short-lived that its lifetime is not adequately resolved by our analysis.
To quantify the impact that this may have on our results, we have also considered one galaxy (g1.37e11) that was simulated with much closer output timesteps.
Using $\Delta t=13\,$Myr allows us to resolve the time spent in the hot ISM, which is larger than $30\,$Myr by construction, since we model SN feedback as in \citet{stinson_etal06}.
 
In Fig.~\ref{fig_mean_time}, the mean lifetimes for g1.37e11 are shown with open squares rather than open circles. The lifetime in the hot ISM is so short ($37\,$Myr) that the yellow square is outside the plotting range:
as soon as gas is allowed to cool, it does it in just $7\,$Myr.
The blue open square and the red open square lie a factor of two below the blue curve and the red curve, respectively, but their orders of magnitude are reproduced correctly.
The black open square is consistent with the black curve.

Our findings for the blue open square and the red open square should not surprise us.
Consider a particle that it is in a same phase at the consecutive output times $t_1$ and $t_2>t_1$; we conclude that this particle has been in that phase for a time $\ge t_2-t_1$.
If a snapshot at an intermediate time $t_3$ becomes available,  the additional information at $t_3$ may support this conclusion but also disprove it (if we find that at $t_3$ the particle is in a different phase).
Hence, a finer time resolution can only shorten the mean time spent uninterruptedly in any phase, and this is what Fig.~\ref{fig_mean_time} shows.
However, the integrated total time spent in any phase robust to this effect.

With the caveat above in mind, Fig.~\ref{fig_mean_time} shows that gas expelled from haloes remains at $r>r_{\rm vir}$ for a very long time (typically, until the simulation ends). In contrast,
{\it cold gas within haloes is usually reaccreted on a timescale of order}  $\sim 1\,$Gyr, which is shorter at high $M_{\rm vir}$.
This behaviour has a simple physical explanation.
Cold outflows are approximately ballistic. Hence, the reaccretion timescale $t_{\rm reaccr}$ is simply the dynamical time $r_{\rm vir}/v_{\rm vir}$ 
(in our cosmology, $r_{\rm vir}/v_{\rm vir}=1.9\,$Gyr at $z=0$) times a factor that depends on $v_0/v_{\rm vir}$ and concentration only, if we assume that the DM dominates the gravitational potential
($v_0$ is the initial speed at which gas is ejected)\footnote{In an NFW halo: 
$$
t_{\rm reaccr} = 2\frac{r_{\rm vir}}{v_{\rm vir}}\int_0^{x_{\rm max}} \frac{{\rm d}x}{\sqrt{2 \frac{c}{f\!(c)}\frac{{\rm ln}(1+x)}{x}}},
$$
where $x_{\rm max}$ is found by solving the energy-conservation equation:
$$
\frac{1}{2}\left(\frac{v_0}{v_{\rm vir}}\right)^2 = \frac{c}{f\!(c)}\left[1-\frac{{\rm ln}(1+x_{\rm max})}{x_{\rm max}}\right]
$$
and $f(c) = {\rm ln}(1+c)-c/(1+c)$.
}. Differences due to concentration are small and the dynamical time is the same for all haloes at a given redshift.
Thus, if $v_0/v_{\rm vir}$ were the same for all haloes, $t_{\rm reaccr}$ would be the same, too.
If, on the contrary, $v_0$ were constant, then $t_{\rm reaccr}$ would be a strongly decreasing function of $v_{\rm vir}$.
Its form would be more complicated than a simple power-law of $v_{\rm vir}$. However, if we assume an NFW gravitational potential with $c\sim 11 - 14$, a universal outflow speed implies $t_{\rm reaccr}\propto v_{\rm vir}^{-3}$ for
$0.4<v_{\rm vir}/v_0<0.8$ and $t_{\rm reaccr}\propto v_{\rm vir}^{-2}$ for $v_{\rm vir}/v_0\gsim 0.8$ ($v_{\rm vir}/v_0=0.8$ corresponds to $t_{\rm reaccr}\sim 0.4\,$Gyr at $z=0$).
Fig.~\ref{fig_vr_distribution} shows an intermediate situation: $v_0$ increases with $v_{\rm vir}$ but not as rapidly as $v_{\rm vir}$.
Hence, the reaccretion time decreases with $M_{\rm vir}$ but not as rapidly as assumed by \citet{henriques_etal13,henriques_etal15}, who used $t_{\rm reaccr}\propto M_{\rm vir}^{-1}$ independent of redshift.

If we fit the blue circles in  Fig.~\ref{fig_mean_time} with a power-law function of $M_{\rm vir}$, we find:
\begin{equation}
t_{\rm reaccr}\simeq 1.1\left(M_{\rm vir} \over 10^{11}\,M_\odot\right)^{-0.18}\,\rm Gyr.
\label{t_cycle_cold}
\end{equation}
(blue dashed line in Fig.~\ref{fig_mean_time}), although it is possible that Eq.~(\ref{t_cycle_cold}) may overestimate the reaccretion time for cold outflows by up to a factor of two 
(see above for a discussion of how time resolution can affect our results).
Eq.~(\ref{t_cycle_cold}) gives $t_{\rm reaccr}\sim 0.7{\rm\,Gyr}$ for $M_{\rm vir}=10^{12}\,M_\odot$, which becomes  $t_{\rm reaccr}\sim 0.4{\rm\,Gyr}$ if we renormalise the blue dashed line
so that it passes through the blue square. These reaccretion times are intermediate between those in \citet{oppenheimer_etal10} and Christensen et al. (2016; $t_{\rm reaccr}\sim 1\,$Gyr),
and those in Angl{\'e}s-Alc{\'a}zar et al. (2017; $t_{\rm reaccr}\sim 0.1 - 0.35\,$Gyr)\footnote{\citet{angles-alcazar_etal17} used a time resolution of $0.1\,$Gyr, higher than the one used to compute the blue circles in Fig.~\ref{fig_mean_time} ($0.216\,$Gyr)
but lower than the one used to compute the open square ($0.013\,$Gyr).}.
For comparison, the semiempirical model by \citet{mitra_etal15} reproduces the observations with $t_{\rm reaccr}\propto 0.52(M_{\rm vir}/10^{12}\,M_\odot)^{-0.45}(1+z)^{-0.32}\,$Gyr.

Discussing the lifetime in the hot CGM requires more subtlety.
Hot outflows can be reaccreted only if they cool down and it is well known (since \citealp{rees_ostriker77}, \citealp{silk77} and \citealp{binney77}) that the cooling time is longer at higher masses.
However, if we interpret the red curve in Fig.~\ref{fig_mean_time} as a measure of the radiative cooling time, then
Fig.~\ref{fig_mean_time} gives results that conflict with our estimate for g1.12e12 (Fig.~\ref{fig_rhoT_motions}a).
The reason is that gas can leave the hot CGM not only because it cools but also and more likely because it escapes from the halo (hot outflows). 
Hence, the typical $1 - 2\,$Gyr lifetime depicted by the red curve in Fig.~\ref{fig_mean_time} measures the time to leave the virial radius 
(a particle with an average speed of $100\,\rm km\,s^{-1}$ travels $100\,\rm kpc$ in about $1\,\rm Gyr$) more than the cooling time.

This conclusion may appear to conflict with our previous finding that most of the hot CGM has radial velocities lower than the escape speed (Fig.~\ref{fig_vr_distribution}). However, the escape speed in Fig.~\ref{fig_vr_distribution} is the escape speed from the center of the halo, while most of the hot CGM is at several galactic radii.

Massive spirals such as g1.12e12 are the only ones that have proper CGM composed of hydrostatic rather than outflowing gas and are not representative of the average NIHAO galaxy in this regard.
For these massive spirals, we have recomputed the mean lifetime in the hot CGM when we retain only particles that cool and return to the ISM\footnote{We stress that this lifetime gives a cooling time 
and not a reaccretion time because after cooling gas must still fall back onto the galaxy. Hence, the reaccretion time will be longer.}.
We expected that this calculation would give us a longer time than the one we found by considering all particles.
On the contrary, when we applied this analysis to g1.12e12 the survival time in the hot CGM decreased from $2\,$Gyr to $0.7\,$Gyr
(in even greater disagreement with the cooling time that we infer from Fig.~\ref{fig_rhoT_motions}a).
Our interpretation is that Fig.~\ref{fig_rhoT_motions}a gives the correct cooling time for the hot CGM of g1.12e12, but this cooling time is an average over two types a particles:
a few particles that cool within the radiative cooling time $(3/2)nKT/{\cal C}\sim 0.7\,$Gyr expected if there were no heating and many more that never do.

The final step of our analysis is to study the different paths through which baryons can be ejected and reaccreted and to compute  the masses processed through each of them.
For this purpose, we introduce the concept of cycle.
A cycle begins when a particle leaves the cold ISM.
It finishes when the particle returns to the cold ISM (closed cycle) or when the simulation reaches $z=0$ (open cycle), whichever occurs first.

The frequency of a type of cycle (e.g., the cycle cold ISM $\rightarrow$ hot ISM $\rightarrow$ hot CGM $\rightarrow$ cold CGM $\rightarrow$ cold ISM) is quantified by the mass processed through the cycle 
divided the total mass $M_{\rm gal,acc}$ processed by the galaxy throughout the lifetime of the Universe.
If all particles that accrete onto the galaxy went through this cycle once, then the cycle's frequency would be unity.
This is is impossible because some particles must form stars for feedback to occur. However, a cycle's frequency could still be equal to one
if half of the accreted particles formed stars and the other half went through the cycle twice. The cycle must be close for that to happen.

\begin{figure}
\includegraphics[width=1.\hsize,angle=0]{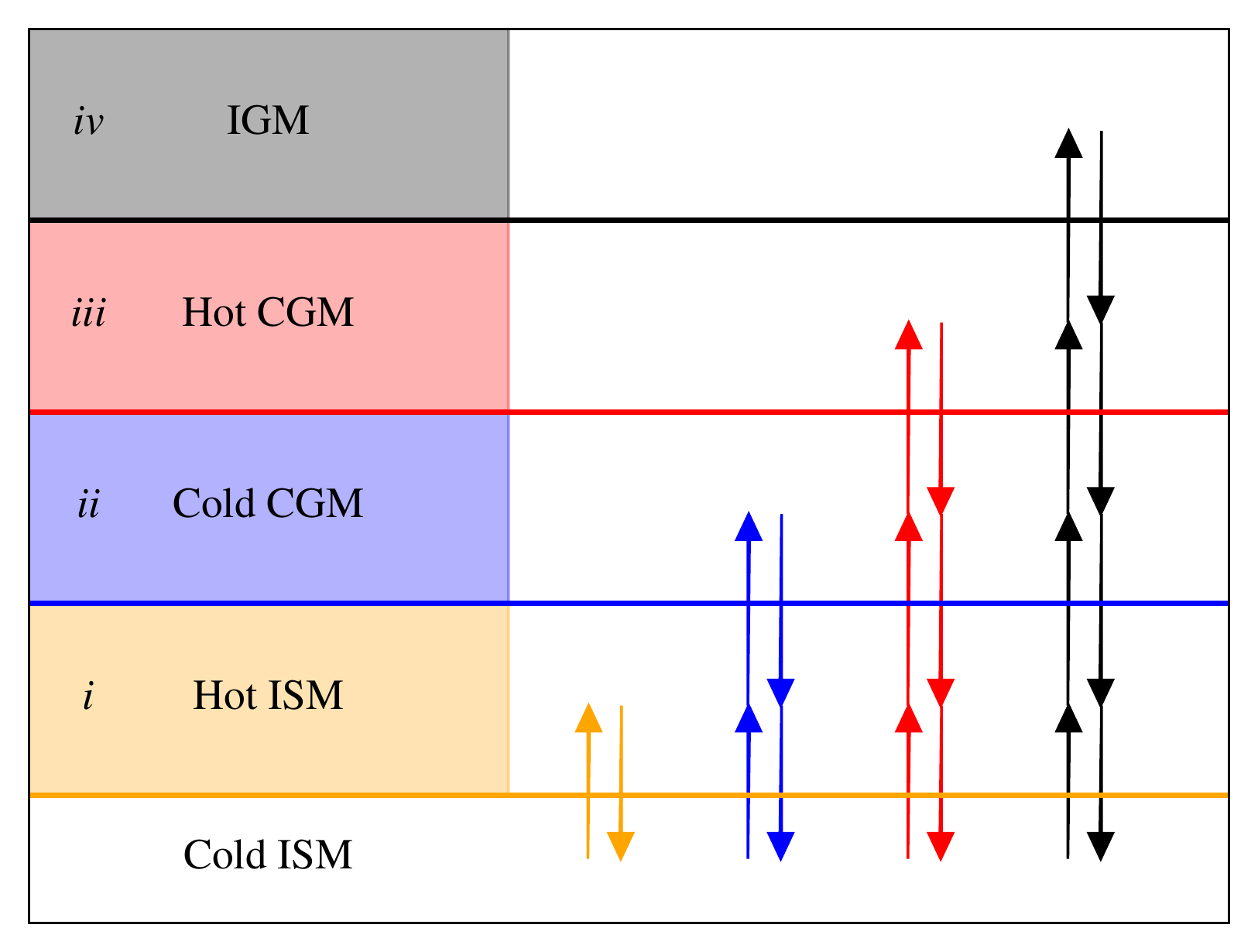}
\caption{Cycles considered in Fig.~\ref{fig_expel_maxphase}. All cycles start in the cold ISM.
They must pass through the cold ISM, the cold CGM, the hot CGM and the IGM to be classified as type {\it i}, {\it ii}, {\it iii} or {\it iv}, respectively,
but may miss intermediate phases. Cycles are subdivided into closed and open.
Closed cycles return to the cold ISM. Open cycles may or may not have a descending branch but have not returned to the cold ISM by $z=0$.}
\label{fig_scheme}
\end{figure}

The hot ISM that vents out of galaxies may end up in either the hot CGM or the cold CGM, and the hot CGM may become cold CGM, but gas cannot move from the cold CGM to the hot ISM or the hot CGM 
(at least not without being reaccreted onto the galaxy)\footnote{In reality, some particles do (Fig.~\ref{fig_rhoT_motions}a) but these transitions are not significant.}. 
Neither is it physically possible that gas escapes from the galaxy to the ISM without passing through either the cold or the hot CGM,
although it can happen in our analysis if the time resolution is inadequate. Hence, cycles can come in many different permutations, but there is a hierarchy of phases that allows to reduce them to four basic types
(Fig.~\ref{fig_scheme}): 
{\it i}) the cold ISM $\rightarrow$ hot ISM $\rightarrow$ cold ISM cycle, 
{\it ii}) the  cold ISM $\rightarrow$ hot ISM $\rightarrow$ cold CGM $\rightarrow$ cold ISM cycle (and its variant without the hot ISM),
{\it iii)} the  cold ISM $\rightarrow$ hot ISM $\rightarrow$ hot CGM $\rightarrow$ cold CGM $\rightarrow$ cold ISM cycle (and its variants without the hot ISM and/or the cold CGM), and 
{\it iv}) any cycle that passes through the IGM.
This classification has been formulated for closed cycles but we can easily generalise it to include open cycles by dropping the requirement that particles should return to the cold ISM.

\begin{figure}
\includegraphics[width=1.\hsize,angle=0]{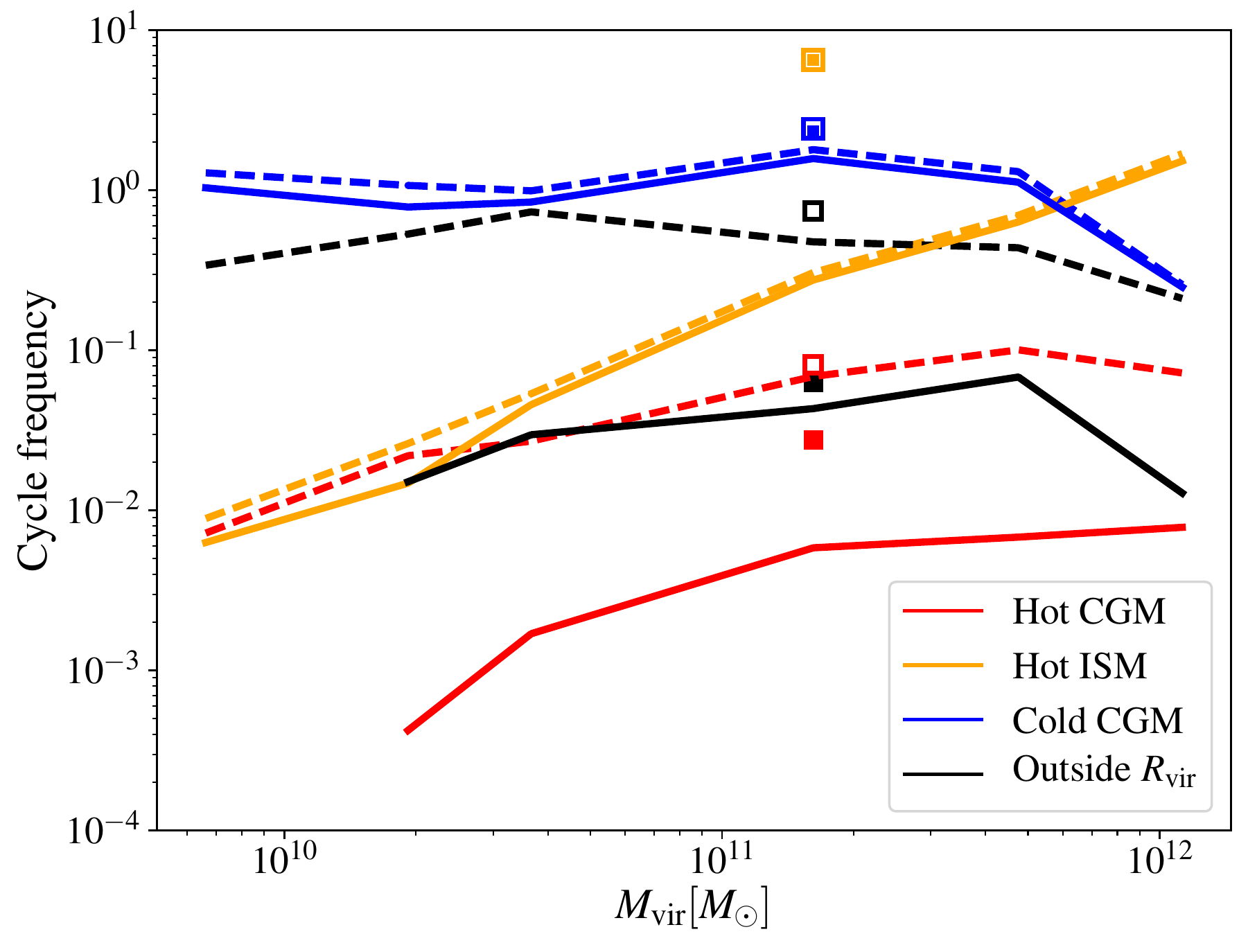}
\caption{Mass-weighted frequency of the different cycles an accreted particle could follow (as per Fig.~\ref{fig_scheme}).
All lines are medians based on following cycles with a time resolution of $0.216\,$Gyr.
Solid lines are for closed cycles. Dashed lines include the contributions of both close and open cycles.
Filled/open squares are equivalent to solid/dashed curves, respectively, for
the galaxy in which the baryon cycle has been followed with a resolution of $0.013\,$Gyr.
}
\label{fig_expel_maxphase}
\end{figure}

Fig.~\ref{fig_expel_maxphase} shows the average number of type {\it i}, {\it ii}, {\it iii} and {\it iv} cycles that an average particle makes as a function of $M_{\rm vir}$.
The solid curves are for closed cycles only. The dashed curve include open cycles. The yellow solid curve and the yellow dashed curve run close to each other.
So do the blue solid curve and the blue dashed curve. The implication is that 
type {\it i} and type {\it ii} cycles are predominantly closed. They correspond to internal heating of the ISM by SNe and the galactic fountain, respectively.
In contrast, the red solid curve and the red dashed curve are separated by about an order of magnitude.
So are the black solid curve and the black dashed curve.
Therefore, type {\it iii} and type {\it iv} cycles are predominantly open ({\it gas that mixes with the hot CGM or leaves the halo is unlikely to be reaccreted}).


\begin{figure}
\includegraphics[width=1.\hsize,angle=0]{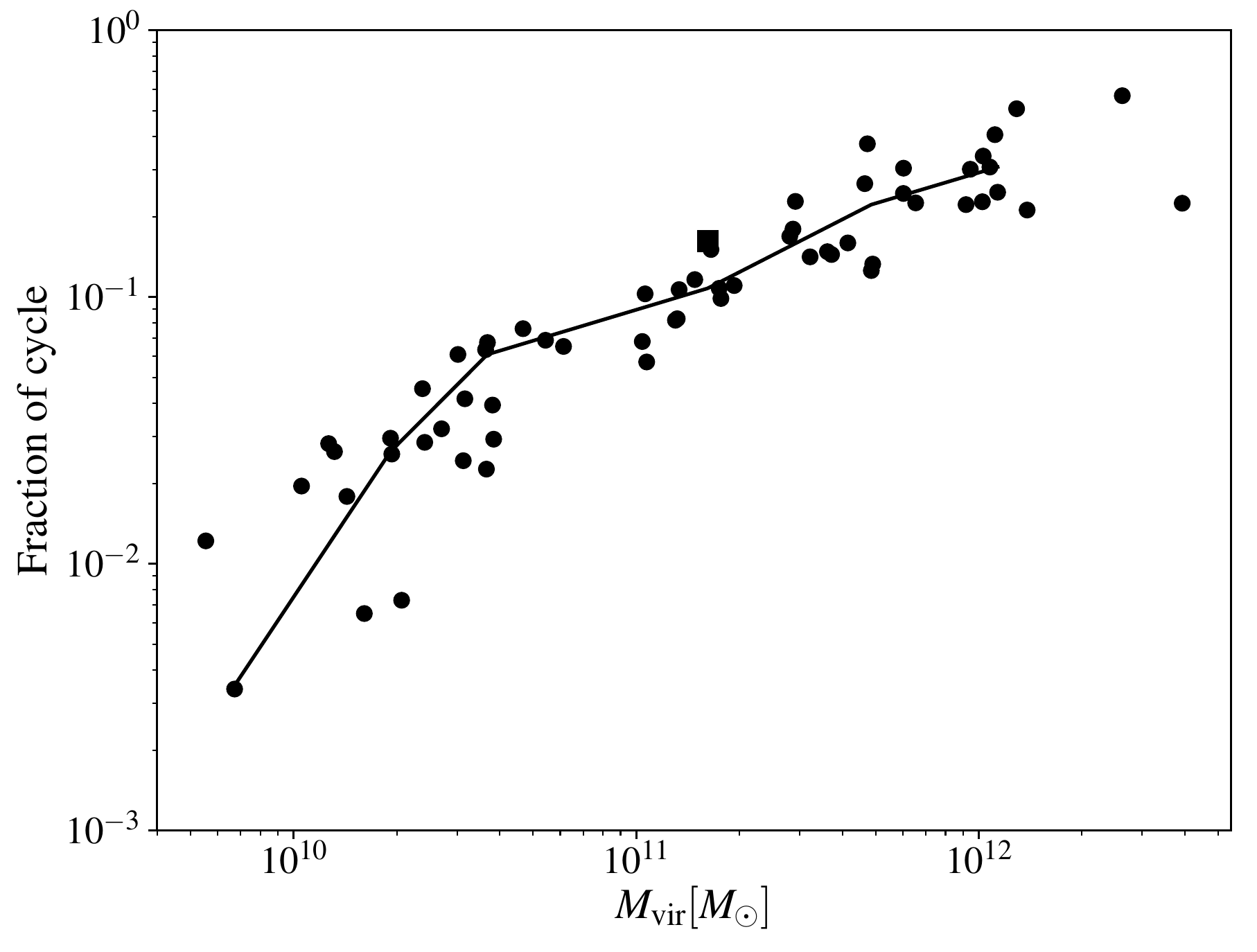}
\caption{Fraction of type {\it ii} cycles with a transient hot-ISM phase.
The square is for the simulation with snapshots separated by $\Delta t = 0.013\,$Gyr.}
\label{fig_type2}
\end{figure}

Type {\it ii} cycles cold ISM $\rightarrow$ (hot ISM) $\rightarrow$ cold IGM $\rightarrow$ cold ISM are the most common at all but the highest masses (Fig.~\ref{fig_expel_maxphase}, blue solid curve).
An average gas particle makes one or two of them after entering the galaxy (\citealp{angles-alcazar_etal17}, too, find that an average gas particle is ejected and reaccreted twice). 
The intermediate hot-ISM phase is missing in most cases but especially at low masses (Fig.~\ref{fig_type2}).
Hence, type {\it ii} cycles trace cold material entrained by the less massive hot component (also see Fig.~\ref{fig_galaxy_wind_edgeon}) rather than hot gas that has expanded and cooled.
As the galaxy analysed with $\Delta t=0.013\,$Gyr (black square in Fig.~\ref{fig_type2}) follows the same mean relation as the other NIHAO galaxies, we are confident
that Fig.~\ref{fig_type2} depicts a real physical result rather than our inability to detect short spells in the hot ISM.

Type {\it iv} cycles are the second most common variety. Their median frequency is about half per particle (Fig.~\ref{fig_expel_maxphase}, black dashed curve).
This result does not conflict with Fig.~\ref{fig_reaccreted_fraction}, which shows that only $10\%$ to $30\%$ of the ejected particles escape to $r>r_{\rm vir}$,
because Fig.~\ref{fig_reaccreted_fraction} refers to one ejection event, whereas Fig.~\ref{fig_expel_maxphase} refers to the lifetime of the Universe.
If we roll a dice once, the probability of an ace is one sixth.  If we roll it several times, the probability to obtain an ace at least once will be larger.

Type {\it i} cycles are rare at low masses but become predominant at $M_{\rm vir}>6\times 10^{11}\,M_\odot$ (Fig.~\ref{fig_expel_maxphase}, yellow solid curve).
Type {\it iii} cycles are rare at all masses but particularly at the lowest.
Even in massive spirals, they affect at most one particle out of ten (Fig.~\ref{fig_expel_maxphase}, red dashed curve).

As we have seen that the time resolution of our analysis is a potentially important issue, we have compared the curves in Fig.~\ref{fig_expel_maxphase} with our findings for g1.37e11, for which we have a snapshot
every $13\,$Myr. Better time resolution allows us to detect a much greater number of 
type {\it i} cycles (compare the yellow squares and the yellow curves), but these cycles are so short that, even collectively, their contribution is not significant anyway.
If it were, the hot ISM would contain a significant fraction of the ISM's total mass. On the contrary, in the NIHAO
simulations, the hot phase make up $\lsim 10\%$ of the mass ot the ISM on average, although the value can rise to $\gsim 20\%$ for massive spirals such as the Milky Way.
The frequencies of the cycles that matter most (those of type {\it ii} and {\it iv}) appear to be robust.
So are the frequencies of type {\it iii} cycles when we lump closed and open cycles together.
The only difference is the frequency of type {\it iii} closed cycles. Type {\it iii} cycles are more likely to be closed
in the simulation analysed with  $\Delta t=0.013\,$Gyr than they are on average in the simulations analysed with $\Delta t=0.216\,$Gyr
(the red filled square is much closer to the red open square than the red solid line is to the red dashed line).
However, type {\it iii} closed cycles are so rare that this difference concerns $2\%$ of the baryons at most.

Fig.~\ref{fig_expel_maxphase} is very important because it summarises and quantifies the two main ways through which the expulsion of matter reduces the masses of galaxies: fountain and ejective feedback.
In a typical galaxy, baryons are ejected and reaccreted once or twice and, in the process, $20\%$ to $80\%$  
(thus, half on average) are blown out of the virial radius and never come back.

\begin{figure}
\includegraphics[width=1.\hsize,angle=0]{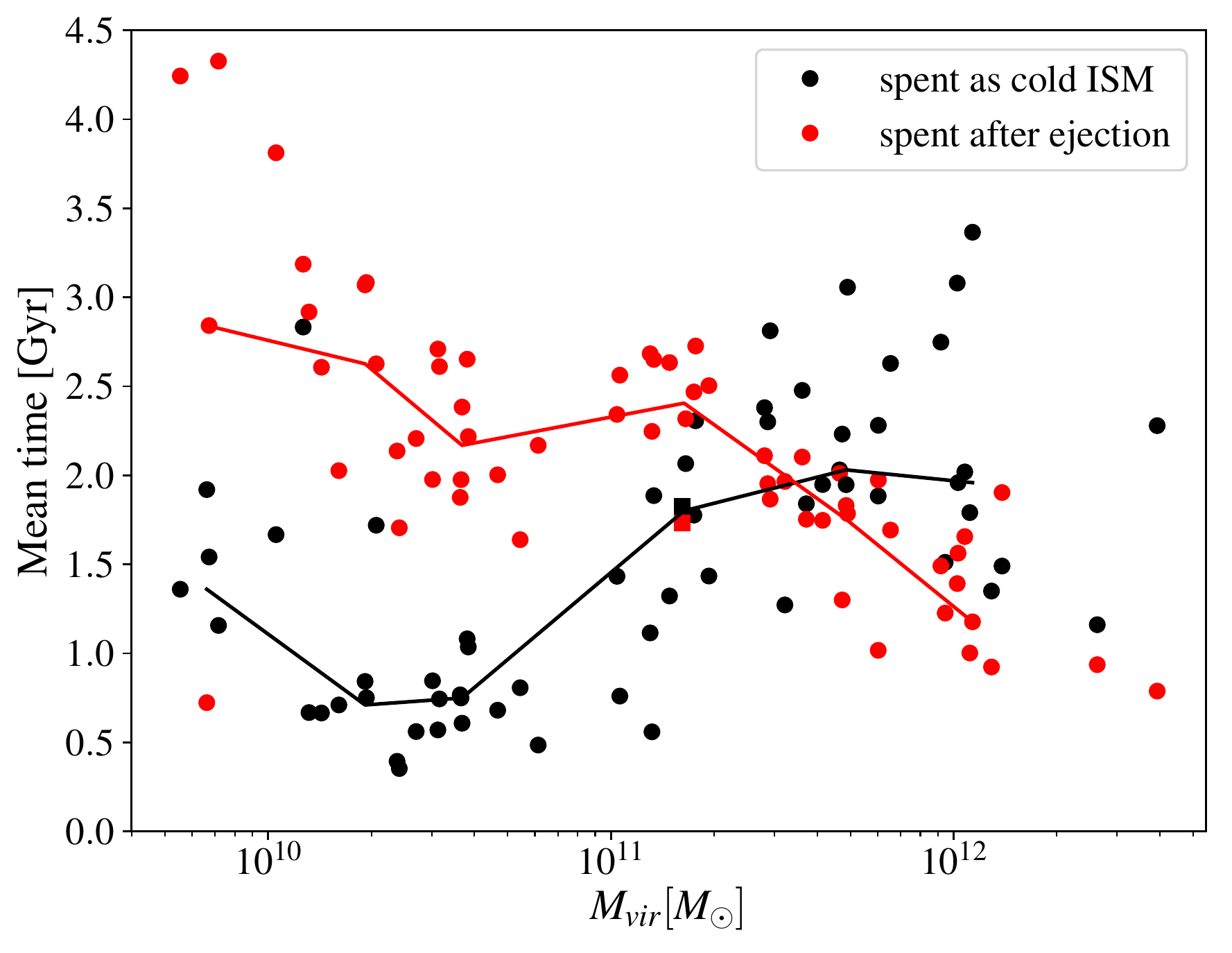}
\caption{Mean time spent in the cold ISM (black circles) and the fountain (red circles) for gas particles that accrete onto galaxies. 
The low $M_\star/M_{\rm vir}$ ratios of galaxies with $10^{10}\,M_\odot<M_{\rm vir}<10^{11}\,M_\odot$ are partly due to the short time spent in the cold ISM.
The galactic fountain reduces the time available for star formation by a factor of two to four.
}
\label{fig_mean_time2}
\end{figure}

Fountain feedback acts by reducing the time gas spends in the cold ISM (the only phase where it may form stars).
To quantify this effect, we have considered all gas particles that accrete onto a galaxy 
from when they were first accreted to their last time in the cold ISM, which can be when they were lasted ejected, when they formed stars, or when the simulation ended.
Fig.~\ref{fig_mean_time2} splits this time interval into two: the time spent in the cold ISM (black circles) and the time spent in closed cycles outside the cold ISM (red circles).
In dwarf galaxies with $10^{10}\,M_\odot<M_{\rm vir}<10^{11}\,M_\odot$, an average gas particle spends $\sim 2.5\,$Gyr in the fountain and less than $1\,$Gyr in the cold ISM.
Hence, it is available for star formation for only a quarter of the time.
At $M_{\rm vir}>3\times 10^{11}\,M_\odot$, the time available for star formation becomes $\gsim 50\%$ but the time spent in the fountain remains significant.
The galaxy that we have analysed at intervals of $\Delta t=0.013\,$Gyr spends $\sim 10\%$ less time in the fountain than an average galaxy of the same mass (based on our analysis with
$\Delta t=0.216$) but the difference is so small that it will not  change anything in our conclusions.

\section{Discussion and conclusion}

The last decades have seen considerable progress in our understanding of the cosmological context in which galaxies form and evolve.
These advancements are due to precise measurements of the cosmological parameters \citep{bennett_etal13,Planck14},
galaxy surveys of unprecedented size and depth (SDSS, 2dF, DEEP2, VVDS, UltraVISTA, CANDELS),
and hydrodynamic computer simulations \citep{hopkins_etal14,schaye_etal15,vogelsberger_etal14} but,
even with today's most powerful supercomputers, the range of scales that a simulation can explore is still limited.
Hence, feedback -- the way processes on the scale of stars and the ISM affect the evolution of galaxies at large --
has emerged as the main outstanding problem in galaxy formation.

All simulations need subgrid models to capture the physics they cannot resolve, but some models are more physical than others.
The FIRE collaboration \citep{hopkins_etal14} has run simulations that include type I and type II SNe, radiation pressure, stellar winds, photoionisation and photoelectric heating.
We cannot claim that our simulations follow the interaction of stars with the ISM to a comparable level, but neither is that our objective.
Our goal is not to study the physics of stellar feedback but rather to explore their effects on the formation and evolution of galaxies.

The NIHAO simulations \citep{wang_etal15} are perfectly suited for that purpose because they contain ninety galaxies that span the entire mass range
from $M_{\rm vir}=4\times 10^9\,M_\odot$ to  $M_{\rm vir}=4\times 10^{12}\,M_\odot$, and 
they have been shown to reproduce several properties of galaxies, which include stellar masses and SFRs \citep{wang_etal15},
disc sizes and the Tully-Fisher relation \citep{Dutton2017}, galaxy velocity function \citep{maccio_etal16} and the metallicity in the CGM \citep{Gutcke2016}


In this article, depletion times for {\sc Hi} and {\sc H}$_2$ \citep{boselli_etal14} and mass-loading factors from {\sc CO} and {\sc CH}$^+$ data \citep{cicone_etal14,falgarone_etal17}
have been added to the list.
The NIHAO simulations' ability to reproduce several observational constraints simultaneously gives us confidence that, even if the physics of stellar feedback are not resolved,
their effects are described correctly. This confidence is important. Without it, there would be no reason to believe that what we learn from our simulations applies to the Universe.

Many of our results confirm findings that were already know from previous studies \citep{munshi_etal13,hopkins_etal14,shen_etal14,mitchell_etal18,angles-alcazar_etal17},
but our work delineates a new coherent picture of stellar feedback as a multiscale and multiphase phenomenon.
Stellar feedback operates on all scales from the ISM to several virial radii.
Outflows have different components that we must separate for a meaningful comparison with observational data.
We summarise our key results proceeding from the smallest scales to the largest ones.

On ISM scales, feedback reduces the efficiency of star formation  via two mechanisms:
it prevents the condensation of atomic gas into molecular clouds and it moves gas from the cold ISM to a warm/hot phase with $T> 1.5\times 10^4\,$K.
The first process, which we call {\it anti-condensation feedback},
 is responsible for the long {\sc Hi}'s depletion times in dwarf galaxies
(\citealp{boselli_etal14} measures values longer than the Hubble time for galaxies with $M_\star\lsim 10^9\,M_\odot$).
The specificity of anti-condensation feedback is that it does not expel or heat the gas.
It simply maintains it to a temperature $T>1000\,$K at which it cannot reach the density required for star formation.
In contrast, the warm/hot ISM with $T>1.5\times 10^4\,$K is not a significant mass reservoir in any galaxies, but even less so in dwarves, where gas vents out as soon as it is heated.
Its significance is rather that it is the phase from which winds originate.

On galactic scales, galaxies with  $M_{\rm vir}\sim 3\times 10^{10}\,M_\odot$ can expel more than $95\%$ of the accreted baryons
but most of this gas  ($50\%$ to $70\%$) cools  very rapidly and falls back in a fountainlike motion
on a timescale of the order of $1\,$Gyr, which is shorter at higher masses (Eq.~\ref{t_cycle_cold}).
{\it Fountain feedback} reduces the masses of galaxies by a factor of two to four by reducing the time during which gas is available to form stars
(in dwarves, gas spends thrice more time in the fountain than it does in the galaxy).

On the scale of the CGM, SN heating offsets the radiative losses and explains the hot CGM's very long cooling time.
This {\it maintenance feedback} is analogous to the one by supermassive black holes in X-ray groups and clusters (e.g., \citealp{cattaneo_etal09} and references therein).

Expulsion from the halo has probability of $10\%$ to $30\%$  per ejection event in massive spirals and dwarves, respectively.
Hence, it is a less likely outcome than reaccretion through the fountain.
However,
after several ejection and reaccretion episodes $20\%$ to $70\%$ of the baryons that accrete onto galaxies escape from the virial radius ({\it ejective feedback}).
These baryons seldom come back.

On megaparsec scales, galactic winds can divert the infalling gas while it is still several virial radii away and prevent its accretion onto galaxies. We call this mechanism {\it pre-emptive feedback}.
The ability of SNe to reduce the gas inflow rate through a sphere of radius $r\sim 0.5 - 1r_{\rm vir}$ at low masses had already been remarked in cosmological simulations at $z\gsim 3$ by 
\citet{mitchell_etal18}.
However, we have been the first to show that outflows can disrupt  inflows and affect the kinematics of gas on scales as large as $6r_{\rm vir}$.
Our work shows that SNe and not photoionisation are responsible for the low baryonic fractions of haloes with  $M_{\rm vir}\lsim 3\times 10^{10}\,M_\odot$.

The low $M_\star/M_{\rm vir}$ ratios of galaxies are the result of not one but all the above mechanisms put together. 
Pre-emptive feedback reduces the baryons that can accrete onto a halo with $M_{\rm vir}\sim 3\times 10^{10}\,M_\odot$ to half of the universal baryon fraction.
Due to ejective feedback, only $30\%$ of these baryons are retained.
Only a quarter of this $30\%$ are in the galaxy because the rest are in the fountain, and, due to anti-condensation feedback,
the baryons in the galaxy form stars on a very long timescale (Fig.~\ref{fig_tsf}).
If an average gas particle spends only $\sim 1\,$Gyr inside the galaxy and the star-formation timescale is $\sim 10\,$Gyr,
then only a tenth of the baryons in the galaxy will be converted into stars.
Only by multiplying all these factors do we find $M_\star/M_{\rm vir}\sim 0.004$, which is what we expect from abundance matching for a galaxy with  $M_{\rm vir}\sim 3\times 10^{10}\,M_\odot$.

At even lower masses, star formation becomes sporadic. Ultradwarves are characterised by occasional bursts of star formation interspersed by long period of quiescence,
during which there are not enough SNe for outflows of any consequence. In these galaxies, anti-condensation feedback and pre-emptive feedback are the main mechanisms.

Massive spirals accrete nearly $100\%$ of the universal baryon fraction and expel only a small fraction of their baryons.
This is due to their deeper gravitational potential wells but also to the appearance of massive hot CGMa that confines winds within the central regions of DM haloes.
The fountain is still operating but the mass it processes is at least four times lower, so that the average gas particle spends more time in the galaxy than it does in the fountain;
and star formation is much more efficient in massive objects, which have a much higher molecular fraction ($t_{\rm sf}\sim 1\,$Gyr).
All these differences combined explain why anticondensation, fountain, ejective and pre-emptive feedback are much less effective in massive spirals than they are in dwarves.
Maintenance is the only process that becomes more important at high masses for the simple reason that low-mass haloes do not contain much hot gas, hence, it makes no difference whether it cools or not.
In massive spirals, we also observe the appearance of a hot ISM, which is practically absent in dwarf galaxies.

Our picture of SN feedback is very different from the frequently portrayed one, in which the ejected gas mixes with the hot CGM,
so that its reaccretion is determined by the hot CGM's radiative cooling time. 
We find that only a small fraction of the baryons that accrete onto galaxies are ejected and reaccreted through this route\footnote{The red solid curve in Fig.~\ref{fig_expel_maxphase}
suggests that this fraction is $<1\%$, although the real value could be larger if the black solid curve included a contribution from hot atmospheres that have spilled just a little outside $r_{\rm vir}$.}.

A physical picture of galactic winds requires considering their multiphase structure.
We propose a model where winds have a hot and a cold component.
Blastwaves from SNe form the hot ISM, which vents out of galaxies if it has enough pressure to overcome the depth of the gravitational potential well and the pressure of the surrounding CGM;
this is the origin of the hot phase.
The cold phase is mainly composed of cold gas that has been entrained by the hot wind.
The cold phase is completely dominant at low masses, since there is no need of powerful winds to lift gas from the shallow potential wells of dwarf galaxies.
At $M_\star\sim 10^{11}\,M_\odot$, however, the hot component is often dominant.
It is only when we separate the cold phase from the hot phase that we are able to reproduce the outflow rates from IRAM \citep{cicone_etal14} and ALMA \citep{falgarone_etal17} 
over a mass range that spans two orders of magnitude in $M_\star$.

Separating cold outflows and hot outflows is important not only for a correct comparison with observations but also because of their different destinies.
The hot component (gas that moves to $r>r_{\rm g}$ with $T>1.5\times  10^4\,$K) usually escapes from the halo and never comes back.
The cold component falls back onto the disc and forms the fountain.
Semianalytic models of galaxy formation could be improved by considering the presence of these two components and by using the NIHAO simulations to calibrate their importance:
\begin{equation} 
\begin{aligned} 
\eta_{\rm gal}=23\left({v_{\rm vir}/67{\rm\,km\,s}^{-1}}\right)^{-4.6} \\
\eta_{\rm gal}^{\rm cold}=23\left({v_{\rm vir}/67{\rm\,km\,s}^{-1}}\right)^{-5.6}
\end{aligned}
\label{eta_gal_conc}
\end{equation}
for  $v_{\rm vir}>67{\rm\,km\,s}^{-1}$ and 
\begin{equation}
\eta_{\rm gal}=\eta_{\rm gal}^{\rm cold}=23\left({v_{\rm vir}/67{\rm\,km\,s}^{-1}}\right)^{-2}
\label{eta_gal_cold_lowv}
\end{equation}
for  $v_{\rm vir}<67{\rm\,km\,s}^{-1}$.

Eqs.~(\ref{eta_gal_conc}) to~(\ref{eta_gal_cold_lowv}) in conjunction with Eq.~(\ref{t_cycle_cold}) for the cold component's reaccretion timescale
are the equations that we recommend as our reference model.
We anticipate, however, that a semianalytic model based on these equations will be successful only if its cooling rate is as low as the one that we measure in the NIHAO simulations,
that is, only if it also includes some form of maintenance feedback.

\section*{Acknowledgments}

We thank G.~A.~Mamon for assistance in computing the stellar masses of M82, NGC~3256, NGC~3628, NGC~253 and NGC~2146.
A.~C. would also like to thank C.~B.~Brook for assistance with {\sc pynbody}.
We thank N.~Bouch{\'e} for his useful comments.
The authors  gratefully acknowledge the Gauss Centre for Supercomputing e.V. (www.gauss-centre.eu) 
for funding this project by providing computing time on the GCS Supercomputer SuperMUC at Leibniz Supercomputing Centre (www.lrz.de) and 
the High Performance Computing resources at New York University Abu Dhabi.

\bibliographystyle{mn2e}

\bibliography{ref_av}

\label{lastpage}
\end{document}